\documentclass[11pt,reqno]{amsart}
\usepackage{xr}
\usepackage{color}
\usepackage{geometry}                % See geometry.pdf to learn the layout options. There are lots.
\usepackage{amsmath,amssymb,amsthm}
\usepackage{setspace}

\usepackage{amsmath,amssymb,amsthm}
\usepackage[authoryear]{natbib}
\usepackage{graphicx}
\usepackage{comment}
\usepackage{natbib}
\usepackage{hyperref}
\hypersetup{pdfstartview=FitB}

\usepackage{framed}

\theoremstyle{plain}
\newtheorem{theorem}{Theorem}[section]	
\newtheorem{lemma}{Lemma}[section]
\newtheorem{corollary}{Corollary}[section]
\newtheorem{proposition}{Proposition}[section]
\newtheorem{assumption}{Assumption}[section]
\theoremstyle{definition}

\DeclareMathOperator{\Cov}{Cov}
\DeclareMathOperator{\Var}{Var}

\newcommand{\Ep}{{\mathrm{E}}}

\numberwithin{equation}{section}

\begin{document}

%%
%% The title of the paper goes here.  Edit to your title.
%%

\title[Factor-Lasso and K-Step Bootstrap]{ The Factor-Lasso and K-Step Bootstrap Approach for
Inference in High-Dimensional Economic Applications}\thanks{Christian Hansen, Booth School of Business, University of Chicago, Chicago, IL 60637. Christian.Hansen@chicagobooth.edu.  Yuan Liao, Department of Economics, Rutgers University, New Brunswick, NJ 08901. yuan.liao@rutgers.edu.  The authors are grateful to Shakheeb Khan, Roger Moon, and seminar participants at the Australasian Meetings of the Econometric Society, University of Chile, National University of Singapore, Xiamen University, University of Toronto, and Stevens Institute of Technology for helpful comments. This material is based upon work supported by the National Science Foundation under Grant No. 1558636 and the University of Chicago Booth School of Business.  First version: June 2016.  This version: \today}

%%
%% Now edit the following to give your name and address:
%% 

\author{Christian Hansen}
\address{Booth School of Business, University of Chicago, Chicago, IL 60637}
\email{Christian.Hansen@chicagobooth.edu} 
%\urladdr{www.math.sc.edu/$\sim$howard} % Delete if not wanted.

\author{Yuan Liao}
\address{Department of Economics, Rutgers University, New Brunswick, NJ 08901}
\email{yuan.liao@rutgers.edu}
%\urladdr{www.math.sc.edu/$\sim$howard} % Delete if not wanted.

%%
%% If there is another author uncomment and edit the following.
%%

%\author{Second Author}
%\address{Department of Mathematics, University of South Carolina,
%Columbia, SC 29208}
%\email{second@math.sc.edu}
%\urladdr{www.math.sc.edu/$\sim$second}

%%
%% If there are three of more authors they are added in the obvious
%% way. 
%%

%%%
%%% The following is for the abstract.  The abstract is optional and
%%% if not used just delete, or comment out, the following.
%%%

\begin{abstract}
\singlespacing
We consider   inference about coefficients on a small number of variables of interest in a linear panel data model with additive unobserved individual and time specific effects and a large number of additional time-varying confounding variables.  We allow the number of these additional confounding variables  to be larger than the sample size,  and suppose that, in addition to unrestricted time and individual specific effects, these confounding variables are generated by a small number of common factors and high-dimensional weakly-dependent disturbances.  We allow that both the factors and the disturbances are related to the outcome variable and other variables of interest. To make informative inference feasible, we impose that the contribution of the part of the confounding variables not captured by time specific effects, individual specific effects, or the common factors can be captured by a relatively small number of terms whose identities are unknown.
 Within this framework, we provide a convenient computational algorithm based on factor extraction followed by lasso regression for  inference about parameters of interest and show that the resulting procedure has good asymptotic properties.   We also provide a simple k-step bootstrap procedure that may be used to construct inferential statements about parameters of interest and prove its asymptotic validity.  The proposed bootstrap may be of substantive independent interest outside of the present context as the proposed bootstrap may readily be adapted to other contexts involving inference after lasso variable selection and the proof of its validity requires some new technical arguments.  We also provide simulation evidence about performance of our procedure and illustrate its use in two empirical applications.
\end{abstract}

\maketitle

%%
%% LaTeX can automatically make a table of contents.  This is done by
%% uncommenting the following:
%%

%\tableofcontents

%%
%%  To enter text is easy.  Just type it.  A blank line starts a new
%%  paragraph. 
%%

%\onehalfspacing

\allowdisplaybreaks{ 
 

  
\section{Introduction}

%\textcolor{red}{Is there a condition on the rate for $q_n$ stated anywhere?  $q_n = .1/log(n)$ is used in the simulation and empirical examples.}

%\textcolor{red}{Is there better notation than $q_n$ for a tuning parameter and $\gamma_d$ and $\gamma_y$ for coefficients?}

%\textcolor{red}{We use $w$ to denote a variety of different things - fixed effects, weights, variables, a place-holder variable, ...  I think the usages are all pretty clear, but we might be able to come up with something better.}

%\textcolor{red}{We use $\Delta_{various}$ in stating the assumptions on the quality of factor estimation.  We use $\Delta_{\gamma}=\gamma_y^0-\tilde\gamma_y$ in Appendix \ref{App: FactorLasso}.  We use $\tilde \Delta_{d}$ in Appendix \ref{App: Thm 3.1} without definition.}

%\textcolor{red}{We need to decide what part of appendices to keep in main paper and what to put in supplement and make this division.}

%\textcolor{red}{We use $\lambda$ for the penalty parameter and $\lambda_{tm}$ to denote rows of $\Lambda_t$.  Is this confusing?}

%\textcolor{red}{I didn't notice a definition of $\|\cdot\|_F$.  Make sure it's somewhere.}

%\textcolor{red}{Make it clear that we are maintaining that the dimension of the factors is fixed early in the main text.}	

%\textcolor{red}{Should probably put in a $\blacksquare$ or some other notation to clearly denote the end of a proof to ease readability of the appendix.}

Data in which there are many observable variables available for each observation, i.e. ``high-dimensional data,'' are increasingly common and available for use in empirical applications.  Having rich high-dimensional data offers many opportunities for empirical researchers but also poses statistical challenges in that regularization or dimension reduction will generally be needed for informative data analysis.  The success of regularized estimation for either forecasting or inference using high-dimensional data relies on using a regularization device that is appropriate for the type of data at hand.  Effective regularization imposes substantive restrictions in estimation, and the resulting estimates can perform very poorly, for example suffering from large biases and missing important explanatory power, when the restrictions provide poor approximations to the underlying data generating mechanism.  It is thus important to employ regularized estimators that accommodate sensible beliefs about the structure of an underlying econometric model.

Two structures which are common in the econometrics literature are sparse structures and factor structures.  To fix ideas, consider the linear regression model
\begin{align}\label{SLR}
y_i = x_i'\theta + \varepsilon_i
\end{align}
where $i \leq n$ indexes individual observations, $y_i$ is the observed outcome of interest, $x_i$ is a $p \times 1$ vector of observed predictor variables with $p \gg n$ allowed, and $\varepsilon_i$ is a regression disturbance.  A sparse structure essentially imposes that the number of non-zero elements in $\theta$ is small.  
 Intuitively, the sparse structure relies on the belief that the majority of the explanatory power in the observed predictor variables concentrates within a small number of the available variables.  Estimators that are appropriate for sparse models, such as the lasso or variable selection procedures, may perform very poorly when the true model is ``dense'' in the sense that there are many non-zero elements in $\beta$ that are moderate in magnitude.

A commonly employed version of a linear factor model employs a different structure where
\begin{align}\label{SFR1}
y_i &= f_i'\xi + \varepsilon_i \\
\label{SFR2}
x_i &= \Lambda f_i + U_i.
\end{align}
Here $f_i$ denotes a latent $K\times 1$ vector of factors with $K \ll n$ that are important in determining both the observed outcome of interest, $y_i$, and the observed $p \times 1$,  with $p \gg n$, vector of observed predictor variables $x_i$.  Within this structure, one may  obtain estimates of the latent factors and build a model for the outcome given the extracted factors; see, e.g. \cite{bai03},   \cite{BN02},   \cite{SW02} and \cite{fan2016sufficient}.  
The basic factor model differs markedly from the sparse linear model (\ref{SLR}).  Importantly, data generated from model (\ref{SFR1})-(\ref{SFR2}) would generally result in a dense coefficient vector $\theta$ in the regression of $y_i$ onto $x_i$, and sparsity based estimation strategies would tend to perform poorly as a result. Of course, if the data generated by the sparse model (\ref{SLR}), common factors will generally not capture the explanatory power, which loads on a small number of the raw regressors, and pure factor-based estimation will perform poorly.

In this paper, we propose a simple model that nests both the sparsity-based and factor-based structures.  The model allows for the observed predictors to have a factor structure but then allows both the factors and the factor residuals, the $U_i$ in equation (\ref{SFR2}), to load in the outcome equation.   That is, we replace (\ref{SFR1}) with
\begin{align}\label{SFLR1}
y_i = f_i'\xi + U_i'\theta + \varepsilon_i
\end{align}
and impose that $\theta$ is sparse.  This model allows for the fact that all of the relevant explanatory power in the predictors may not be captured entirely by the factors but imposes that any predictive power not captured by the factors concentrates on only a few elements of the high-dimensional covariate vector.   (\ref{SFLR1}) clearly reduces to (\ref{SLR}) when there is no factor structure in $x$ and reduces to (\ref{SFR1}) when $\theta = 0$.  We note that this model shares much in common with factor augmented regression models, e.g. \cite{BN06} and \cite{BBE:FAVAR}, with the key points of departure being that we do not assume the identity of the additional variables to include in the model is known and that $U$ is not observable.
\cite{HMC:PFM} consider a model that shares the essential structure of (\ref{SFLR1}) and (\ref{SFR2}) from a Bayesian standpoint.   They show that forecasts obtained from their Bayesian estimator of this model tend to outperform forecasts obtained based on either pure sparsity or pure factor based models.

The first key contribution of the present paper is offering a practical estimation and inference procedure that is appropriate for inference in a panel generalization of the model given by equations (\ref{SFLR1}) and (\ref{SFR2}) and providing a formal treatment of the procedure's theoretical properties.  Specifically, we proceed by first running a factor extraction step and taking residuals from regressing each observed variable on the estimated factors.  Using these residuals, we then follow the lasso-based estimation and inference procedures of \cite{BCHK:FE}.  We show that the resulting estimator of parameters of interest specified ex ante by the researcher is asymptotically normal with readily estimated asymptotic variance under sensible conditions.   These conditions allow for errors in selection of the elements of the covariate vector that load after controlling for the factors but maintain sufficiently strong conditions to allow oracle selection of the number of factors.  The theoretical analysis is substantially complicated by the fact that factors and factor-residuals are not observed and must be extracted from the data.  The estimation error in this extraction then enters the second step nonlinear and non-smooth lasso problem.  Due to this complication, the theoretical results in this paper make use of arguments that, to our knowledge, are not implied by results existing in the current factor modeling literature or the current lasso literature. These results may be of some interest outside of the context of establishing the properties of our proposed inferential procedure.

By addressing estimation and inference in an interesting high-dimensional factor augmented regression model appropriate for panel data, our paper contributes to the rapidly growing literature dealing with obtaining valid inferential statements following regularized estimation.  See, for example, \cite{belloni2012sparse,BCK-LAD,BCY-honest,BCFH:Policy,belloni2014inference,BCHK:FE}, \cite{BerkEtAl13}, %\cite{CHS:AnnRev,DoubleML}, 
\cite{DoubleML}, \cite{dezeure2016high}, \cite{fan2001variable}, \cite{FL11}, \cite{Farrell:JMP}, \cite{GautierTsybakovHDIV}, \cite{AdaptiveGraphLasso}, \cite{FST14}, \cite{JM:ConfidenceIntervals}, \cite{Kozbur:FS}, \cite{LeeTaylorScreening}, \cite{LeeEtAlLasso}, \cite{LockhartEtAlLasso}, \cite{LoftusStepwise}, \cite{TaylorEtAlAdaptive}, \cite{vdGBRD:AsymptoticConfidenceSets}, \cite{WagerAthey:RandForTE}, and \cite{ZhangZhang:CI} for approaches to obtaining valid inferential statements in a variety of different high-dimensional settings.  

As a second main contribution, we offer a new, computationally convenient bootstrap method for inference. Specifically, we consider a bootstrap where we apply our main procedure, including extraction of factors and lasso estimation steps, within each bootstrap replication.  As computation of the lasso estimator within each bootstrap sample may be demanding, we explicitly consider a k-step bootstrap following \cite{andrews2002higher} where we start at the lasso solution from the full sample and then iterate a numeric solution algorithm for the lasso estimator for k-steps.  We make use of solution algorithms for which the updates are available in closed form which leads to fast computation.  We provide high-level conditions under which the procedure provides asymptotically valid inference for parameters of interest and provide specific examples with lower level conditions.  The k-step bootstrap we propose complements other bootstrap procedures that have been proposed for lasso-based inference, for example, \cite{BCFH:Policy}, \cite{chatterjee2011bootstrapping}, \cite{CCK:AOS13}, and \cite{dezeure2016high}.  In particular, the approach we take is something of a middle ground between  \cite{CCK:AOS13}, which uses resampling of model scores to avoid recomputation of the lasso estimator, and \cite{dezeure2016high} which fully recompute the lasso solution within each bootstrap replication.  The former approach is extremely computationally convenient and asymptotically valid but does not capture any finite sample uncertainty introduced in the lasso selection, while the latter may be computationally cumbersome due to fully recomputing the lasso solution within each iteration.  We note that the bootstrap procedure could be easily applied outside of the specific model considered in this paper and that the technical analysis here is new and may be of interest outside of the present context. 

The remainder of this paper is organized as follows.  In Section \ref{Sec: Model}, we describe the panel factor-lasso model and outline the basic algorithm we will employ for inference.  We present formal results for the proposed procedure in Section \ref{Sec: Asymptotics}, providing regularity conditions under which the estimator of parameters of interest is asymptotically normal and valid confidence statements may be obtained.  Section \ref{Sec: Bootstrap} describes the k-step bootstrap approach in detail and provides a formal analysis establishing the validity of the resulting bootstrap inference.  Section \ref{Sec: PC} discusses the factor extraction part of the problem in more detail and provides examples with accompanying low-level conditions that are sufficient for the high-level conditions stated in Section \ref{Sec: Asymptotics}.  We then provide simulation and empirical examples that motivate the model we consider and illustrate the use of the estimation procedure in Section \ref{Sec: Examples}.  Key proofs are collected in an appendix with additional results provided in a supplementary appendix.

Throughout the paper, we use  $\|\beta\|_1$ and $\|\beta\|_2$ to respectively denote the $\ell_1$- and $\ell_2$- norms of a vector $\beta$; use $\|A\|$  and $\|A\|_F$ to respectively denote the spectral  and Frobenius norms  of a matrix $A$. In addition,  denote by  $|J|_0$ as the cardinality of a finite set $J$. Finally, for two positive sequences $a_n, b_n$, we write $a_n\asymp b_n$ if $a_n=O(b_n) $ and $b_n=O(a_n).$

\section{Panel Factor-Lasso Model and Algorithm}\label{Sec: Model}

\subsection{Panel Partial Factor Model}

Consider the linear panel model defined by 
\begin{align}\label{PPFM:y}
y_{it} &= \alpha d_{it} + \xi_t'f_i + U_{it}'\theta + g_i + \nu_t + \epsilon_{it} \\
\label{PPFM:d}
d_{it} &= \delta_{dt}'f_i + U_{it}'\gamma_d + \zeta_i + \mu_t + \eta_{it} \\
\label{PPFM:x}
X_{it} &= \Lambda_t f_i +w_i+\rho_t+U_{it}
\end{align}
where $i \leq n$ indexes cross-sectional observations, $t \leq T$ indexes time series observations, $X_{it}$ are observed potentially confounding variables, and $d_{it}$ is an a priori specified ``treatment'' variable of interest.\footnote{Our results will immediately apply to the case where $d_{it}$ is an $r \times 1$ vector with $r$ fixed.  The analysis could also be extended to handle unbalanced panels where observations are missing at random.  We omit both cases for convenience.}  
$f_i$ is a $K \times 1$ vector of latent factors with time-varying $K \times 1$ factor loading vectors $\xi_t$, $\delta_{dt}$ and $p \times K$ dimensional factor-loading matrix $\Lambda_t$.   We will take asymptotics where $\dim(X_{it})=p \rightarrow \infty$, $n \rightarrow \infty$, and $T$ is either fixed or growing slowly relative to $n$ and $p$ when stating our formal results, and we explicitly allow for scenarios where $p \gg nT$.  $K$ is assumed fixed throughout the paper. Our object of interest is  the parameter $\alpha$ on the variable of interest $d_{it}$.  Following \cite{HMC:PFM}, we refer to the model (\ref{PPFM:y})-(\ref{PPFM:x}) as the ``panel partial factor model'' (PPFM).\footnote{\cite{HMC:PFM} consider a similar structure to (\ref{PPFM:y})-(\ref{PPFM:x}) which excludes the individual and time effects and imposes that the $\epsilon_{it}$ are i.i.d. Gaussian innovations.  They refer to this model as a partial factor model.}
 
 In each equation, we also allow for additive unobserved individual effects, $(g_i,\zeta_i,w_i')$, and time specific effects, $(\nu_t,\mu_t,\rho_t')$, where $g_i$, $\zeta_i$, $\nu_t$, and $\mu_t$ are scalars and $w_i$ and $\rho_t$ are $p \times 1$ vectors.  We do not impose structure over the individual or time specific effects and thus treat them as fixed effects.  This treatment differentiates the common factors, $f_i$, from the additive heterogeneity $(g_i,\zeta_i,w_i')$ and $(\nu_t,\mu_t,\rho_t')$ as we impose that the $f_i$ are common to each observed series with common, time-varying loadings. %while we allow additive unobserved heterogeneity to be essentially arbitrarily different across each series.  
Term $U_{it}$ represents the part of the observed $X_{it}$ that is orthogonal to the factors and unobserved time and individual specific heterogeneity.  We allow  $U_{it}$ to be correlated to both the outcome and variable of interest after controlling for the factors and individual and time fixed effects.  Because $p \gg nT$, we assume that $\theta$ and $\gamma_d$ are approximately sparse vectors.  We assume that observed right-hand side variables are strictly exogenous so that $\Ep[\eta_{it}|X_{i1},...,X_{iT}] = 0$ and $\Ep[\epsilon_{it}|X_{i1},...,X_{iT},d_{i1},...,d_{iT}] = 0$.  We will assume that data are iid across $i$ but allow for dependence across time periods, $t$.  Finally, we note that while we treat the PPFM defined in (\ref{PPFM:y})-(\ref{PPFM:x}) in the formal analysis, the results clearly apply to models without additive fixed effects or to a single cross-section.\footnote{We consider a cross-sectional instrumental variables version of the model in both a simulation and an empirical example.}

As noted in the Introduction,  the PPFM generalizes the high-dimensional sparse fixed effects model examined in \cite{BCHK:FE} and conventional large-dimensional factor models and factor augmented regression models; e.g. \cite{BN06}.  The PPFM is also related to, but distinct from, interactive fixed effects models as in, for example, \cite{bai,bai2014theory}, \cite{MW10,MW11},   \cite{pesaran} and \cite{Su13}.\footnote{See also \cite{bonhomme:manresa} for a distinct but related approach based on a grouped fixed effects model.}  A simple version of the interactive fixed effects model analogous to (\ref{PPFM:y}) is
\begin{align*}
y_{it} = \alpha d_{it} + z_{it}'\beta + \lambda_t f_i + \epsilon_{it}.
\end{align*}
In this model, $z_{it}$ represents a known, low-dimensional set of variables that must be controlled for in addition to the factors in $f_i$.  There appear to be three key distinctions between the high-dimensional PPFM and interactive fixed effects approaches.   First, we relax the assumption that one knows the exact identity of the variables that should appear in the model, $z_{it}$, by allowing for a high-dimensional set of observed potential confounds in $X_{it}$.  Second,  we allow for the fact that  the relevant explanatory power in the predictors may not be captured entirely by the factors, but impose that any predictive power not captured by the factors concentrates on only a few elements of the high-dimensional vector $U$.  Third, we directly extract estimates of the factors and $U$ from $X$ which can proceed even when $T$ is small.  Approaches to estimating the interactive fixed effects structure rely on having a large number of observations in both the time series and cross-sectional dimensions.  We thus view the PPFM and interactive fixed effects approaches as complementary where one may prefer one or the other depending on the nature of the data at hand.

\subsection{Estimation Algorithm} 
     
To estimate $\alpha$, we begin by taking the within transformation of all observed variables to remove the fixed effects.  To this end, let
$$
\tilde  z_{it}=z_{it}-\bar z_{\cdot t}- \bar z_{i \cdot} +\bar{\bar z}
$$  
for any variable $z_{it}$ where $\bar z_{\cdot t} = \frac{1}{n}\sum_{i=1}^{n} z_{it}$, $\bar z_{i \cdot} = \frac{1}{T}\sum_{t=1}^{T} z_{it}$, and $\bar{\bar z} = \frac{1}{nT}\sum_{i = 1, t = 1}^{n,T} z_{it}$.
We can then define a demeaned model as
\begin{align}\label{dPPFM:y}
\tilde y_{it} &= \alpha \tilde d_{it} + \tilde \xi_t'\tilde f_i + \tilde U_{it}'\theta + \tilde \epsilon_{it} \\
\label{dPPFM:d}
\tilde d_{it} &= \tilde \delta_{dt}'\tilde f_i + \tilde U_{it}'\gamma_d  +\tilde  \eta_{it},    \\
\label{dPPFM:x}
\tilde X_{it} &= \tilde\Lambda_t \tilde f_i + \tilde U_{it}.
\end{align}

After removing the additive unobserved heterogeneity, we estimate the (demeaned) latent factors as well as the (demeaned) idiosyncratic components from the model $\tilde X_{it}=\tilde\Lambda_t'\tilde f_i+\tilde U_{it}$.\footnote{We note that recovering the untransformed $f_i$ and $U_{it}$ would only be possible with large $n$ and $T$ due to the presence of the unrestricted fixed effects.  Fortunately, recovering these quantities is unnecessary within the model with common coefficients $\theta$, $\gamma_d$, and $\alpha$ as only $\tilde f_i$ and $\tilde U_{it}$ appear in the equations of interest.  This simplification would not generally occur if we allowed heterogeneity in $\theta$, $\gamma_d$, or $\alpha$ over time or across individuals, and we would need to consider incidental parameters bias introduced by removing the additive fixed effects.  We leave exploration of this issue to future research.}
Let $\widehat F=(\widehat f_1,...,\widehat f_n)'$ be  the $n\times K$ matrix of estimated factors. We shall discuss some examples of $\widehat F$ in Section \ref{Sec: PC}.
Given $\widehat F$, we estimate $\tilde\Lambda_t$ and $\tilde U_{it}$ by least squares: 
\begin{align}\label{Eq: hatU}
 \widehat\Lambda_t=\sum_{i=1}^n\tilde X_{it}\widehat f_i'(\widehat F'\widehat F)^{-1},\quad \widehat U_{it}=\tilde X_{it}-\widehat\Lambda_t \widehat f_i,\quad i\leq n, \ t\leq T.
\end{align}
Substituting (\ref{dPPFM:d}) to (\ref{dPPFM:y}), we obtain
\begin{eqnarray*}
\tilde y_{it} &=& \alpha(\tilde \delta_{dt}'\tilde f_i + \tilde U_{it}'\gamma_d  +\tilde  \eta_{it}) + \tilde \xi_t'\tilde f_i + \tilde U_{it}'\theta + \tilde \epsilon_{it}\cr
&:=&\tilde\delta_{yt}'\tilde f_i+\tilde U_{it}'\gamma_y+\tilde e_{it}.
\end{eqnarray*}
Now let $\tilde Y_t = (\tilde y_{1t},...,\tilde y_{nt})'$ and $\tilde D_t = (\tilde d_{1t},...,\tilde d_{nt})'$ denote the vectors of outcome and treatment variable within each time period $t$.  We next regress $\tilde Y_t$ and $\tilde D_t$ onto the extracted factors $\widehat F$ time period by time period to obtain $\{\widehat \delta_{yt}\}_{t=1}^{T}$ and $\{\widehat \delta_{dt}\}_{t=1}^{T}$ for
\begin{align}\label{Eq: deltas}
\widehat \delta_{yt}=(\widehat F'\widehat F)^{-1}\widehat F'\tilde Y_t \ \text{and} \ \widehat \delta_{dt}=(\widehat F'\widehat F)^{-1}\widehat F'\tilde D_t.
\end{align}
We then run the lasso with the residuals from each of these factor regressions as dependent variable and the estimated factor disturbances $\widehat U_{it}$ as predictors.  That is, we obtain
\begin{align}\label{lassoY}
\tilde \gamma_y &= \arg\min_{\gamma\in\mathbb{R}^{p}}\frac{1}{nT}\sum_{t=1}^T\sum_{i=1}^n( \tilde y_{it}-   \widehat\delta_{yt}'\widehat f_i-\widehat U_{it}'\gamma)^2+\kappa_n\|\widehat\Psi^y\gamma\|_1, \\
\label{lassoD}
\tilde \gamma_d &= \arg\min_{\gamma\in\mathbb{R}^{p}}\frac{1}{nT}\sum_{t=1}^T\sum_{i=1}^n( \tilde d_{it}-   \widehat\delta_{dt}'\widehat f_i-\widehat U_{it}'\gamma)^2+\kappa_n\|\widehat\Psi^d\gamma\|_1.
\end{align}
where the tuning parameter $\kappa_n$ is chosen as, for some $c_0>1$ and $q_n \rightarrow 0$, 
\begin{equation}\label{tuning}
\kappa_n=\frac{2c_0}{\sqrt{nT}}\Phi^{-1}(1-q_n/(2p)),\quad 
\log(q_n^{-1})=O( \log p)
\end{equation}
and $\widehat\Psi^y$ and $\widehat\Psi^d$ are diagonal penalty loading matrices.  Given the fixed effects panel structure, we use the clustered penalty loadings of \cite{BCHK:FE} which have diagonal elements defined as
\begin{align}\label{LoadingsY}
[\widehat\Psi^y]_{j,j} &= \sqrt{\frac{1}{nT}\sum_{i=1}^{n}\sum_{t=1}^{T}\sum_{t'=1}^{T} \widehat U_{it,j} \widehat U_{it',j} \widehat e_{it} \widehat e_{it'} }     \\
\label{LoadingsD}
[\widehat\Psi^d]_{j,j} &= \sqrt{\frac{1}{nT}\sum_{i=1}^{n}\sum_{t=1}^{T}\sum_{t'=1}^{T} \widehat U_{it,j}\widehat U_{it',j}  \widehat \eta_{it} \widehat \eta_{it'} }
\end{align}
where $\widehat e_{it}$ is an estimator of $\tilde e_{it} = \tilde y_{it} - \tilde \delta_{yt}'\tilde f_i- \tilde U_{it}'\gamma_y$ and $\widehat \eta_{it}$ is an estimator of $\tilde\eta_{it} = \tilde d_{it} - \tilde \delta_{dt}'\tilde f_i - \tilde U_{it}'\gamma_d$.   \footnote{We obtain $\widehat e_{it}$ and $\widehat \eta_{it}$ through an iterative algorithm similar to that of \cite{belloni2014inference}, which starts from a preliminary estimate. In addition,   we use $c_0 = 1.1$ and $q_n = .1/\log(n)$ in the simulation and empirical examples.  }    

For the final step, we adopt the post-double-selection procedure of \cite{belloni2014inference}.  Let $\widehat J = \{j \leq p: \tilde \gamma_{y,j} \neq 0 \} \cup \{j \leq p: \tilde \gamma_{d,j} \neq 0\}$, and let $\widehat U_{it, \widehat J}$ be a subvector of $\widehat U_{it}$ whose elements are $\{\widehat U_{it, j}: j \in \widehat J\}$.  We then run the regression of 
$\tilde y_{it}-\widehat\delta_{yt}'\widehat f_i$ on $\widehat U_{it, \widehat J}$ and $\tilde d_{it} - \widehat\delta_{dt}'\widehat f_i$ on $\widehat U_{it, \widehat J}$ and obtain  
\begin{align}\label{PDSy}
\widehat\gamma_y &= ( \sum_{i=1}^n\sum_{t=1}^T\widehat U_{it,\widehat J}\widehat U_{it,\widehat J}')^{-1} \sum_{i=1}^n\sum_{t=1}^T\widehat U_{it,\widehat J}(\tilde y_{it}-\widehat\delta_{yt}'\widehat f_i), \\
\label{PDSd}
\widehat\gamma_d &= ( \sum_{i=1}^n\sum_{t=1}^T\widehat U_{it,\widehat J}\widehat U_{it,\widehat J}')^{-1} \sum_{i=1}^n\sum_{t=1}^T\widehat U_{it,\widehat J}(\tilde d_{it}-\widehat\delta_{dt}'\widehat f_i).
\end{align}
The final estimator of $\alpha$ is then given by 
\begin{align}\label{alphahat}
\widehat\alpha=(\sum_{i=1}^n\sum_{t=1}^T\widehat\eta_{it}^2)^{-1}\sum_{i=1}^n\sum_{t=1}^T\widehat\eta_{it}\widehat e_{it}
\end{align}
where $\widehat e_{it} = \tilde y_{it}- \widehat\delta_{yt}'\widehat f_i-\widehat U_{it,\widehat J}'\widehat\gamma_y$ and $\widehat \eta_{it} = \tilde d_{it}- \widehat\delta_{dt}'\widehat f_i-\widehat U_{it,\widehat J}'\widehat\gamma_d$
are the residuals from the regressions specified in (\ref{PDSy}) and (\ref{PDSd}).

The estimator $\widehat\alpha$ can be expressed more compactly in matrix form.  Write   
\begin{align*}
\tilde Y = \left (\begin{matrix}
\tilde Y_1 \\
\vdots \\
\tilde Y_T
\end{matrix}\right)_{nT\times 1}, \quad \tilde D= &\left (\begin{matrix}
\tilde D_1\\
\vdots\\
\tilde D_T
\end{matrix}\right)_{nT\times 1}, \quad\widehat U_{\widehat J}=\left(\begin{matrix}
\widehat U_{1, \widehat J}\\
\vdots\\
\widehat U_{T, \widehat J}
\end{matrix}\right)_{nT\times |\widehat J|_0}, \\
\widehat e=\begin{pmatrix}
\widehat e_1\\
\vdots\\
\widehat e_T
\end{pmatrix}_{nT\times 1},& \ \text{and} \ 
\quad  \widehat \eta=\begin{pmatrix}
\widehat \eta_1\\
\vdots\\
\widehat \eta_T
\end{pmatrix}_{nT\times 1}.
\end{align*}
 In addition, for a matrix $A$, define $M_A=I- A(A'A)^{-}A'$, where $(A'A)^{-}$ represents a generalized inverse of $A'A$. Then it is straightforward to verify that 
$$\widehat e = M_{\widehat U_{\widehat J}}(I_T\otimes M_{\widehat F})\tilde Y, \ \text{and} \
\widehat\eta = M_{\widehat U_{\widehat J}}(I_T\otimes M_{\widehat F})\tilde D
$$
are the estimated residuals $(\tilde e_1',...,\tilde e_T')'$ and $(\tilde \eta_1',...,\tilde \eta_T')'$ defined above.
Then
$$
\widehat\alpha=(\widehat\eta'\widehat\eta)^{-1}\widehat\eta'\widehat e.
$$

Note that the estimator $\widehat\alpha$ is numerically equivalent to the coefficient on $\tilde d_{it}$ in the regression of $\tilde y_{it}$ on $\tilde d_{it}$, $\widehat f_{i}$ interacted with time dummy variables, and $\widehat U_{it, \widehat J}$.  In Theorem \ref{t3.1} of  the next section, we verify that inference for $\widehat\alpha$ can proceed using the output from this OLS regression as long as clustered standard errors (e.g. \cite{arellano:feinf}, \cite{bdm:cluster}, and \cite{hansen:cluster}) are used.

The following algorithm summarizes the estimation strategy detailed above.

\begin{framed}

\noindent Algorithm (\textbf{Factor-Lasso Estimation of $\alpha$}.) 

\begin{enumerate}

\item Obtain $\{\widehat f_i, \widehat U_{it}\}_{i\leq n, t\leq T}$ by extracting factors from the model $\tilde X_{it}=\tilde\Lambda_t'\tilde f_i+\tilde U_{it}$. 

\item For $\widehat \delta_{yt}$ and  $\widehat \delta_{dt}$ defined in (\ref{Eq: deltas}), run the cluster-lasso programs (\ref{lassoY}) and (\ref{lassoD}) to obtain $\tilde \gamma_y$ and $\tilde \gamma_d$.  

\item Obtain the estimator $\widehat\alpha$ and corresponding estimated standard error as the coefficient on $\tilde d_{it} - \widehat\delta_{dt}'\widehat f_i$ and associated clustered standard error from the regression of $ \tilde y_{it}- \widehat\delta_{yt}'\widehat f_i-\widehat U_{it,\widehat J}'\widehat\gamma_y$ on $ \tilde d_{it}- \widehat\delta_{dt}'\widehat f_i-\widehat U_{it,\widehat J}'\widehat\gamma_d$ where $\widehat U_{it, \widehat J}$ is the subvector of  $\widehat U_{it}$ whose elements are $\{\widehat U_{it, j}: j\in\widehat J\}$.  

\end{enumerate}

\end{framed}

\section{Assumptions and Asymptotic Theory}\label{Sec: Asymptotics}

In this section, we present a set of sufficient conditions under which we establish asymptotic normality of $\widehat\alpha$ and provide a consistent estimator of its asymptotic variance.   Throughout we consider sequences of data generating processes (DGPs) where $p$ increases as $n$ and $T$ increase and where 
%$\Lambda_t$, $\theta$, and $\gamma_d$ 
model parameters are allowed to depend on $n$ and $T$.  We suppress this dependence for notational simplicity.  We use the term ``absolute constants'' to mean given constants that do not depend on the DGP.
 
\subsection{Regularity Conditions}

Write  $\epsilon_t=(\epsilon_{1t},...,\epsilon_{nt})'$, $\eta_t=(\eta_{1t},...,\eta_{nt})'$, and $U_{t}=(U_{1t}',...,U_{nt}')'.$  Similarly, let $\epsilon_i=(\epsilon_{i1},...,\epsilon_{iT})'$ and $\eta_i=(\eta_{i1},...,\eta_{iT})'$,  $U_i=(U_{i1}',...,U_{iT}')'$.

Our first two conditions collect various restrictions on dependence, tail behavior, and moments of the unobserved features of the model.   We assume there are positive absolute constants $C_1,C_2$ and $C_3$ such that the following assumption holds. 

\begin{assumption}[DGP]\label{a3.2} 
(i) $\{f_i, \eta_{i}, \epsilon_{i}, U_{i}\}_{i\leq n}$ are  independent and identically distributed  across $i=1,2,..., n$ and satisfy
$$
E(\eta_{i}|\epsilon_{i}, U_{i}, f_i)=0,\quad E(\epsilon_{i}| \eta_{i}, U_{i},f_i )=0,\quad E(U_{i}|\eta_{i}, \epsilon_{i}, f_i)=0.
$$
In addition,  given $\{f_i\}_{i\leq n}$, the sequence   $\{U_{i}, \eta_{i}, \epsilon_{i}\}_{i\leq n, t\leq T}$ is  also  conditionally independent across $i$.

(ii)   Given $\{f_i\}_{i\leq n}$, the sequence    $\{U_{t}, \eta_{t}, \epsilon_{t}\}_{t\leq T}$ is   stationary across $t$, and satisfies a strong-mixing condition.  That is, there exists an  absolute constant $r>0$  such that  for all $T\in\mathbb{R}^+$,
 $$\sup_{A\in\mathcal{F}_{-\infty}^0, B\in\mathcal{F}_{T}^{\infty}}|P(A)P(B)-P(AB)|\leq \exp(-C_1T^{r}),$$%
 where $\mathcal{F}_{-\infty}^0$ and $\mathcal{F}_{T}^{\infty}$ denote the $\sigma$-algebras generated by $\{(U_t, \eta_t,\epsilon_t): -\infty\leq t\leq 0\}$ and  $\{(U_t, \eta_t,\epsilon_t): T\leq t\leq \infty\}$ respectively.  

(iii) %Weak conditional dependence:  There is an absolute constant $C>0$ such that, 
Almost surely,
 $$\max_{i\leq n, m\leq p,t\leq T}\sum_{k=1}^p\sum_{s=1}^T|E( U_{it,k}\  U_{is,m}|f_i,\epsilon_i,\eta_i)|<C_2.$$

(iv) %Sub-Gaussian tail:   
For any $s>0$, $i\leq n$, $j\leq p$ and $k\leq K$,
\begin{eqnarray*}&&P(|U_{it,j}|>s)\leq\exp(-C_3s^{2}),\quad 
P(|f_{ik}|>s)\leq\exp(-C_3s^{2}),\cr
&&P(| \eta_{it} |>s)\leq\exp(-C_3s^{2}),\quad P(| \epsilon_{it} |>s)\leq\exp(-C_3s^{2}).
\end{eqnarray*} 
 
% $ \max_{it} \sum_s|E(\eta_{it}\eta_{is}|F)|=O(1)$,  $ \max_{it} \sum_s|E(\epsilon_{it}\epsilon_{is}|F)|=O(1)$,   almost surely in $F$. 
 
 (v)    Let $\theta_m$ and $\gamma_{d,m}$ be the $m^{\text{th}} $ entries of $\theta$ and $\gamma_d$, and $\lambda_{\text{tm}}'$ be the $m^{th}$ row of $\Lambda_t$. 
 $$
 |\alpha|+ \max_{t\leq T}(\|\xi_t\|+\|\delta_{dt}\|)+\max_{m\leq p}(|\theta_m|+|\gamma_{d,m}|)+\max_{m\leq p, t\leq T}\|\lambda_{tm}\|<C_2.
 $$
 
\end{assumption}

Assumption \ref{a3.2} collects reasonably standard regularity conditions that restrict the dependence across observations and tail behavior of random variables.  These conditions impose that the unobserved variables in the model are cross-sectionally independent, are weakly dependent and stationary in the time series, and have sub-Gaussian tails.  %The restrictions on behavior in the time series are stronger than those adopted in the sparse high-dimensional linear panel model in \cite{BCHK:FE} but are important in allowing us to accommodate the presence of latent factors $f_i$. 
 Assumption \ref{a3.2}.(iii) further imposes weak conditional dependence in the factor residuals, $U_{it}$.  In the simple case where $U_{it}$ is independent of $f_i$, $\eta_i$, and $\epsilon_i$ for all $t$, this condition reduces to weak intertemporal correlation and no strong dependence among the columns of $U_{it}$.  Importantly, it does not imply that all correlation among the observed $X_{it}$ is captured by factors but allows for the presence of a rich covariance structure in the part of $X_{it}$ that is not linearly explained by the factors.  The condition also allows for some dependence between ``control'' variables $U_{it}$ and structural unobservables $\eta_i$ and $\epsilon_i$ but restricts the magnitude of any such dependence so that it is asymptotically negligible. Finally, condition (v) requires that all the low dimensional parameters are well bounded. 

Recall that $e_{it}=\alpha\eta_{it}+\epsilon_{it}$. 
\begin{assumption}[Moment bounds] \label{a3.3} 
For $m\leq p, i\leq n, t\leq T$, define 
$$
W_{im}=\frac{1}{\sqrt{T}}\sum_{t=1}^T(U_{it,m}-\bar U_{i\cdot, m})(e_{it}-\bar e_{i\cdot}).
$$ 
There are absolute constants $c,C>0$, such that \\
(i)
$
\max_{i\leq n, m\leq p} E|W_{im}|^3 \leq C$ and $  c<\min_{i\leq n, m\leq p}EW_{im}^2\leq \max_{i\leq n, m\leq p}EW_{im}^2 <C $, and
$$
\Var \left(\frac{1}{\sqrt{nT}}\sum_{i=1}^n\sum_{t=1}^T( \eta_{it}-\bar \eta_{i\cdot})(\epsilon_{it}-\bar\epsilon_{i\cdot}) \right) >c.
$$
 (ii) almost surely in $F=(f_1,...,f_n)'$, 
 $$\max_{m\leq p, t\leq T} \frac{1}{n}\sum_{i=1}^n  E( U_{it, m} ^8|F)<C,\quad \max_{t\leq T}\frac{1}{n}\sum_{i=1}^nE(e_{it}^8|F)<C.
 $$
 
\end{assumption}

Assumption \ref{a3.3} collects additional high-level moment bounds.  The bounds on moments of normalized sums in Condition (i) could be established under a variety of sufficient lower level conditions.   Condition (ii) places restrictions on the dependence between $\{U_{it},e_{it}\}_{i=1,t=1}^{n,T}$ and $\{f_i\}_{i=1}^{n}$.  

Before stating the next assumption, we decompose the high dimensional coefficients as
$$
\gamma_y=\underbrace{\gamma_y^0}_{\text{exactly sparse}}+\underbrace{R_y}_{\text{remainder}} \quad \text{and} \quad 
\gamma_d=\underbrace{\gamma_d^0}_{\text{exactly sparse}}+\underbrace{R_d}_{\text{remainder}}
$$
where $\gamma_y^0$ and $\gamma_d^0$ are sparse vectors that approximate the potentially dense true coefficient vectors $\gamma_y$ and $\gamma_d$ and $R_y$ and $R_d$ represent approximation errors.
Let  
$
J=\{j\leq p: \gamma_{y,j}^0\neq 0\}\cup\{j\leq p: \gamma_{d,j}^0\neq 0\}
$ be the union of the support of the exactly sparse components.

 \begin{assumption}[Rate Conditions] \label{a4.1}
%   $\log p+\log q_n^{-1}=o(n^{1/3})$
(i)  $\|R_d\|_1+\|R_y\|_1=o(\sqrt{\frac{\log p}{nT}})$.    

  (ii)   $|J|_0^2\log^{3}(p)=O(n)$. %and    $|J|_0^2\log ^{1/2}(p) \log T=o(n)$.
  
  (iii)    $|J|_0^2T=o(n)$. In addition, the number of factors, $K$, stays constant.
    \end{assumption}

%  \vspace{-1em}

Assumption \ref{a4.1} collects restrictions on the quality of the approximation provided by $\gamma_y^0$ and $\gamma_d^0$ and rates of growth of model complexity as measured by $J$ and $p$ and sample sizes in the cross-sectional and time series dimension.  %Treating the sparse coefficient vectors $\gamma_y^0$ and $\gamma_d^0$ as approximations allows us to accommodate approximately sparse models where, for example, all entries in the vectors $\gamma_y$ and $\gamma_d$ in the DGP are non-zero.  Assumption \ref{a4.1} requires that an approximating model of size $|J|_0$ satisfying growth restrictions given in conditions (ii) and (iii) provides a high-enough quality approximation to $\gamma_y$ and $\gamma_d$ that the approximation errors are small in the sense captured by condition (i).   Condition (ii) is standard and allows for the dimension, $p$, of the observed variables to be much larger than the sample size. Primitive examples of approximately sparse models are given in \cite{belloni2014inference}.  
Condition (iii) imposes the somewhat nonstandard requirement that $T$ be much smaller than $n$.  The need for this condition arises from the fact that we need to obtain high-quality estimates of the idiosyncratic term in the factor equation, $U_{it}$, which depends on accurately estimating both the unknown factors and the loadings.  Estimating the loading matrix $\Lambda_t$ well for any given $t$ requires a relatively large $n$, and we thus require $T$ to be smaller than $n$ as the number of unknown loading  matrices $\{\Lambda_t\}_{t\leq T}$ is $O(T)$.

Our next assumption restricts the covariance matrix of the within-transformed factor residuals $\tilde U_{it}$.

\begin{assumption}\label{a3.4}
For any $\delta\in\mathbb{R}^{p}/\{0\}$, write  
$$\mathcal{R}(\delta)=\frac{ \delta' \frac{1}{nT}\sum_{i=1}^n\sum_{t=1}^T\tilde U_{it}\tilde U_{it}'\delta}{\delta'\delta}.
$$
 Define restricted  and sparse eigenvalue constants:
\begin{eqnarray*}
\underline{\phi}(m)&=& \inf_{\delta\in\mathbb{R}^{p}: \|\delta_{J^c}\|_1\leq m\|\delta_{J}\|_1}\mathcal{R}(\delta),\cr
\phi_{\min}(m)&=&\inf_{\delta\in\mathbb{R}^p: 1\leq \|\delta\|_0\leq m}\mathcal{R}(\delta),\cr
  \phi_{\max}(m)&=&\sup_{\delta\in\mathbb{R}^p: 1\leq \|\delta\|_0\leq m}\mathcal{R}(\delta).
\end{eqnarray*}

(i) (restricted eigenvalue) For any $m>0$ there is an absolute constant $\underline{\phi}>0$ so that with probability approaching one,
$$
 \underline{\phi}(m)>\underline{\phi}.
$$

(ii) (sparse eigenvalue) There is a sequence of absolute constants $l_T\to\infty$ and $c_1, c_2>0$ so that with probability approaching one,
$$
c_1< \phi_{\min}(l_T|J|_0)\leq \phi_{\max}(l_T|J|_0)<c_2.
$$

\end{assumption}

 {Maintaining Assumptions \ref{a3.2}-\ref{a4.1},} a simple sufficient condition for Assumption \ref{a3.4} is that all the eigenvalues of $\frac{1}{nT}\sum_{i}\sum_tE( U_{it}-\bar U_{i,\cdot})(U_{it}-\bar U_{i\cdot})'$ are well bounded. This is a typical  condition   in high-dimensional approximate factor models (e.g., \cite{bai03, SW02}). It ensures that the idiosyncratic components are weakly dependent and therefore the decomposition   $\tilde X_{it}=\tilde \Lambda_t'\tilde f_i+\tilde U_{it}$ is asymptotically identified (as $p\to\infty$). %We formally state this simple sufficient condition in Lemma \ref{l4.1} below. 

%Finally, the unknown factors should be estimated accurately.  We present high-level conditions on the accuracy of $\widehat F$ in Assumption \ref{ca.1} in Appendix \ref{sa} to avoid unnecessarily complicating the presentation as the conditions are somewhat long and technical.  The high-level conditions potentially allow for many estimators of the factors, and we verify that these conditions hold under more primitive assumptions for the case of estimating the factors using PCA.  We also note that verification of these conditions in the case of PCA does not follow from results in the existing literature but makes use of some new arguments that may be of interest in other contexts.

Finally, we present high-level conditions on the accuracy of $\widehat F$ in Assumption \ref{ca.1}.  The high-level conditions potentially allow for many estimators of the factors, and we verify that these conditions hold under more primitive assumptions for the case of estimating the factors using PCA in Appendix \ref{sf}.

\begin{assumption}[Quality of Factor Estimation in Original Data] \label{ca.1} Suppose  there is an invertible $\dim(f_i)\times \dim(f_i)$ matrix $H$  with   $\|H\|+\|H^{-1}\|=O_P(1)$,  and  non-negative sequences $\Delta_F$, $\Delta_{eg}$,  $\Delta_{ud}$,   $\Delta_{fum}$, $ \Delta_{fe},   \Delta_{\max}$,   so that for $\tilde z_{it}\in\{\tilde \epsilon_{it}, \tilde \eta_{it}\}$, $\tilde w_{tm}\in\{\tilde\Lambda_t'\gamma_d, \tilde\Lambda_t'\gamma_y, \tilde\delta_{dt},  \tilde \delta_{yt}, \tilde \lambda_{tm}\}, $ $\tilde h_{tk}\in\{\tilde \delta_{dt}, \tilde \delta_{yt},  \tilde\lambda_{tk}\}$, and $\gamma\in\{\gamma_d,\gamma_y\},$
 \begin{eqnarray*}
 &&\max_{i\leq n}\|\widehat f_i-H'\tilde f_i\|_2=O_P(\Delta_{\max}),\quad
\frac{1}{n}\sum_{i=1}^n\|\widehat f_i-H'\tilde f_i\|_2^2=O_P(\Delta_F^2)\cr
&& \frac{1}{T} \sum_{t=1}^T\|\frac{1}{n}\sum_{i=1}^n (\widehat  f_i-H'\tilde f_i)\tilde z_{it}\|^2_2
 =O_P(\Delta_{fe}^2),\cr
 %\quad \frac{1}{T} \sum_{t=1}^T\|\frac{1}{n}\sum_{i=1}^n (\widehat  f_i-H'\tilde f_i)\tilde U_{it}'\gamma\|^2_2 =O_P(\Delta_{fu}^2)\cr
   &&\max_{m\leq p} \| \frac{1}{{nT}}\sum_{i=1}^n\sum_{t=1}^T(\widehat  f_i -H'\tilde f_i) \tilde z_{it}  \tilde w_{tm} '\|_F =O_P(\Delta_{eg}),\cr
&&
  \max_{m, k\leq p}\|\frac{1}{nT} \sum_{i=1}^n  \sum_{t=1}^T(\widehat  f_i -H'\tilde f_i) \tilde U_{it,m}  \tilde h_{tk}'  \|_F=O_P(\Delta_{ud}), \cr
    && \max_{m\leq p, t\leq T}\|\frac{1}{n}\sum_{j=1}^n(\widehat f_j -H'\tilde f_j)\tilde U_{jt, m}\|_2=O_P(\Delta_{fum}).
\end{eqnarray*}
These sequences satisfy the following restrictions:
\begin{align*} 
&\sqrt{nT}|J|^2_0\Delta_F^2=o(1), \ \Delta_{eg}=o(\frac{1}{\sqrt{nT}}), \ \Delta_{ud}=o(\sqrt{\frac{\log p}{nT}}), \ |J|_0^2 \sqrt{\log p } \Delta_{ud}=o(1), \\
&\Delta_{fum}^2=o(\frac{\log p}{T|J|^2\log(pT)}), \ \Delta_{fe}^2=o(\frac{\log p}{T\log (pT)}), \ \Delta_{\max}^2=O(\log(n)), \ \text{and} \\
&\Delta_{\max}^2|J|_0^2T(\lambda^2_n|J|_0+ \Delta_F^2|J|^2_0+ {\frac{|J|_0}{n}})=o(1).
\end{align*}
  
\end{assumption}
 
 One of the major technical tasks of this paper is to show that the effects of estimating the latent factor and idiosyncratic terms are stochastically dominated by the  plug-in  tuning parameter $\kappa_n$ in (\ref{tuning}). Since $\kappa_n\asymp\sqrt{\frac{\log p}{nT}}$, this is a strong requirement, and gives rise to Assumption \ref{ca.1} (and Assumption \ref{a6.2} below for the bootstrap sample). Technically,  existing results in the literature on estimating factors models are not directly applicable to verify these conditions.
In Appendix \ref{sf}, we show that  
\begin{align*}
\frac{1}{n}\sum_{i=1}^n\|\widehat f_i-H'\tilde f_i\|_2^2&=O_P(\frac{1}{pT}+\frac{1}{n^2}+\frac{1}{nT^2})
\end{align*}
when $\widehat f_t$ is estimated via PCA.  While this result is essentially standard and allows conditions involving $\Delta_F$ to be directly verified, however, it does not imply the uniform convergence condition $\max_{t\leq T}\|\widehat f_t-H'\tilde f_t\|_2$. Nor is this result sufficient to verify the other stated conditions because other terms,  e.g. $\Delta_{eg}, \Delta_{fum}, \Delta_{fe}$, involve ``weighted averages" of $\{\widehat f_i- H'\tilde f_i\}$ whose rates of convergence can be derived and shown to be faster than that of $\Delta_F=\frac{1}{pT}+\frac{1}{n^2}+\frac{1}{nT^2}$. For instance, if we use a simple Cauchy-Schwarz inequality to bound $\Delta_{ud}$, we would have
$$
\max_{m, k\leq p}\|\frac{1}{nT} \sum_{i=1}^n  \sum_{t=1}^T(\widehat  f_i -H'\tilde f_i) \tilde U_{it,m}  \tilde h_{tk}'  \|_F^2\leq \frac{1}{n}\sum_{i=1}^n\|\widehat f_i-H'\tilde f_i\|_2^2 \max_{m,k\leq p}\frac{1}{n}\sum_{i=1}^n\|\frac{1}{T}\sum_{t=1}^T \tilde U_{it,m}  \tilde h_{tk}\|_2^2.
$$
It can be shown that $\max_{m,k\leq p}\frac{1}{n}\sum_{i=1}^n\|\frac{1}{T}\sum_{t=1}^T \tilde U_{it,m}  \tilde h_{tk}\|_2^2=O_P(\frac{\log p}{T})$, so this crude bound gives us $\Delta_{ud}=\Delta_F\sqrt{\frac{\log p}{T}}$. Unfortunately, this bound is not sharp enough to verify the condition $ \Delta_{ud}=o(\sqrt{\frac{\log p}{nT}})$ unless $n=o(pT)$. In the special case that $T$ is fixed, requiring $n=o(p)$ is a restrictive condition. Rather than relying on these crude bounds, we achieve sharper bounds by directly deriving the rate of convergence for each required term in Appendix \ref{sf} which relies on some novel technical work.  These conditions only require  $n=o(p^2T)$ which provides much more freedom on the ratio $n/p$.

\subsection{Main results}
 
The asymptotic variance of $\widehat\alpha$ will depend on the quantities
$$
\sigma_{\eta\epsilon}=\Var \left(\frac{1}{\sqrt{nT}}\sum_{i=1}^n\sum_{t=1}^T( \eta_{it}-\bar \eta_{i\cdot})(\epsilon_{it}-\bar\epsilon_{i\cdot}) \right) \quad \text{and} \quad
\sigma_{\eta}^2=\frac{1}{nT}\sum_{i=1}^n\sum_{t=1}^T\Var(\eta_{it}-\bar \eta_{i\cdot})
$$
for which 
$$
\widehat \sigma_{\eta\epsilon}=\frac{1}{nT}\sum_{i=1}^n \left(\sum_{t=1}^T\widehat\eta_{it}\widehat\epsilon_{it}\right)^2 \quad \text{and} \quad
\widehat \sigma_{\eta}^2=\frac{1}{nT}\sum_{i=1}^n\sum_{t=1}^T\widehat\eta_{it}^2
$$
are natural estimators.  Note that $\widehat\sigma_{\eta\epsilon}$ is just the usual clustered covariance estimator with clustering at the individual level.

\begin{theorem}\label{t3.1} Suppose $n,p\to\infty$, and $T$ is either fixed or growing.  Under  Assumptions \ref{a3.2}-\ref{ca.1},
%\ref{a3.2}-\ref{a3.4} and Assumption \ref{ca.1},
$$
\sqrt{nT}\sigma_{\eta\epsilon}^{-1/2}\sigma_{\eta}^2(\widehat\alpha-\alpha)\to^d\mathcal{N}(0,1),
$$
In addition,
$$ 
\sqrt{nT}\widehat \sigma_{\eta\epsilon}^{-1/2}\widehat \sigma_{\eta}^2(\widehat\alpha-\alpha)\to^d\mathcal{N}(0,1).
$$
\end{theorem}

\begin{corollary}\label{co3.1}
Let $\mathcal{P}$ be a collection of all DGP's such that the assumptions of Theorem \ref{t3.1} hold uniformly over all the DGP's in $\mathcal{P}$. Let $\zeta_{\tau}=\Phi^{-1}(1-\tau/2).$ Then as $n,p\to\infty$, and $T$ is either fixed or growing with $n$,  uniformly over $P\in \mathcal{P}$, 
$$
\lim_{n,p\to\infty} P\left(\alpha\in[\widehat\alpha\pm \frac{\zeta_{\tau}}{\sqrt{nT}}\widehat\sigma_{\eta\epsilon}^{1/2}\widehat\sigma_{\eta}^{-2}]\right)=1-\tau.
$$
\end{corollary}
 
The main implication of Theorem \ref{t3.1} and Corollary \ref{co3.1} is that $\widehat\alpha$ converges at a $\sqrt{nT}$ rate and that inference may proceed using standard asymptotic confidence intervals and hypothesis tests.  Importantly, the inferential results hold uniformly across a large class of approximately sparse models which include cases where perfect selection over which elements of $\tilde U_{it}$ enter the model is impossible even in the limit.  It is also important to highlight that the conditions on estimation of the factors do rule out the presence of weak factors, and the inferential results do not hold uniformly over sequences of models in which perfect selection of the number of factors and fast convergence of the factors and factor loadings do not hold.  The difficulty with handling weak factors arises due to the entry of the estimation errors of the factors in the cluster-lasso problems (\ref{lassoY}) and (\ref{lassoD}) and the non-smooth and highly nonlinear nature of this problem.  Extending the results to accommodate the presence of weak factors and imperfect selection of the number of factors would be an interesting direction for further research.

\section{\texorpdfstring{$k$}{k}-Step Bootstrap}\label{Sec: Bootstrap}

In this section, we present a computationally tractable bootstrap procedure that can be used in lieu of the plug-in asymptotic inference formally presented in Theorem \ref{t3.1} and Corollary \ref{co3.1}.  While well-developed in low-dimensional settings, there are relatively few formal treatments of bootstrap procedures in high-dimensional settings, though see \cite{chatterjee2011bootstrapping}, \cite{CCK:AOS13}, \cite{BCFH:Policy}, and \cite{dezeure2016high} for important existing treatments.  In the following, we consider a bootstrap procedure which only approximately solves the cluster-lasso problem within each bootstrap replication and thus may remain computationally convenient while also intuitively capturing the sampling variation introduced in the lasso selection.
  
\subsection{The k-Step Bootstrap}\label{sks}
  
Let $D^*=\{\tilde y_{it}^*, \tilde d_{it}^*, \tilde X_{it}^*\}_{i\leq n, t\leq T}$ denote a sample of bootstrap data, and let $\widehat\alpha^*$ be the estimator obtained by applying the factor-lasso estimator with data $D^*$.  Let $B$  denote the number of bootstrap repetitions.

A potential computational problem with bootstrap procedures for lasso estimation is that one needs to solve $B$ lasso problems where $B$ will typically be fairly large.  To circumvent this problem, we adopt the approach of \cite{andrews2002higher} by using the fact that the complete lasso estimator based on the original data, denoted by $\tilde \gamma_{lasso}$, should be close to the complete lasso estimator based on bootstrapped data $D^*$, denoted by $\tilde\gamma^*_{lasso}$.  Hence, within each bootstrap replication, we can use $\tilde \gamma_{lasso}$ as the initial value for solving the lasso problem and iteratively update the estimator for $k$ steps.   Denote the resulting $k$-step bootstrap lasso estimator by $\tilde\gamma^*$. We simply use $\tilde\gamma^*$ in place of $\tilde\gamma^*_{lasso}$ wherever the solution to a lasso problem shows up in the factor-lasso problem.  The main result of this section is showing that the $k$-step bootstrap procedure is first-order valid for statistical inference about $\alpha$ as long as the minimization error after $k$ steps is less than the statistical error (i.e. $o_{P^*}(({nT})^{-1/2})$.

The substantive difference between the present context and \cite{andrews2002higher} is that \cite{andrews2002higher} makes use of Newton-Raphson updates for the k-steps while face a regularized optimization problem at each iteration.  Tractability relies on the fact that there are a variety of procedures for updating within the lasso problem that are available in closed form.  Using these analytic updates greatly reduces  the overall computational task and makes a k-step bootstrap procedure attractive within the lasso context.
  
Specifically, consider the following lasso problems on the bootstrap data.
Let   
\begin{align}\label{Eq: bootlassoY}
\begin{split}
\tilde \gamma_{y,lasso}^*&=\arg\min_{\gamma\in\mathbb{R}^{p}}\mathcal{L}^*_y(\gamma)  +\kappa_n\|\widehat\Psi^y\gamma\|_1, \\
\tilde \gamma^*_{d,lasso}&=\arg\min_{\gamma\in\mathbb{R}^{p}} \mathcal{L}^*_d(\gamma)+\kappa_n\|\widehat\Psi^d\gamma\|_1,
\end{split}
\end{align}
where
\begin{align*}
\mathcal{L}^*_y(\gamma)&= \frac{1}{nT}\sum_{t=1}^T\sum_{i=1}^n( \tilde y_{it}^*-   \widehat\delta_{yt}^{*'}\widehat f_i^*-\widehat U_{it}^{*'}\gamma)^2, \\
\mathcal{L}^*_d(\gamma)&= \frac{1}{nT}\sum_{t=1}^T\sum_{i=1}^n( \tilde d_{it}^*-   \widehat\delta_{dt}^{*'}\widehat f_i^*-\widehat U_{it}^{*'}\gamma)^2. 
\end{align*}
The definitions of $\{\tilde y_{it}^*, \tilde d_{it}^*,  \widehat\delta_{yt}^{*},  \widehat\delta_{dt}^{*}, \widehat f_i^*, \widehat U_{it}^{*} \}_{i\leq n, t\leq T}$ will be formally given below.  
Let $\tilde\gamma_{y}$  and $\tilde \gamma_d$ be the lasso solutions obtained from the original data.  Also, note that we fix the value of $\kappa_n$ and of the penalty loadings $\widehat\Psi^y$ and $\widehat\Psi^d$ to the same values as used to obtain the solutions $\tilde\gamma_y$ and $\tilde\gamma_d$ in the original data.

Within each bootstrap replication, we then approximately solve the lasso problems (\ref{Eq: bootlassoY})  by applying the following procedure.  The maximum number of steps $k$ to be taken should be determined on a case-by-case basis according to the available computational capacity.\footnote{In applications where obtaining the full lasso solution is not too burdensome, one may simply iterate to convergence.} 

\begin{framed}

\noindent Algorithm (\textbf{$k$-Step Lasso Iteration}.) 

Set $k$ to be a pre-determined number of iterations.
\begin{itemize}

\item[(A1)] Set $l=0$ and  initialize at  $ \gamma_{y,0}=\tilde\gamma_{y}$, $ \gamma_{d,0}=\tilde\gamma_{d}$.

\item[(A2)] Determine  one-step iteration mappings $ \mathcal{S}_y, \mathcal{S}_d:\mathbb{R}^p\to\mathbb{R}^p$.    Let 
\begin{equation}\label{eq4.1}
 \gamma_{y,l+1}=\mathcal{S}_y(\gamma_{y,l}),\quad  \gamma_{d,l+1}=\mathcal{S}_d(\gamma_{d,l})
\end{equation}
Set $l=l+1. $

\item[(A3)] Repeat (A2) until $l=k$. Let the $k$-step lasso estimators be
$$
\tilde\gamma_y^*=\gamma_{y,k},\quad   \tilde\gamma_d^*=\gamma_{d,k}.
$$ 

\end{itemize}

\end{framed}

There are a variety of iteration mappings that can be used in Step (A2) of the k-step lasso problem.  
A commonly used and simple mapping is the ``coordinate descent method,''  also known as the ``shooting method,'' studied by \cite{fu1998penalized}.\footnote{Another commonly used iterative scheme that could readily be applied in the present setting is the ``composite gradient method'' (e.g. \cite{nesterov2007gradient} and \cite{agarwal2012fast}).  We choose to focus on the coordinate descent method as our  concrete example as it does not rely on additional tuning parameters and performed well numerically in preliminary simulation experiments. In addition, coordinate descent requires weaker regularity conditions than the composite gradient method for our theoretical analysis.} 
For solving problem (\ref{Eq: bootlassoY}), write the solution after the $l^{\text{th}}$ iteration as  $\gamma_{y,l}=(\gamma_{y,l,1},...,\gamma_{y,l,p})'$.   
The coordinate descent method updates $\gamma_{y,l+1}$  by  iteratively cycling through all coordinates. Specifically, we solve the following one-dimensional optimization problem for $m=1,...,p$, 
\begin{align}\label{eq4.2}
\begin{split}
 \gamma_{y,l+1,m}&=\arg\min_{g\in\mathbb{R}}\frac{1}{nT}\sum_{i,t}(\tilde y_{it}^*-\widehat\delta_{yt}^{*'}\widehat f_i^*-\widehat U_{it,m^-}^{*'}\gamma_{y,l+1, m^-}-\widehat U_{it,m^+}^{*'}\gamma_{y,l, m^+}-\widehat U_{it,m}^* g)^2 \\
& \qquad \qquad \qquad +\kappa_n|\widehat\Psi^y_{m}g|.
\end{split}
\end{align}
 Here $m^-=\{j: j< m\}$; and  $\gamma_{y,l+1,m^-}$ and $\widehat U_{it, m^-}^{*} $ are  $\mathbb{R}^{m-1}$ dimensional vectors whose   components   are respectively those of  $\{\gamma_{y,l+1, j}: j<m\}$ and $\{\widehat U^*_{it, j}: j<m\}$.  Similarly,  $m^+=\{j: j>m\}$;
 and $\gamma_{y,l,m^+}$ and $\widehat U_{it,m^+}^{*} $ are  $\mathbb{R}^{p-m}$ dimensional vectors whose   components   are respectively those of  $\{\gamma_{y,l, j}: j>m\}$ and $\{\widehat U^*_{it, j}: j>m\}$. When $m=1$, $m^-$ is empty; and when $m=p$, $m^+$ is empty.  In these cases, the corresponding subvectors, $\gamma_{y,l+1,m^-}$ and $\widehat U_{it, m^-}^{*} $ or $\gamma_{y,l,m^+}$, and $\widehat U_{it,m^+}^{*} $, are defined as zero.  Note that when $\gamma_{y, l+1,m}$ is being updated the previous $m-1$ elements have already been updated, while the remaining $p-m$ elements are yet to be updated. Thus, $\gamma_{y, l+1,m^-}$ is a subvector of $\gamma_{y, l+1}$, but $\gamma_{y, l,m^+}$ is a subvector of $\gamma_{y, l}$. Denote by $\gamma_{y, l+1}^{(m)}:=(\gamma_{y, l+1, m^-}, \gamma_{y, l+1,m}, \gamma_{y, l, m^+})'$ the vector that results immediately after the $m^{\text{th}}$ coordinate has been updated during the $(l+1)^{\text{th}}$ iteration. When $m=p$, all the components have been updated; and we obtain $\gamma_{y, l+1}:=\gamma_{y, l+1}^{(p)}.$  
 
Importantly, (\ref{eq4.2}) is a one-dimensional $\ell_1$-penalized quadratic problem which has an analytical solution given by the soft thresholding operation:
 \begin{align}
\begin{split}\label{eq4.3}
 \gamma_{y, l+1,m}&=\left[sgn\left(\frac{1}{nT}\sum_{i=1}^T\sum_{t=1}^TZ_{it, l,m}^*\widehat U^*_{it,m}\right)\right] \\
& \qquad \times \left( \left|  \frac{1}{nT}\sum_{i=1}^T\sum_{t=1}^TZ_{it, l,m}^*\widehat U^*_{it,m}\right|-
 \frac{1}{2}\kappa_n\widehat\Psi_{m}^y
   \right)_+ \left( \frac{1}{nT}\sum_{i=1}^T\sum_{t=1}^T\widehat U_{it,m}^{*2}\right)^{-1},
	\end{split}
 \end{align}
 where 
 $
 Z_{it, l,m}^*:=\tilde y_{it}^*-\widehat\delta_{yt}^{*'}\widehat f_i^*-\widehat U_{it,m^-}^{*'}\gamma_{y,l+1, m^-}-\widehat U_{it,m^+}^{*'}\gamma_{y,l, m^+},
 $
 $(x)_+=\max\{x,0\}$, and $sgn(x)$ takes the sign of $x$. Therefore, the mappings in (\ref{eq4.1})  are given by
 $$
 \mathcal{S}_y(\gamma_{y,l})=(\gamma_{y, l+1, 1},...,\gamma_{y, l+1, p})', \quad \text{where each $\gamma_{y, l+1, m}$ is given in (\ref{eq4.3}).}
 $$
$\mathcal{S}_d(\gamma_{d,l})$ is obviously defined similarly.

With the k-step lasso program defined, we now state the complete algorithm for the proposed k-step bootstrap procedure.  We make use of a wild residual bootstrap to generate the data at each bootstrap replication. 

\begin{framed}
 
\noindent Algorithm (\textbf{$k$-Step Wild Bootstrap}.) 
  
Let $\{\widehat f_i, \widehat U_{it}, \widehat\Lambda_t\}_{i\leq n, t\leq T}$ denote the estimates of the features of the factor model using the original data.  Let $\widehat\alpha, \widehat\delta_{dt}, \widehat\delta_{yt}, \widehat\gamma_d, \widehat\gamma_y$ be the estimated coefficients from the original data, defined in (\ref{Eq: hatU}) through (\ref{alphahat}). Also, let
\begin{eqnarray*}
\widehat\xi_t&=&\widehat\delta_{yt}-\widehat\alpha\widehat\delta_{dt}, \quad t=1,...,T, \ \text{and}
\cr
\widehat\theta&=&\widehat\gamma_y-\widehat\alpha\widehat\gamma_d.
\end{eqnarray*}
  
\begin{enumerate}

\item For each $i=1,...,n,$  let  $w_i^x$  ($x=U, Y, D$)  be mutually independent random variables, where $\{w_i^x\}_{i\leq n}$ are i.i.d. with mean zero and variance one. Let 
$$
\tilde U_{it}^*=w_i^U\widehat U_{it},\quad \tilde \eta_{it}^*=w_i^D\widehat\eta_{it}, \quad \tilde \epsilon_{it}^*=w_i^Y\widehat\epsilon_{it}, \quad t=1,...,T.
$$
Define $\{\tilde y_{it}^*, \tilde d_{it}^*, \tilde X_{it}^* \}_{t\leq T}$ as
\begin{align*}
\tilde y_{it}^* &= \widehat \alpha \tilde d_{it}^* +  \widehat\xi_t'\widehat f_i + \tilde U_{it}^{*'}\widehat \theta + \tilde \epsilon_{it}^* \\
\tilde d_{it}^* &=\widehat\delta_{dt}'\widehat f_i + \tilde U_{it}^{*'}\widehat\gamma_d  +\tilde  \eta_{it}^*,    \\
\tilde X_{it}^* &= \widehat\Lambda_t \widehat f_i + \tilde U_{it}^*.
\end{align*}

\item Apply the Factor-Lasso Algorithm to the bootstrap data $\{\tilde y_{it}^*, \tilde d_{it}^*, \tilde X_{it}^*\}_{i\leq n,t\leq T}$ to obtain an estimated alpha $\widehat\alpha^*$ replacing the lasso estimation in Step (2) of the Factor-Lasso Algorithm with steps (A1)-(A3) from the $k$-Step Lasso Iteration defined above.

\item Repeat the above steps (1)-(2) $B$ times to obtain $\{\widehat\alpha^*_{b}\}_{b\leq B}$.

\end{enumerate}
  
Let $q_{\tau}^*$ be the $\tau^{\text{th}}$ upper quantile of $\{\sqrt{nT}|\widehat\alpha_b^*-\widehat\alpha|  \}_{b\leq B}$, so that
$$
P^*(\sqrt{nT}|\widehat\alpha_b^*-\widehat\alpha|\leq q_{\tau}^*)=1-\tau.
$$
Construct the bootstrap confidence interval:
$$
\left[\widehat\alpha\pm \frac{q_{\tau}^*}{\sqrt{nT}}\right].
$$
    
\end{framed} 
 
\subsection{Validity of k-Step Bootstrap Confidence Interval}
   
In the following, we present conditions under which we verify that  the  bootstrap confidence  intervals are asymptotically valid:
$$
P\left(\alpha\in  \left[\widehat\alpha\pm \frac{q_{\tau}^*}{\sqrt{nT}}\right]\right)\to 1-\tau.
$$
   
The first assumption imposes high-level conditions that will admit the use of general updating rules in (\ref{eq4.1}) of the $k$-Step Lasso Iteration. The assumption provides high-level conditions on the computational properties and sparsity of the solution resulting after taking $k$ iterations in the solution of the lasso problem.  Recall that $ \tilde\gamma_y^*=\gamma_{y,k}$ and $\tilde\gamma_d^*=\gamma_{d,k}.$  
\begin{assumption}\label{a6.1}  The following conditions hold for $x\in\{y, d\}$:\\
(i) Minimization Error: There is a deterministic sequence $a_n$ such that $a_n\sqrt{nT}=o(1)$,
  %$a_n|J|_0\log p=o(1)$,
and a $K_0>0$, such that when $k>K_0$,  
$$\mathcal{L}^*_x(\tilde \gamma_x^*)+\kappa_n\|\widehat\Psi^x\tilde\gamma_x^*\|_1 \leq \mathcal{L}^*_x(\tilde\gamma_{x,lasso}^*)+\kappa_n\|\widehat\Psi^x\tilde\gamma_{x,lasso}^*\|_1+O_{P^*}(a_n).
$$
 %for $x=y,d$.
\\
(ii)  Sparsity: $|\widehat J^*|=O_{P^*}(|J|_0)$, where $\widehat J^*=\{j\leq p: \tilde\gamma_{dj}^*\neq 0\}\cup \{j\leq p: \tilde\gamma_{yj}^*\neq 0\}.$
\end{assumption}

Condition (i) requires  that the minimization error should be negligible compared to the statistical error  after $k$ iteration steps.  Condition (ii) guarantees the sparsity of the  iterated solutions.   As a concrete example, we verify both conditions for the coordinate descent method. We note that, to the best of our knowledge, showing the $|J|_0$-sparsity of the $k$-step iterated coordinate descent estimator has not been done previously when $p$ is potentially much larger than $n$ and may be of some independent interest.
   
\begin{proposition} \label{p6.1}
The coordinate descent iteration as given in (\ref{eq4.3}) satisfies Assumption \ref{a6.1}.
\end{proposition}

We next impose a fairly standard notion of regularity on the high-dimensional component $\tilde U_{it}$ which shows up in the infeasible lasso problem with known factors.
\begin{assumption}[Restricted Strong Convexity]\label{a6.2new}  
There is a constant  $c>0$, and a sequence $\tau_n=o(|J|_0^{-1})$ so that for all $\delta\in\mathbb{R}^{p}$, 
$$
\delta' \frac{1}{nT}\sum_{i=1}^n\sum_{t=1}^T\tilde U_{it}\tilde U_{it}'\delta\geq \frac{c}{2}\|\delta\|_2^2-O_P(\tau_n)\|\delta\|_1^2.
$$
\end{assumption}

This assumption has been discussed  by many authors, and various sufficient conditions have been provided (e.g., \cite{Raskutti:2010} and \cite{loh2015regularized}).  The following lemma provides a simple sufficient condition for both this assumption and the restricted/sparse eigenvalue assumption.
  
\begin{lemma}\label{l4.1}
Suppose Assumption \ref{a3.2} holds. Let $\lambda_1\leq...\leq \lambda_p$ be the eigenvalues of $\frac{1}{nT}\sum_{i}\sum_tE\left[( U_{it}-\bar U_{i,\cdot})(U_{it}-\bar U_{i\cdot})'\right]$. Suppose for some $0<c<C$,
$$
c<\lambda_1\leq\lambda_p<C.
$$
Then Assumptions \ref{a3.4}  and \ref{a6.2new} are satisfied. 
\end{lemma}
As we described earlier, even if $p/n\to\infty$, requiring that the eigenvalues of the population covariance matrix are well-bounded is not a stringent condition.  Note that this condition is imposed only on the factor-residuals, $U_{it}$, and that similar conditions on the population covariance matrix of factor residuals are typically imposed in the formal analysis of large approximate factor models. 

The following conditions are imposed on the bootstrap weights.
\begin{assumption}\label{a4.3}
For $x=U,Y,D$, $Ew_i^x=0$ and $\Var(w_i^x)=1$. In addition,  there exist  $L,r >0$, such that for any $s>0$, $i\leq n$,
$$P(|w^x_{i}|>s)\leq\exp(-Ls^{r}).
$$
\end{assumption}
The sub-exponential condition for the bootstrap weights enables us to bound many stochastic processes uniformly in $m\leq p$ and $t\leq T$.  In our numerical studies, we follow \cite{mammen1993:bootstrap} and use $w_i^x=\zeta_{1,i}^x/\sqrt{2}+((\zeta_{2,i}^x)^2-1)/2$ where $\zeta_{1,i}^x$ and $\zeta_{2,i}^x$ are independent standard normals and $x \in \{U,Y,D\}$. 

Finally, we impose further regularity on the quality of estimation of the factors in the bootstrap data.

\begin{assumption} [Quality of Factor Estimation in Bootstrap Data] \label{a6.2} Suppose  there is an invertible $\dim(f_i)\times \dim(f_i)$ matrix $H^*$  with   $\|H^*\|+\|H^{*-1}\|=O_{P^*}(1)$,    
and  non-negative sequences $\Delta_F^*$, $\Delta_{eg}^*$,  $\Delta_{ud}^*$,    $ \Delta_{fe}^*$, so that for  $\tilde z_{it}^*  \in\{\tilde \eta_{it}^*, \tilde \epsilon_{it}^* \},   $    
  $\widehat g_{tm}\in\{\widehat\Lambda_t'\widehat\gamma_d, \widehat\Lambda_t'\widehat\gamma_y, \widehat\delta_{dt},\widehat\delta_{yt} , \widehat\lambda_{tm}\}$, and  $\widehat h_{tm}\in\{ \widehat\delta_{dt},\widehat\delta_{yt} , \widehat\lambda_{tm}\}$,
   \begin{eqnarray*}
&&  \frac{1}{n}\sum_{i=1}^n\|\widehat f_i^*-H^{*'}\widehat f_i\|_2^2=O_P(\Delta_F^{*2})\cr
%&&\frac{1}{T}\sum_{t=1}^T\|\frac{1}{n}\sum_{i=1}^n\tilde z_{it}^*  	 ( \widehat  f_i^{*}- H^{*'}\widehat f_i)\|_2  ^2=O_{P^*}(\Delta_{fe}^{*2})\cr
&&\max_{m\leq p}\| \frac{1}{{nT}}\sum_{i=1}^n\sum_{t=1}^T \tilde z_{it}^*  \widehat g_{tm}(\widehat  f_i ^*-H{^*}\widehat f_i)'\|_F=O_{P^*}(\Delta_{eg}^*)\cr
 %  && \max_{m\leq p, t\leq T}\|\frac{1}{n}\sum_{j=1}^n(\widehat f_j^*-H^{*'}\widehat f_j)\tilde U^*_{jt, m}\|_2=O_P(\Delta_{fum}^*).\cr
   &&  \max_{m, k\leq p}\|\frac{1}{nT} \sum_{i=1}^n  \sum_{t=1}^T(\widehat f_i^*-H^{*'}\widehat f_i) \tilde U_{it,m}^*  \widehat h_{tk}'  \|_F=O_P(\Delta^*_{ud}).
\end{eqnarray*}
 These sequences satisfy the following restrictions:
\begin{align*}
&\sqrt{nT}|J|^2_0\Delta_F^{*2}=o(1), \ \Delta_{eg}^*=o(\frac{1}{\sqrt{nT}}), \ \Delta_{ud}^*=o(\sqrt{\frac{\log p}{nT}}), \ |J|_0^2 \sqrt{\log p } \Delta_{ud}^*=o(1), \\ 
&\Delta_{F}^{*2}=o(\frac{\log p}{T \log (pT)}), \ \text{and} \Delta_{\max}^2|J|_0^2T\Delta_{F}^{*2}=o(1).
\end{align*} 
\end{assumption}

As with Assumption \ref{ca.1}, we show that  
\begin{align*}
\frac{1}{n}\sum_{i=1}^n\|\widehat f_i-H'\tilde f_i\|_2^2&=O_{P^*}(\frac{1}{pT}+\frac{1}{n^2}+\frac{1}{nT^2})
\end{align*}
when $\widehat f_t^*$ is estimated using PCA in Appendix \ref{sf} which allows direct verification of conditions involving $\Delta_F^*$.  We handle the remaining terms by directly deriving the rates of convergences for each required term in Appendix \ref{sf}.

Under these additional conditions, we are able to verify that the confidence interval resulting from application the $k$-step bootstrap procedure has asymptotically correct coverage.

\begin{theorem}\label{t6.1} Suppose $n,p\to\infty$, and $T$ is either fixed or growing.  Under  Assumptions 
%\ref{a3.2}-\ref{a3.4}, \ref{a6.1}-\ref{a4.3},  and Assumptions \ref{ca.1} and \ref{a6.2},
\ref{a3.2}-\ref{ca.1} and \ref{a6.1}-\ref{a6.2},
$$  
\sqrt{nT}\sigma_{\eta\epsilon}^{-1/2}\sigma_{\eta}^2( \widehat\alpha^*-\widehat\alpha)\to^{d^*}\mathcal{N}(0,1).
$$
In addition, %suppose the assumptions of Theorem  hold uniformly over $P\in \mathcal{P}$, then  uniformly over $P\in \mathcal{P}$, 
$$
P(\sqrt{nT}|\widehat\alpha-\alpha|\leq q_{\tau}^*)\to1-\tau.
$$

\end{theorem}

\section{Estimating Factors Using Principal Components Analysis}\label{Sec: PC}

In this section, we discuss estimation of factors and factor residuals using principal components (PC).\footnote{We choose to focus on the PC estimator as a concrete example because it is relatively simple and is free of tuning parameters. One could consider other options which would also satisfy our assumed high-level conditions.  For example, the weighted PC estimator (e.g. \cite{choi, BL13}), can be more efficient than the standard PC estimator but requires additional tuning parameters for practical application.}  We also provide low-level conditions under which the high-level conditions used in establishing Theorem \ref{t3.1} are satisfied for PC. 

\subsection{Principal Components Estimator}
Let
$$
\tilde X=\begin{pmatrix}
\tilde X_{i1}&\cdots &\tilde X_{n1}\\
\vdots&&\vdots\\
\tilde X_{iT}&\cdots &\tilde X_{nT}
\end{pmatrix}_{pT\times n},
 \quad  \tilde \Lambda=\begin{pmatrix}
\tilde \Lambda_1\\
\vdots \\
\tilde\Lambda_T
\end{pmatrix}_{pT\times K},\quad \tilde F=\begin{pmatrix}
\tilde f'_1\\
\vdots \\
\tilde f'_n
\end{pmatrix}_{n\times K},
$$
and define $\tilde U$ similarly.  The matrix form of the factor model is then
$$
\tilde X=\tilde\Lambda \tilde F'+\tilde U,
$$
where the individual and time effects have already been removed.

One of the most commonly used factor estimators is based on the PC of the $n\times n$ matrix $\tilde X'\tilde X$. Let $\widehat F$ denote the $n\times K$ matrix of the estimated factors.  The columns of $\widehat F/\sqrt{n}$ are the eigenvectors of the first $K$ eigenvalues of  $\tilde X'\tilde X/(npT)$. Let $V$ be the $K$ by $K$ diagonal matrix consisting of the first $K$ eigenvalues. Then the PC estimator estimates $\tilde F$ up to a $K\times K$ rotation matrix (e.g., \cite{SW02} and \cite{bai03}) $H$ defined by 
$$
H=\frac{1}{npT}\tilde\Lambda'\tilde\Lambda \tilde F'\widehat FV^{-1}.
$$
The factor estimator in the bootstrap sampling space is defined similarly with $\widehat F^*$ denoting the $n\times K$ matrix of the estimated factors whose columns are $\sqrt{n}$ times the first $K$ eigenvectors of $\tilde X^{*'}\tilde X^*/(npT)$. 
Finally, it is important to note that we do not need to estimate the factors but only need to estimate the space spanned by the factors for Theorem \ref{t3.1} to hold.

\subsection{Regularity Conditions}

We now present additional regularity conditions which are sufficient to verify that the PC estimator satisfies the conditions given in Assumption \ref{ca.1}.  These conditions are standard for high-dimensional approximate factor models.

 \begin{assumption}[Pervasiveness]\label{pervasive}
 There are $c, C>0$ so that
 $$
c< \frac{1}{T}\sum_{t=1}^T\frac{1}{p}\tilde\Lambda_t'\tilde\Lambda_t<C.
 $$
 \end{assumption}
 
Assumption \ref{pervasive} effectively implies that the factors do not load on a small number of series but rather are related to a large number of the available $X$-variables.  The use of this assumption in high-dimensional factor models provides part of the motivation for the factor-lasso approach where at least some forms of association between factors that are not pervasive but instead load on only a few elements in $X$ and an outcome can be captured through the presence of the factor residuals in the equations of interest.

  \begin{assumption}[Second Order Weak Dependence]\label{adep}
There is $C>0$,
\begin{eqnarray*} 
&&\max_{mti}    \sum_{s=1}^T \sum_{v=1}^p\Cov(U_{it,m} ^2, U_{is,v} ^2 )<C,\cr
&& \max_{imstv}\sum_{h=1}^T\sum_{l=1}^p   |\Cov( U_{it,v}  U_{is,m}, U_{ih,l}  U_{is,m})|<C,\cr
&&\max_{i} \frac{1}{T^2p} \sum_{m,l\leq p} \sum_{t,s, h,v\leq T} \Cov (U_{it,m}U_{is, m}, U_{ih,l}U_{iv, l})<C,\cr
&&\max_{im} \frac{1}{T^2p} \sum_{k,l\leq p} \sum_{t,s, h,v\leq T} |\Cov (U_{it,k}U_{is, m}, U_{ih,l}U_{iv, m})|<C.
\end{eqnarray*}
 
 \end{assumption}

%Assumption \ref{adep} imposes restrictions on the strength of dependence in the factor residuals $U_{it}$.  
The  left hand side of the third condition equals $\max_i \Var(\frac{1}{\sqrt{p}}  \sum_{m=1}^p(\sqrt{T}\bar U_{i\cdot, m})^2)$. In addition, if we ignore the absolute value, then the left hand side of the fourth condition equals $\max_{im}  \Var (\frac{1}{\sqrt{p}} (\sqrt{T}\bar U_{i\cdot ,k})(\sqrt{T}\bar U_{i\cdot, m}))$. 
Hence the third condition means that the variance of the standardized squared average should be bounded, and 
  the fourth  condition is slightly stronger than requiring  $\max_{im}  \Var (\frac{1}{\sqrt{p}} (\sqrt{T}\bar U_{i\cdot ,k})(\sqrt{T}\bar U_{i\cdot, m}))<C$. In the special case when $\{U_t\}$  is serially independent across $t$, all of the four conditions in Assumption \ref{adep} can be directly verified under various notions of weak cross-sectional dependence.  %By ``weak cross-sectional dependence", we mean: 
% $$ \max_{m\leq p}|A(m)|_0=O(1),$$
% where  $$ A(m)=\{k\leq p:   U_{t,k}  \text{ is NOT independent of }  U_{t,m}  \}.$$

%$$\frac{1}{n}\sum_{i=1}^n \Var(\frac{1}{p}  \sum_{m=1}^p\bar U_{i\cdot, m}^2) =\frac{1}{p^2n}\sum_{i=1}^n \sum_{l=1}^p  \sum_{m=1}^p   \frac{1}{T}\sum_{t=1}^T\frac{1}{T}\sum_{s=1}^T \frac{1}{T}\sum_{h=1}^T\frac{1}{T}\sum_{v=1}^T \Cov(U_{it,m}  U_{is,m},U_{ih,l}  U_{iv,l}  ) $$

The following   proposition show that the high-level Assumptions \ref{ca.1}  and \ref{a6.2} are satisfied by the PC estimator.

\begin{proposition}\label{p4.1} Further assume $|J|_0^4=o(nT^3)$,  $|J|_0^4n=o(p^2T)$ and $|J|_0^2\log n=o(p)$. 
Then Assumptions  \ref{ca.1} and  \ref{a6.2} about $\widehat F $ and $\widehat F^*$ are satisfied. 
\end{proposition}

%\begin{proposition}\label{p6.2} Consider the PC estimator $\widehat F$.  Further suppose that $|J|_0^4=o(nT^3)$, $|J|_0^4n=o(p^2T)$ Then Assumption  is satisfied. 
%\end{proposition}

%\textcolor{red}{Should probably just combine both of these into one statement.}\textcolor{blue}{ The only difference was: the boostrap version requires an additional $J^2\log n=o(p)$, which was not needed for the original data. But since this additional condition is not essential, we can definitely combine the two propositions.}
 
The conditions  $|J|_0^4n=o(p^2T)$  and $|J|_0^2\log n=o(p)$ require lower bounds on the growth of $p$.  These conditions differ from those used in the literature on inference in purely sparse high-dimensional, e.g. \cite{belloni2014inference}, in that lower bounds on $p$ are not required in the purely sparse setting.  These lower bounds arise since accurately  estimating the unknown factors using PCA requires a large number of observed series.  Indeed, the ``average rate of convergence" is 
$$
\frac{1}{n}\sum_{i=1}^n\|\widehat f_i-H'\tilde f_i\|_2^2=O_P(\frac{1}{n^2}+ 
\frac{1}{nT^2}+\frac{1}{pT}),
$$
where the product $pT$ is the dimension of $\tilde X_i$. In the special case $|J|_0=O(1)$, these conditions require 
$$
T \ll n \ll p^2T, \ \log ^{3}p=O(n), \ \text{and} \ \log n=o(p).
$$
The results developed in this paper will thus be inappropriate in settings where $p$ is quite small relative to $n$.  Of course, in the setting with $p$ small relative to $n$, a simple approach is to just use all of the available variables without dimension reduction.

Finally, though we have been assuming the number of factors, $K$, is known a priori, our procedure admits data-dependent methods (e.g., \cite{BN02} or \cite{AH}) for selecting $K$. Under mild conditions such as those employed in \cite{BN02} or \cite{AH}, $\widehat K$, the estimator for $K$, is consistent. All the preceding results hold can then be shown to hold following first-step estimation of $K$ by conducting the theoretical analysis conditional upon the event that $\widehat K = K$ and then arguing that the results asymptotically hold unconditionally as $P(\widehat K=K) \to 1$.
%\footnote{One can use a conditional argument: first conduct the theoretical analysis conditioning on the event $\widehat K=K$, then argue that the results still hold unconditionally as $P(\widehat K=K)\to1$.}

\section{Numerical Studies and Examples}\label{Sec: Examples}

We now present simulation and empirical results in support of the formal analysis presented in the previous sections.  The first simulation example is based directly on the PPFM given in (\ref{PPFM:y})-(\ref{PPFM:x}).  The second simulation example is based on a purely cross-sectional model that allows for instrumental variables estimation of the parameter on an endogenous variable in the presence of a low-dimensional set of instrumental variables (IVs) and a large number of potential control variables.\footnote{The formal development in the IV case with a small number of instruments is a notationally burdensome but straightforward extension of the results developed in this paper.}  Following the simulation experiments, we then present results from two empirical applications.  In the first, we apply the developed procedure to estimate the effects of gun prevalence on crime following \cite{cook:ludwig:guns} using the data from \cite{BCHK:FE}.  In the second example, we apply the instrumental variables strategy of \cite{acemoglu:colonial} to try to estimate the effect of institutions on growth.%; see also \cite{BCH:JEP} for a similar analysis based on the sparse high-dimensional linear IV model.

\subsection{Simulation Examples}

\subsubsection{Panel Partial Factor Model Simulations}

In our first set of simulations, we report results for estimation and inference on $\alpha$ with data generated according to
\begin{align*}
y_{it} &= \alpha d_{it} + (c_{\xi} \xi_t)'f_i + U_{it}'(c_{\theta} \theta) + g_i + \nu_t + \epsilon_{it} \\
d_{it} &= (c_{\delta}\delta_{dt})'f_i + U_{it}'(c_{\gamma} \gamma_d) + \zeta_i + \mu_t + \eta_{it} \\
X_{it} &= (c_{\Lambda}\Lambda_t) f_i +w_i+\rho_t+U_{it}
\end{align*}
with $n = 100$, $T = 10$, $K = 3$, and $p = 100$.  We take $\epsilon_{it} \sim N(0,1)$, $\eta_{it} \sim N(0,1)$, and $U_{it} \sim N(0_{p},\Sigma_U)$ where $0_{p}$ is a $p \times 1$ vector of zeros, $\Sigma_U$ has $(r,s)$ element given by $[\Sigma_{U}]_{[r,s]} = .7^{|r-s|}$, and $\epsilon_{it}$, $\eta_{it}$, and $U_{it}$ are i.i.d. over $i$ and $t$ and jointly independent of each other.  We generate unobserved individual-specific and time-specific heterogeneity by taking $n$ i.i.d. draws, one for each individual, $(g_i,\zeta_i,w_i) \sim N(0_{p+2},I_{p+2})$ where $I_{p+2}$ is a $(p+2) \times (p+2)$ identity matrix and taking $T$ i.i.d. draws, one for each time period, $(\nu_t,\mu_t,\rho_t) \sim N(0_{p+2},I_{p+2})$.    The latent factors, $f_i$, are generated as i.i.d. draws from $ N(0_{K},I_K)$.  The factor loading vectors $\xi_t$ and $\delta_{dt}$ and factor loading matrix $\Lambda_t$ are drawn independently over time with each entry generated as an independent draw from a standard normal random variable.  The individual-specific, time-specific heterogeneity terms and factor loadings  are drawn once, and the same values are used in each simulation replication.

We set the $j^{\text{th}}$ entry of $\theta$ and $\gamma_d$ as $\theta_j = \gamma_{d,j} = \frac{1}{j^2}$.  $c_{\Lambda}$, $c_{\delta}$, $c_{\gamma}$, $c_{\xi}$, and $c_{\theta}$ are scalars that are set to alter the relative strength of $f_i$ and $U_{it}$ in each equation.  We choose $c_{\Lambda}$ so that the average $R^2$ from the $p$ regressions of $X_{it,j}$ on $f_i$ is 0.5.  We choose $(c_{\delta},c_{\gamma})$ so that the $R^2$ of the infeasible regression of $d_{it} - \zeta_i - \mu_t$ on $(c_{\delta}\delta_{dt})'f_i + U_{it}'(c_{\gamma} \gamma_d)$ is 0.7 and the factors account for 0\%, 25\%, 50\%, 75\%, or 100\% of the explanatory power in this regression.  We similarly choose $(c_{\xi},c_{\theta})$ so that the $R^2$ of the infeasible regression of $y_{it} - \alpha d_{it} - g_i - \nu_t$ on $(c_{\xi}\xi_{t})'f_i + U_{it}'(c_{\theta} \theta)$ is 0.7 and the factors account for 0\%, 25\%, 50\%, 75\%, or 100\% of the explanatory power in this regression.  Finally, we set $\alpha = 1$.  

We compare the performance of the procedure developed in this paper to several benchmarks.  Because we consider a design with $p < nT$, ordinary least squares of $y_{it}$ on $d_{it}$, $X_{it}$ and a full set of individual and time dummy variables is feasible (OLS).  We also consider estimating $\alpha$ based on the assumption that confounding is entirely captured by latent factors.  To implement this procedure, we extract factors, $\widehat f_i$, from $\tilde X_{it}$ by PCA as discussed in Section \ref{Sec: PC}.  We then regress $y_{it}$ on $d_{it}$, $\widehat f_i$ interacted with a complete set of time dummy variables, and a full set of individual and time dummy variables to obtain the estimator for $\alpha$ (Factor).  For our third procedure, we directly apply   the fixed effects double-selection procedure of \cite{BCHK:FE}  which is appropriate for a sparse high-dimensional model with fixed effects (Double Selection).  We then consider two \textit{ad hoc} variants of the double-selection approach.  In the first, we extract the first 20 principal components and interact these with a full set of time dummies.  We then apply the fixed effects double-selection procedure of \cite{BCHK:FE} to the data $(Y,D,X^*)$ where $X^*$ denotes the original $X$ variables augmented to include the interactions of principal components with time dummies (Double Selection F).  The second \textit{ad hoc} procedure extracts factors  from $\tilde X_{it}$ by PCA.  We then obtain estimates $\widehat U_{it}$ as in (\ref{Eq: hatU}) and apply the fixed effects double-selection procedure of \cite{BCHK:FE} to the data $(Y,D,\hat{U}^*)$ where $\hat{U}^*$ denotes the matrix formed by combining $\widehat U$ with the interactions of principal components with time dummies (Double Selection U).  Finally, we directly apply the factor-lasso approach outlined in this paper (Factor Lasso). We use the \cite{AH} procedure to select the number of factors to use in obtaining the Factor, Double Selection U, and Factor Lasso results.

Figure \ref{Fig: PPFM RMSE} gives simulation RMSEs for the estimator of $\alpha$ resulting from applying each procedure.  The RMSEs are truncated at 0.1 for readability of the figure.  The most striking feature of Figure \ref{Fig: PPFM RMSE} is that only the proposed factor lasso procedure delivers uniformly good performance regardless of the relative strength of the factors and factor residuals in this simulation design.  Each of the other procedures exhibits behavior that depends strongly on the exact strength of the factors in the different equations.  In terms of RMSE, the factor-lasso procedure uniformly dominates regular OLS, Double Selection ignoring the factor structure, and the \textit{ad hoc} procedure Double Selection F within the design considered.  The factor-lasso estimator of $\alpha$ is outperformed by the pure factor model in the case where all of the explanatory power in the outcome equation is contained in the factors, which corresponds to the case where the pure factor model is correctly specified and there is no additional confounding based on the factor residuals, and the Double Selection U procedure when the factors have no explanatory power in the treatment (D) equation but all explanatory power in the Y equation.  It is also important to note that the performance loss is small in these few cases where the factor lasso is outperformed.  A final interesting point to note is that the conventional lasso-based double selection procedure is outperformed by the factor lasso even when the factors do not load in either the treatment or outcome equation.  It seems likely that the loss in this case is due to the presence of the factors in the observed explanatory variables which leads to strong correlation among these variables.  This strong correlation among the $X$'s is well-known to pose challenges for lasso-type estimators.

We report size of 5\% level tests based on standard asymptotic approximations for each of the six procedures considered in Figure \ref{Fig: PPFM SIZE} where the sizes are truncated at 0.3 for readability of the figure.   In each panel, we report the rejection frequency of the standard t-test of the null hypothesis that $\alpha = 1$ with standard errors clustered at the individual level.  The most striking feature of the figure is again the uniformly good performance of tests based on the proposed factor lasso procedure.  Tests based on the factor-lasso procedure effectively control size, with size ranging between 3.3\% and 5.3\% across the design parameters considered in the simulation.  This behavior is in sharp contrast to the other procedures considered which may have large size distortions depending upon exactly how large the relative contribution of the factors is in the $D$ and $Y$ equations.  Importantly, this good behavior does not come at the cost of using an inferior estimator as evidenced by the RMSE results.

We conclude this discussion by looking at the performance of the k-step bootstrap.  In Figure \ref{Fig: PPFM BOOT}, we report size of 5\% level tests using the factor-lasso estimator and the asymptotic approximation provided in Theorem \ref{t3.1}, the k-step bootstrap, and a score bootstrap based on \cite{BCFH:Policy}.
The k-step bootstrap and asymptotic approximation have similar performance that keeps size close to the promised level.  Interestingly, the score-based bootstrap that does not reestimate the factors or the lasso parts of the model exhibits mild size distortions across all of the design settings in this example.

\subsubsection{Instrumental Variables Model Simulations}

We supplement the simulation results from the PPFM with additional simulations in a cross-sectional version of the model generalized to allow for an endogenous variable.  Specifically, we generate data from the model

\begin{align*}
y_{i} &= \alpha d_{i} + (c_{\xi} \xi)'f_i + U_{i}'(c_{\theta} \theta) + \nu + \epsilon_{i} \\
d_{i} &= \pi z_{i} + (c_{\delta_d}\delta_{d})'f_i + U_{i}'(c_{\gamma_d} \gamma_d) + \mu + \eta_{i} \\
z_{i} &= (c_{\delta_z}\delta_{z})'f_i + U_{i}'(c_{\gamma_z} \gamma_z) + \zeta + v_{i} \\
X_{i} &= (c_{\Lambda}\Lambda) f_i + \rho + U_{i}
\end{align*}
with $n = 100$, $K = 2$, and $p = 100$.  Within this model, $d_i$ is an endogenous variable with coefficient of interest $\alpha$ and $z_i$ is an instrumental variable.  We generate $\epsilon_{i} \sim N(0,1)$ and $\eta_{i} \sim N(0,1)$ with $\Ep[\epsilon_i\eta_i] = .8$ i.i.d. across $i$ and independent of all other random variables.  We generate i.i.d. draws for $U_{i}$   as before, and  $v_i \sim N(0,1)$ independently from  $U_i$.  We also generate $(\nu,\mu,\rho, \xi,\delta_d,\Lambda, c_{\Lambda}, c_{\gamma_z}, c_{\delta_d}, c_{\xi}, c_{\theta})$ as before. 
  We set $\theta = \gamma_d = \gamma_z$ to be vectors with $j^{\text{th}}$ entry given by $\theta_j = \gamma_{d,j} = \gamma_{z,j} = \frac{1}{j^2}$.   To control the strength of the instrument,  we choose $(c_{\delta_z},c_{\gamma_z})$ so that the $R^2$ of the infeasible regression of $z_{i} - \zeta$ on $(c_{\delta_z}\delta_{z})'f_i + U_{i}'(c_{\gamma_z} \gamma_z)$ is 0.7 and the factors account for 50\% of the explanatory power in this regression.  
 We set $\pi$ so that the fraction of variation accounted for by $z_i$ in the regression of $d_i$ on $z_i$, $f_i$ and $U_i$ is 25\%.  Finally, we set $\alpha = 1$. 

We again estimate $\alpha$ using six different IV procedures similar to those implemented in the previous simulation with one exception.  As the number of features is equal to the sample size in these simulations, we consider an infeasible ``oracle'' estimator that estimates $\alpha$ from IV regression of $y_{i} - (c_{\xi} \xi)'f_i - U_{i}'(c_{\theta} \theta) - \nu$ on $d_{i} - (c_{\delta_d}\delta_{d})'f_i - U_{i}'(c_{\gamma_d} \gamma_d) - \mu$ using $z_{i} - (c_{\delta_z}\delta_{z})'f_i - U_{i}'(c_{\gamma_z} \gamma_z) - \zeta$ as instrument (Oracle).  This  estimator provides a type of best-case benchmark and allows us to ascertain that instruments are strong enough that the usual asymptotic approximation provides a reasonable approximation in the idealized scenario where one is able to perfectly remove the effect of confounding from all variables.

Figure \ref{Fig: IV RMSE} gives simulation RMSEs for the estimator of $\alpha$ resulting from applying each procedure.  The RMSEs are truncated at 0.1 for readability of the figure.\footnote{Theoretically, the MSE of the IV estimator does not exist in this context.  We report root mean truncated squared error with a truncation point of 1.}  Again, we see that the factor lasso procedure delivers good performance regardless of the relative strength of the factors and factor residuals in this simulation design.  Each of the other procedures exhibits behavior that depends strongly on the exact strength of the factors in the different equations.  It might be noted that the dominance of the factor-lasso estimator, in terms of RMSE, over the ``Oracle'' procedure is due to the definition of the oracle that we use which fully removes the variation in each variable due to factors and factor residuals even in situations in which some of these variables produce no confounding.  For example, one need not remove the variation in the instruments due to the factors in cases where the factors have zero loadings in the outcome equation, but this variation is always removed due to the way we have defined the oracle model.

We report size of 5\% level tests based on standard asymptotic approximations for each of the six procedures considered in Figure \ref{Fig: IV SIZE} with size truncated at 0.3 for readability of the figure.  In each panel, we report the rejection frequency of the standard t-test of the null hypothesis that $\alpha = 1$ using heteroscedasticity robust standard errors.  Here, we see that the only procedure that uniformly controls size is the infeasible oracle.  Among the feasible procedures, the proposed factor lasso approach performs relatively well in keeping size distortions small across the majority of combinations of relative strengths of the factors.  In this case, we do see that the factor-lasso procedure suffers from reasonably large size distortions when the factors account for all of the confounding in the outcome equation and a moderate amount of counfounding in the treatment equation.  We also see that the pure factor model controls size well in this case, but performs very poorly once all variation in the outcome equation is not due to the factors.

We again conclude by looking at the performance of the k-step bootstrap in Figure \ref{Fig: IV BOOT}. %we report size of 5\% level tests using the factor-lasso estimator and the asymptotic approximation provided in Theorem \ref{t3.1}, the k-step bootstrap, and a score bootstrap based on \cite{BCFH:Policy}.  For the bootstrap procedures, we construct a 95\% level confidence interval using the bootstrap quantiles as described in Section \ref{Sec: Bootstrap} and reject if the true value falls outside of the interval. 
We see that there is a modest, but clearly visible, improvement from using the k-step bootstrap relative to the asymptotic approximation.  The score based bootstrap, on the other hand, lines up reasonably well with the asymptotic approximation.  

\subsubsection{Summary of Simulation Results}

Overall, the results from the two simulation experiments are supportive of the asymptotic theory.  We see that the factor-lasso approach delivers estimators with good properties relative to other feasible procedures that leverage either a pure factor structure or a pure sparse structure in partial factor model settings.  We see that both point estimation properties, measured in terms of RMSE, and inferential quality, as measured by size of tests, are competitive or much better than the other procedures considered in our simulation design.  The results also suggest that the proposed k-step bootstrap procedure works relatively well and may offer some gains relative to the asymptotic Gaussian approximation.

\subsection{Empirical Examples}

\subsubsection{Estimating the Effects of Gun Prevalence on Crime}

In this example, we follow \cite{BCHK:FE} who build upon the work of \cite{cook:ludwig:guns} and attempt to estimate   the effect of gun prevalence on crime in a setting with a high-dimensional set of potential controls.  As in \cite{BCHK:FE}, we focus exclusively on trying to measure the effect of gun prevalence on homicide rates.  An important difficulty with estimating the effect of gun prevalence in the United States is that exact gun-ownership numbers are difficult to obtain.  Due to this difficulty, \cite{cook:ludwig:guns} use the fraction of suicides committed with a firearm (abbreviated FSS) within a county to proxy for county-level gun ownership rates.  \cite{cook:ludwig:guns} provide a series of arguments and evidence from secondary data sources supporting the claim that FSS provides a useful proxy for gun ownership.  For the analysis in this paper, we simply take it as given that estimating a causal effect of FSS on crime measures is worthwhile and abstract from any further measurement or data issues surrounding the use of this proxy.

Both \cite{cook:ludwig:guns} and \cite{BCHK:FE} estimate linear fixed effects models of the form
\begin{align}\label{CLModel}
\log Y_{it} = \alpha \text{log FSS}_{it-1} + X_{it}'\beta + g_i + \nu_t + \epsilon_{it}
\end{align}
where $g_i$ and $\nu_t$ are treated as parameters to be estimated, $X_{it}$ are control variables, and $Y_{it}$ is one of three dependent variables: the overall homicide rate within county $i$ in year $t$, the firearm homicide rate within county $i$ in year $t$, or the non-firearm homicide rate within county $i$ in year $t$.  \cite{cook:ludwig:guns} use the four variables percent African American, percent of households with female head, nonviolent crime rates, and percent of the population that lived in the same house five years earlier as their set of controls $X_{it}$.  \cite{BCHK:FE} maintain the assumption of approximate sparsity and employ their variable selection approach using a much larger set of potential controls generated by taking variables compiled by the US Census Bureau as $X_{it}$.  Their variables include county-level measures of demographics, the age distribution, the income distribution, crime rates, federal spending, home ownership rates, house prices, educational attainment, voting patterns, employment statistics, and migration rates along with interactions of the initial (1980) values of all control variables with a linear, quadratic, and cubic term in time.  

Rather than adopt the approximately sparse model in (\ref{CLModel}), we employ the PPFM, (\ref{PPFM:y})-(\ref{PPFM:x}), and factor-lasso approach to estimate $\alpha$ using 909 variables in $X_{it}$ constructed as in \cite{BCHK:FE}.\footnote{The exact identities of the variables are available upon request.  The data is from the U.S. Census Bureau USA Counties Database, http://www.census.gov/support/USACdataDownloads.html.}  The PPFM model seems very appropriate for this data as it directly incorporates a mechanism to accommodate the concern that there are features of counties that are not directly observed, the $f_i$, but are related to the evolution of the outcome and treatment variable of interest, which is captured by the time-varying factor loadings.  Obviously, exclusion of these factors would then lead to omitted variables bias in any estimator of $\alpha$ that fails to capture them.  Concern about the existence of such factors is common in empirical applications involving aggregate panel data.

The key assumption that we leverage to allow us to simply accommodate these latent factors is that the same correlated unobserved factors that lead to confounding are related to the evolution of other observed county-level aggregates and that we have access to a large number of these auxiliary aggregates.  While this key assumption is strong, the PPFM also naturally provides some robustness to the presence of shocks ($U_{it}$) that are related to movements of the observed $X_{it}$ series as well as movements in the variable of interest and outcome.  Such shocks may be motivated, for example, by the factor structure being misspecified, by the presence of variables that are not strongly related to factors but are confounded with the treatment and outcome, and simply by the presence of local shocks not captured by the factors that are related to the observed series.   

We present estimation results in Table \ref{CLTableResults} with results for each dependent variable presented across the columns and rows corresponding to different estimation approaches.  As a baseline, we report numbers taken directly from the first row of Table 3 in \cite{cook:ludwig:guns} in the first row of Table 1 (``Cook and Ludwig (2006) Baseline'').   We report results obtained from our data the remaining rows.\footnote{All results are based on weighted regression where we weight by the within-county average population over 1980-1999.}  For these results, we first report the point estimate and estimate of the asymptotic standard error obtained by clustering by county.  Immediately below these results, we report the 95\% confidence interval obtained from applying the k-step bootstrap procedure in brackets.  The rows labeled ``Post Double Selection'' apply the procedure of \cite{BCHK:FE}.  The rows labeled ``Factor'' are based on a pure factor model;  the rows labeled ``Factor-Lasso'' use the proposed factor-lasso procedure. All  factors are estimated using PCA and the number of factors is selected using \cite{AH}.

We see that the estimates and inferential statements produced for the firearm homicide rate (``Gun'') and the non-firearm homicide rate (``non-Gun'') are broadly consistent with each other.  In all cases, there is a fairly large positive point estimate for the effect on the firearm homicide rate with corresponding 95\% confidence intervals that exclude zero, suggesting positive association between the used measure of gun prevalence and gun homicides.  For the non-firearm homicide rate, all point estimates are negative and modest and confidence intervals include both positive and negative values.  The broad results for the overall homicide rate (``Overall'') are slightly more mixed. The baseline results for \cite{cook:ludwig:guns} and results from a pure factor model suggest a strongly significant, positive effect of gun prevalence on the overall homicide rate.  Assuming sparsity and applying \cite{BCHK:FE} yields a positive estimate of the effect which is statistically insignificant at the 5\% level.  Finally, the factor-lasso estimator is similar in magnitude to the sparsity-based estimator but borderline significant at the 5\% level using the bootstrap confidence interval.

A more interesting comparison can be made by looking more closely and considering the variable and factor selection results.  The ``Post Double Selection'' procedure ends up selecting three variables for estimating the effect on overall homicide rates, three variables for gun homicide rates, and two variables for non-gun homicide rates.  The pure factor model uses one factor.  The factor-lasso approach then uses one factor in all cases but selects eight additional variables for estimating the effect on the overall homicide rate, eight additional variables for the gun homicide rate, and five additional variables for the non-gun homicide rate.  These results suggest that the ``Post Double Selection'' and ``Factor'' results may be based on models that fail to adequately capture the effect of potential confounds.  We also see that the ``Factor'' estimates are substantially shifted away from the ``Factor Lasso'' estimates relative to standard errors and that the factor-lasso estimates are the most precise in the sense of having the shortest confidence intervals.  Both findings are consistent with the asymptotic theory and with the simulation results.

\subsubsection{Estimating the Effects of Institutions on Output}

We revisit the example considered in \cite{acemoglu:colonial}.  \cite{acemoglu:colonial} are interested in the parameter $\alpha$ in a structural model of the form
\begin{align*}
\log(\textrm{GDP per capita}_i) = \alpha (\textrm{Protection from Expropriation}_i) + x_i'\beta + \varepsilon_i
\end{align*}
based on aggregate country level data where ``Protection from Expropriation'' is a measure of the strength of individual property rights that is used as a proxy for the strength of institutions and $x_i$ is a set of variables that are meant to control for geography.  \cite{acemoglu:colonial} adopt an IV strategy where they instrument for institution quality using early European settler mortality to estimate $\alpha$ as institutions are clearly potentially endogenous.  They point out that their instrument would be invalid if there were other factors that are highly persistent and related to the development of institutions within a country and to the country's GDP.  A leading candidate for such a factor that they discuss is geography.  To address this possibility, \cite{acemoglu:colonial} control for the distance from the equator in their baseline specifications and consider different sets of geographic controls such as continent dummies within their robustness checks.\footnote{E.g. \cite{acemoglu:colonial} Table 4.}

%\cite{acemoglu:colonial} argue that this instrument is associated with institutions because settlers were more likely to set up better institutions in places where they were going to establish long term settlements and the likelihood of establishing a long term settlement was related to mortality at the time of initial colonization.  They then argue that present institutions are related to these initial conditions because institutions are extremely sticky.  The exclusion restriction is then motivated by the argument that shocks to GDP are unlikely to be so persistent that any shocks around the time of European colonization that were correlated to early settler mortality are strongly related to current GDP except through their influence on institutions.

There are, of course, many other ways to measure geography besides distance to the equator or continent where a country is found.  Rather than \textit{ex ante} choose a small number of variables to proxy for geography, we put a large number of variables that potentially capture geography in $x_i$ and then use the data to reduce dimension.  Specifically, we consider dummies for Africa, Asia, North America, and South America as well as longitude, renewable water, land boundary, land area, amount of coastline, territorial seas, amount of arable land, average temperature, average high temperature, average low temperature, average precipitation, elevation of highest point, elevation of lowest point, fraction of area that is low-lying, latitude, and spherical distance from London.  

We adapt the analysis of \cite{acemoglu:colonial} to the present setting by considering estimation of a partial factor instrumental variables model 
\begin{align*}
\log(\textrm{GDP per capita}_i) &= \alpha (\textrm{Protection from Expropriation}_i) + f_i'\xi + U_i'\theta + \varepsilon_i \\
\textrm{Protection from Expropriation}_i &= \pi \textrm{Early Settler Mortality}_i + f_i'\delta_d + U_i'\gamma_d + \eta_i \\
\textrm{Early Settler Mortality}_i &= f_i'\delta_z + U_i'\gamma_z + v_i \\
x_i = \Lambda f_i + U_i
\end{align*}
using our 20 geography measures as $x_i$ and the 64 countries from the original \cite{acemoglu:colonial} data.  The factor-lasso approach seems quite sensible in this setting.  Each of the observed geography measures could reasonably be taken as a noisy proxy for a country's geography.  This relationship is likely to be complicated and uneven with the chief features leading to association between the geography proxies plausibly being only weakly related to the notions of geography that are important predictors of mortality and institutions.  The factor-lasso approach, by allowing a small number of elements of $U_i$ to enter the equation of interest in addition to any common geography factors, readily accommodates this latter possibility in a parsimonious, data-dependent way.

We report estimation results for the first stage coefficient on the instrument in Table \ref{AJRTableResults}.  We report results from the factor-lasso approach in the row ``Factor-Lasso''. For comparison, we also report results from a few natural alternative models.  The row labeled ``Latitude'' uses the single variable distance from the equator to control for geography as in the baseline results from \cite{acemoglu:colonial}.  We report results from applying OLS using all 20 available geographic controls without dimension reduction in the row labeled ``All Controls.'' We apply the double selection approach of \cite{belloni2014inference} which would be appropriate if the relationship between geographic controls and the variables of interest were well-approximated by a sparse linear model in ``Double Selection.''  Finally, ``Factor'' reduces dimension through positing a conventional factor model.  All factors are estimated using PCA with number of factors selected by applying the procedure from \cite{AH}.

The first-stage results using only the latitude control suggest there is a fairly strong relationship between the instrument and endogenous variable if latitude is a sufficient control for geography.  The first stage F-statistic using just latitude is 10.9 which many would take to indicate that the instrument is sufficiently strong to identify the effect of interest.\footnote{A benchmark that is commonly used in the applied literature to assess whether there is sufficient variation in the instrument to identify the effect of interest is to compare the first stage F-statistic to 10, with smaller values indicating weak identification.}  The results change in a potentially substantive way after allowing for the possibility that geography is not adequately captured by latitude.  For each of the remaining approaches considered, the first-stage F-statistic drops substantially below 10, with all methods besides applying the pure factor model returning first-stage coefficients that are statistically insignificant at the 5\% level.  

One might dismiss the lack of significance after including all controls without dimension reduction as it seems likely that a model with 20 covariates in addition to the variables of interest and only 64 observations is overfit.  The next strongest result is from the pure factor model which makes use of a single extracted component and produces a first-stage F-statistic of 7.5.  As evidenced in the simulation example, inference results based on a pure factor model may be highly misleading when elements of $U_i$ also have explanatory power.  It is then interesting that the double-selection approach and the factor-lasso approach deliver almost identical results indicating a weak association between the endogenous variable and instrument after controlling parsimoniously for geography.  The double-selection procedure selects four variables\footnote{These variables are the Africa dummy, average temperature, average high temperature, and amount of arable land.}, and the factor-lasso approach uses one factor and two additional variables.\footnote{The two selected variables in addition to the factor are the Africa and Asia dummies.}  One might take this to mean that the four variables selected in the double-selection procedure approximately capture the same information as the single factor and two variables used in the factor-lasso results.  In either case, the results suggest that, at best, identification of the structural effect of institutions as measured by ``Protection from Expropriation'' using settler mortality as instrument is weak after geography is controlled for in a parsimonious, data-dependent way.  Given this apparent weak identification, we do not report second stage estimates of the structural effect.\footnote{We note that it would be straightforward to adapt the weak-identification robust procedure of \cite{ch:WeakId} to the present setting.  We do not pursue this extension for brevity.}

\subsubsection{Summary of Empirical Examples}

We believe the two empirical examples illustrate the potential applicability of partial factor models and the associated factor-lasso approach in applied economics.  The model provides a natural generalization to standard factor models and sparse high-dimensional models and seems appropriate for many economic applications, especially those that make use of aggregate panel or cross-sectional data.  The results in the first example based on \cite{cook:ludwig:guns} roughly line up with the original results, though they demonstrate the potential for efficiency gains from adopting the methods developed in this paper.  In the second example, we draw substantively different conclusions about the strength of identification than one would draw following the approach in \cite{acemoglu:colonial} due to the ability to control more flexibly for the leading candidate for confounding.  Overall, the results suggest that application of the proposed methods may usefully complement the sensitivity analyses performed in empirical economics and also have the potential to strengthen the plausibility of any conclusions drawn.

\small
\appendix

\section{Proof of Theorem \ref{t3.1} and Corollary \ref{co3.1}}\label{App: Thm 3.1}
 
 Define $(KT)\times 1$ matrices $\tilde\Xi = (\tilde\xi_1',...,\tilde\xi_T')'$ and $\tilde\Delta_d = (\tilde\delta_{d1}',...,\tilde\delta_{dT}')'$. Note that 
 \begin{eqnarray*}
 \tilde Y&=&\tilde D\alpha+(I_T\otimes \tilde F)\tilde\Xi+\tilde U\theta+\tilde \epsilon\cr
 \tilde D&=&(I_T\otimes \tilde F)\tilde \Delta_d+\tilde U\gamma_d+\tilde\eta.
 \end{eqnarray*}

Note that
$
\widehat\eta=M_{\widehat U_{\widehat J}}(I_T\otimes M_{\widehat F})\tilde D.$
Hence, 
 \begin{eqnarray*}
\widehat\alpha&=&
(\widehat\eta'\widehat\eta)^{-1}\widehat\eta'M_{\widehat U_{\widehat J}}(I_T\otimes M_{\widehat F})\tilde Y
\cr
&=&\alpha
+(\widehat\eta'\widehat\eta)^{-1}\widehat\eta'M_{\widehat U_{\widehat J}}(I_T\otimes M_{\widehat F})[(I_T\otimes \tilde F)\tilde\Xi+\tilde U\theta+\tilde \epsilon]
\cr
%&=&\alpha+(\widehat\eta'\widehat\eta)^{-1}\widehat\eta'M_{\widehat U_{\widehat J}}(I_T\otimes M_{\widehat F})\tilde \epsilon+(\widehat\eta'\widehat\eta)^{-1}\widehat\eta'M_{\widehat U_{\widehat J}}(I_T\otimes M_{\widehat F})[(I_T\otimes \tilde F)\tilde\Xi+\tilde U\theta] \cr
&=&\alpha
+(\widehat\eta'\widehat\eta)^{-1}(\widehat\eta-\tilde\eta)'M_{\widehat U_{\widehat J}}(I_T\otimes M_{\widehat F})\tilde \epsilon
+(\widehat\eta'\widehat\eta)^{-1}\tilde\eta'M_{\widehat U_{\widehat J}}(I_T\otimes M_{\widehat F})\tilde \epsilon
\cr
&&+(\widehat\eta'\widehat\eta)^{-1}\widehat\eta'M_{\widehat U_{\widehat J}}(I_T\otimes M_{\widehat F}\tilde F) \tilde\Xi +(\widehat\eta'\widehat\eta)^{-1}\widehat\eta'M_{\widehat U_{\widehat J}}(I_T\otimes M_{\widehat F})
\tilde U\theta.
\end{eqnarray*}
 Note that 
$ \tilde\eta'M_{\widehat U_{\widehat J}}(I_T\otimes M_{\widehat F})\tilde \epsilon
=\tilde\eta'  \tilde \epsilon-\tilde\eta' (I_T\otimes P_{\widehat F})\tilde \epsilon- \tilde\eta'P_{\widehat U_{\widehat J}} \tilde \epsilon+\tilde\eta'P_{\widehat U_{\widehat J}}(I_T\otimes P_{\widehat F})\tilde \epsilon
.$
 Hence, 
 \begin{eqnarray}\label{eb.1}
&&\sqrt{nT}\left(\frac{1}{nT}\widehat\eta'\widehat\eta\right)( \widehat\alpha-\alpha)= \frac{1}{\sqrt{nT}} \tilde\eta'  \tilde \epsilon+\sum_{i=1}^6A_i
 \end{eqnarray}
  where
 \begin{align*}
 A_1&=  \frac{1}{\sqrt{nT}}
 (\widehat\eta-\tilde\eta)'M_{\widehat U_{\widehat J}}(I_T\otimes M_{\widehat F})\tilde \epsilon,  &A_2&=\frac{1}{\sqrt{nT}} \widehat\eta'M_{\widehat U_{\widehat J}}(I_T\otimes M_{\widehat F}\tilde F) \tilde\Xi \\
 A_3&=-\frac{1}{\sqrt{nT}}\tilde\eta' (I_T\otimes P_{\widehat F})\tilde \epsilon, &A_4&=\frac{1}{\sqrt{nT}}\widehat\eta'M_{\widehat U_{\widehat J}}(I_T\otimes M_{\widehat F})
\tilde U\theta \\
A_5&=-\frac{1}{\sqrt{nT}} \tilde\eta'P_{\widehat U_{\widehat J}} \tilde \epsilon, &A_6&=\frac{1}{\sqrt{nT}}\tilde\eta'P_{\widehat U_{\widehat J}}(I_T\otimes P_{\widehat F})\tilde \epsilon=0.
  \end{align*}

We shall prove that  $A_i=o_P(1)$ for $i=1,...,6$ and $\frac{1}{nT}\widehat\eta'\widehat\eta-\frac{1}{nT}\tilde\eta'\tilde\eta=o_P(1)$.  So 
 \begin{eqnarray*}\widehat\eta& =&M_{\widehat U_{\widehat J}}(I_T\otimes M_{\widehat F})\tilde D
 =M_{\widehat U_{\widehat J}}(I_T\otimes M_{\widehat F})((I_T\otimes \tilde F)\tilde\Delta_d +\tilde U\gamma_d+\tilde\eta)\cr
 &=&M_{\widehat U_{\widehat J}}(I_T\otimes M_{\widehat F}\tilde F) \tilde\Delta_d +M_{\widehat U_{\widehat J}}(I_T\otimes M_{\widehat F})\tilde U\gamma_d+M_{\widehat U_{\widehat J}}(I_T\otimes M_{\widehat F})\tilde\eta.
   \end{eqnarray*}
Using the fact that $M_{\widehat F}\widehat F=0$, it can be proven that 
\begin{align}\label{eb.2ag}
\begin{split}
\frac{1}{\sqrt{nT}}(\widehat\eta -\tilde\eta)
&= \frac{1}{\sqrt{nT}}M_{\widehat U_{\widehat J}}(I_T\otimes M_{\widehat F}(\tilde FH-\widehat F)H^{-1}) \tilde\Delta_d \\
& \qquad + \frac{1}{\sqrt{nT}}M_{\widehat U_{\widehat J}}(I_T\otimes M_{\widehat F})\tilde U\gamma_d \\
& \qquad - \frac{1}{\sqrt{nT}}P_{\widehat U_{\widehat J}} \tilde\eta
  - \frac{1}{\sqrt{nT}}  M_{\widehat U_{\widehat J}}(I_T\otimes P_{\widehat F})\tilde\eta.
\end{split}	
\end{align}
       
In the subsequent subsections, we provide bounds for $A_i$ for $i=1,...,6$ and for $\frac{1}{\sqrt{nT}}\|\widehat\eta-\tilde\eta\|_2.$

\subsection{Bounding \texorpdfstring{$\widehat\eta-\tilde\eta$}{e-hat - e-tilde}}

Write 
\begin{eqnarray}\label{eb.3a}
\psi_n:=\kappa_n|J|^{1/2}_0+\|R_y\|_1+\Delta_F|J|_0+  \sqrt{\frac{|J|_0}{n}}.
\end{eqnarray}
\begin{proposition}\label{pb.1}
$
\frac{1}{\sqrt{nT}}\|\widehat\eta-\tilde\eta\|_2=O_P( \psi_n).
$
\end{proposition}

\proof  Note that $\|\tilde\Delta_d \|_2=O(\sqrt{T})$. %\textcolor{blue}{Do we need to make explicit that we are assuming all coefficient sequences are strictly bounded?  Did we already state that we are taking $K$ as fixed?  Double check.} 
 Hence  by Lemma \ref{lb.1},
\begin{align*}
&\|\frac{1}{\sqrt{nT}}M_{\widehat U_{\widehat J}}(I_T\otimes M_{\widehat F}(\tilde FH-\widehat F)H^{-1}) \tilde\Delta_d \|_2\leq O_P(1) \frac{1}{\sqrt{nT}}  \|\tilde FH-\widehat F\|_F\|\tilde\Delta_d \|_2=O_P(\Delta_F) \\
&\|\frac{1}{\sqrt{nT}}M_{\widehat U_{\widehat J}} \tilde U\gamma_d\|_2= O_P\left(\kappa_n|J|^{1/2}_0+\|R_y\|_1+\Delta_F|J|_0+  \sqrt{\frac{|J|_0}{n}}\right),
\\
&\|\frac{1}{\sqrt{nT}}M_{\widehat U_{\widehat J}}(I_T\otimes P_{\widehat F})\tilde U\gamma_d\|_2\leq \|\frac{1}{\sqrt{nT}} (I_T\otimes P_{\widehat F})\tilde U\gamma_d\|_2=O_P\left(\sqrt{\frac{|J|_0}{n}}+\Delta_{F}|J|_0\right),\\
&\|  \frac{1}{\sqrt{nT}}  M_{\widehat U_{\widehat J}}(I_T\otimes P_{\widehat F})\tilde\eta\|_2\leq \|  \frac{1}{\sqrt{nT}}  (I_T\otimes P_{\widehat F})\tilde\eta\|_2=O_P\left(\frac{1}{\sqrt{n}}+\Delta_{F}\right),\\
&\|\frac{1}{\sqrt{nT}}P_{\widehat U_{\widehat J}} \tilde\eta\|_2=O_P\left(\sqrt{|J|_0\frac{\log p}{nT}}\right).
\end{align*}
Hence, equation (\ref{eb.2ag}) implies
$
\frac{1}{\sqrt{nT}}\|\widehat\eta-\tilde\eta\|_2=O_P( \psi_n).
$
 $\blacksquare$

\subsection{Showing \texorpdfstring{$A_1, A_3, A_5, A_6=o_P(1)$}{A1, A3, A5, A6 = op(1)}}\label{sc2}

By equation (\ref{eb.2ag}), Lemma \ref{la0}, $P_{\widehat U_{\widehat J}}(I_T\otimes P_{\widehat F})=0$, and noting that $P_{\widehat U_{\widehat J}}M_{\widehat U_{\widehat J}}=0$, we  can show that 
\begin{eqnarray*}
A_1&=& \frac{1}{\sqrt{nT}}
 (\widehat\eta-\tilde\eta)'M_{\widehat U_{\widehat J}}(I_T\otimes M_{\widehat F})\tilde \epsilon\cr
&=&     \tilde \epsilon'\frac{1}{\sqrt{nT}}M_{\widehat U_{\widehat J}}(I_T\otimes M_{\widehat F})\tilde U\gamma_d
+\tilde \epsilon'\frac{1}{\sqrt{nT}}M_{\widehat U_{\widehat J}}(I_T\otimes M_{\widehat F}(\tilde FH-\widehat F)H^{-1}) \tilde\Delta_d. 
\end{eqnarray*}
It then follows from Lemma \ref{lb.2} (i)(v) that $A_1=o_P(1)$. 
 
We can also immediately apply Lemma \ref{lb.2} (iii) to establish that $A_3=o_P(1)$. 

Also, it follows from Lemma \ref{lb.1} (iv) that 
 $$
| A_5|=\left|\frac{1}{\sqrt{nT}} \tilde\eta'P_{\widehat U_{\widehat J}} \tilde \epsilon\right| \leq \sqrt{nT}\left\|\frac{1}{\sqrt{nT}}P_{\widehat U_{\widehat J}} \tilde\epsilon\right\|_2\left\|\frac{1}{\sqrt{nT}}P_{\widehat U_{\widehat J}} \tilde\eta\right\|_2
=O_P\left(\frac{|J|_0\log p}{\sqrt{nT}}\right)=o_P(1)
$$
since $|J|_0^2\log ^2p=o(nT)$.
 
Finally, it follows immediately from  Lemma \ref{la0} that $A_6=0$. $\blacksquare$

\subsection{Showing \texorpdfstring{$A_2=o_P(1)$}{A2 = op(1)}}\label{sc3}
 
By (\ref{eb.2ag}),
\begin{align}
\nonumber
A_2 &= \frac{1}{\sqrt{nT}} \widehat\eta'M_{\widehat U_{\widehat J}}(I_T\otimes M_{\widehat F}(\tilde FH-\widehat F)H^{-1}) \tilde\Xi \\
\label{sc3.a1}
& = \frac{1}{\sqrt{nT}} \tilde\eta'M_{\widehat U_{\widehat J}}(I_T\otimes M_{\widehat F}(\tilde FH-\widehat F)H^{-1}) \tilde\Xi \\
\label{sc3.a2}
& \qquad + \tilde\Xi'(I_T\otimes H^{'-1}(\tilde FH-\widehat F)'M_{\widehat F}) \frac{1}{\sqrt{nT}}M_{\widehat U_{\widehat J}}(I_T\otimes M_{\widehat F}(\tilde FH-\widehat F)H^{-1}) \tilde\Delta_d \\
\label{sc3.a3}
& \qquad - \tilde\Xi'(I_T\otimes H^{'-1}(\tilde FH-\widehat F)'M_{\widehat F})  \frac{1}{\sqrt{nT}}  M_{\widehat U_{\widehat J}}(I_T\otimes P_{\widehat F})\tilde\eta \\
\label{sc3.a4}
& \qquad + \tilde\Xi'(I_T\otimes H^{'-1}(\tilde FH-\widehat F)'M_{\widehat F}) 
 \frac{1}{\sqrt{nT}}M_{\widehat U_{\widehat J}}(I_T\otimes M_{\widehat F})\tilde U\gamma_d.
\end{align}
 It follows from Lemma \ref{lb.2}(i) that (\ref{sc3.a1}) is $o_P(1)$. By the Cauchy-Schwarz inequality and under the assumption that $\sqrt{nT}\Delta_F^2=o(1)$, (\ref{sc3.a2}) is bounded by
\begin{align*}
| \tilde\Xi'(I_T & \otimes H^{'-1}(\tilde FH-\widehat F)'M_{\widehat F}) \frac{1}{\sqrt{nT}}M_{\widehat U_{\widehat J}}(I_T\otimes M_{\widehat F}(\tilde FH-\widehat F)H^{-1}) \tilde\Delta_d| \\
&\leq \frac{1}{\sqrt{nT}}\max_{G=\tilde\Xi, \tilde\Delta_d} \| G'(I_T\otimes H^{'-1}(\tilde FH-\widehat F)'M_{\widehat F}) M_{\widehat U_{\widehat J}}\|_2^2 \\
&\leq \frac{1}{\sqrt{nT}}\max_{G=\tilde\Xi, \tilde\Delta_d} \| G'(I_T\otimes H^{'-1}(\tilde FH-\widehat F)'M_{\widehat F})  \|_2^2 \\
&\leq \frac{1}{\sqrt{nT}}\max_{g_t=\tilde\xi_t, \tilde\delta_{dt}} \sum_t\|g_t'H^{'-1}(\tilde FH-\widehat F)'M_{\widehat F}\|_2^2\\
&\leq O_P(\frac{\sqrt{T}}{\sqrt{n}})\|\tilde FH-\widehat F\|_F^2=O_P(\sqrt{nT}\Delta_F^2)=o_P(1).
\end{align*}

Term (\ref{sc3.a3}) equals  
\begin{align*}
-\frac{1}{\sqrt{nT}} \tilde\Xi'(I_T &\otimes H^{'-1}(\tilde FH-\widehat F)'M_{\widehat F})   M_{\widehat U_{\widehat J}}(I_T\otimes P_{\widehat F})\tilde\eta \\
&= -\frac{1}{\sqrt{nT}} \tilde\Xi'(I_T\otimes H^{'-1}(\tilde FH-\widehat F)'M_{\widehat F})   (I_T\otimes P_{\widehat F})\tilde\eta =0
\end{align*}
where the first equality is due to $ P_{\widehat U_{\widehat J}}(I_T\otimes P_{\widehat F})=0$ and the second equality is due to  $M_{\widehat F}P_{\widehat F}=0$ and the fact that the kronecker product satisfies $(A\otimes B)(C\otimes D)=AC\otimes BD.$
      
Finally, using $M_{\widehat F}P_{\widehat F}=0$ and $ P_{\widehat U_{\widehat J}}(I_T\otimes P_{\widehat F})=0$, (\ref{sc3.a4}) equals
\begin{align}\label{eb.3}
\begin{split}
\tilde\Xi'(I_T &\otimes H^{'-1}(\tilde FH-\widehat F)'M_{\widehat F}) 
 \frac{1}{\sqrt{nT}}M_{\widehat U_{\widehat J}}(I_T\otimes M_{\widehat F})\tilde U\gamma_d \\
&= \tilde\Xi'(I_T\otimes H^{'-1}(\tilde FH-\widehat F)'M_{\widehat F}) 
 \frac{1}{\sqrt{nT}}M_{\widehat U_{\widehat J}} \tilde U\gamma_d
 \\
& \qquad -\tilde\Xi'(I_T\otimes H^{'-1}(\tilde FH-\widehat F)'M_{\widehat F}) 
 \frac{1}{\sqrt{nT}}M_{\widehat U_{\widehat J}}(I_T\otimes P_{\widehat F})\tilde U\gamma_d \\
&= \tilde\Xi'(I_T\otimes H^{'-1}(\tilde FH-\widehat F)'M_{\widehat F}) 
 \frac{1}{\sqrt{nT}}M_{\widehat U_{\widehat J}} \tilde U\gamma_d \\
&= o_P(1)
\end{split}
\end{align}
where the last equality follows from Lemma \ref{lb.2} (vi).
    
Hence, $A_2=o_P(1)$. $\blacksquare$

\subsection{Showing \texorpdfstring{$A_4=o_P(1)$}{A4 = op(1)}}\label{sc4}
 
\begin{align}
\nonumber
A_4 &= \frac{1}{\sqrt{nT}}\widehat\eta'M_{\widehat U_{\widehat J}}(I_T\otimes M_{\widehat F})
\tilde U\theta \\
\label{sc4.1}
&= \frac{1}{\sqrt{nT}}\tilde\eta'M_{\widehat U_{\widehat J}}(I_T\otimes M_{\widehat F})
\tilde U\theta \\
\label{sc4.2}
& \qquad + \frac{1}{\sqrt{nT}}\theta'\tilde U'(I_T\otimes M_{\widehat F}) 
M_{\widehat U_{\widehat J}}(I_T\otimes M_{\widehat F})\tilde U\gamma_d \\
\label{sc4.3}
& \qquad + \frac{1}{\sqrt{nT}}\theta'\tilde U'(I_T\otimes M_{\widehat F}) 
M_{\widehat U_{\widehat J}}(I_T\otimes M_{\widehat F}(\tilde FH-\widehat F)H^{-1}) \tilde\Delta_d.
\end{align}
  
It follows from Lemma \ref{lb.2} (iii) that term (\ref{sc4.1}) is $o_P(1)$. 

By Lemma \ref{lb.1} (i) and (ii), we can bound term (\ref{sc4.2}) by
\begin{align*}
\frac{1}{\sqrt{nT}}\theta'&\tilde U'(I_T\otimes M_{\widehat F}) 
M_{\widehat U_{\widehat J}}(I_T\otimes M_{\widehat F})\tilde U\gamma_d \\
&\leq \sqrt{nT}\max_{g=\theta,\gamma_d}\left\|\frac{1}{\sqrt{nT}} M_{\widehat U_{\widehat J}}(I_T\otimes M_{\widehat F})\tilde Ug\right\|_2^2 \\
&\leq 2\sqrt{nT}\max_{g=\theta,\gamma_d}\left\|\frac{1}{\sqrt{nT}} M_{\widehat U_{\widehat J}} \tilde Ug\right\|_2^2
+2\sqrt{nT}\max_{g=\theta,\gamma_d}\left\|\frac{1}{\sqrt{nT}}  (I_T\otimes P_{\widehat F})\tilde Ug\right\|_2^2 \\
&\leq \sqrt{nT}O_P\left(\|R_y\|_1^2+\kappa_n^2|J|_0+|J|_0^2\Delta_F^2+\frac{|J|_0}{n} \right)=o_P(1),
%&\leq& O_P(\kappa_n^2|J|_0+\|R_y\|_1^2+\Delta_F^2|J|_0^2+  \frac{\log (pT)}{n}|J|_0^2+\frac{|J|_0}{n}+\Delta_{fu}^2)\sqrt{nT}=o_P(1)
\end{align*}
 under the assumption $
 \left(\kappa_n^2|J|_0+\|R_y\|_1^2+\Delta_F^2|J|_0^2+  \frac{|J|_0}{n}\right)\sqrt{nT}=o(1).
 $

The same argument as that employed in the bound given by equation (\ref{eb.3}) yields that term (\ref{sc4.3}) is $o_P(1)$. $\blacksquare$

\subsection{Proof of Theorem \ref{t3.1}}

(i) Write $\iota_{it}:=(\eta_{it}-\bar \eta_{i\cdot})^2$.

\textbf{Step 1: } Show $|\frac{1}{nT}\widehat\eta'\widehat\eta-\frac{1}{nT}\sum_{it}E\iota_{it}|=o_P(1)$.

It follows from Proposition \ref{pb.1} that $|\frac{1}{nT}\widehat\eta'\widehat\eta-\frac{1}{nT} \tilde\eta '\tilde\eta|=o_P(1)$.  Also, 
$$\frac{1}{nT}\tilde\eta'\tilde\eta=\frac{1}{nT}\sum_{i,t}\tilde\eta_{it}^2=\frac{1}{nT}\sum_{i,t}(\eta_{it}-\bar \eta_{i\cdot})^2-   \frac{1}{T}\sum_{t}   \bar\eta_{\cdot t}^2+\bar{\bar\eta}^2.
$$
We have that
$
E\left[\frac{1}{T}\sum_{t}   \bar\eta_{\cdot t}^2\right]=  \frac{1}{T}\sum_{t}  \frac{1}{n^2}\sum_iE\left[ \eta_{it}^2\right]=O(1/n)
$ and that $\bar{\bar \eta}^2=o_P(1)$.  Hence,  
$$
\frac{1}{nT}\tilde\eta'\tilde\eta=\frac{1}{nT}\sum_{it}(\eta_{it}-\bar \eta_{i\cdot})^2+o_P(1)=\frac{1}{nT}\sum_{it}\iota_{it}+o_P(1).
$$
Note that 
$$
\Var(\frac{1}{nT}\sum_{it}\iota_{it})= \frac{1}{n^2T^2}\sum_i\Var(\sum_{t}\iota_{it})=O(1/n).
$$
Hence, $|\frac{1}{nT}\sum_{it}\iota_{it}-\frac{1}{nT}\sum_{it}E\iota_{it}|=o_P(1).$
We then have  
\begin{equation}\label{eb.2new}
|\frac{1}{nT}\tilde\eta'\tilde\eta-\frac{1}{nT}\sum_{it}E\iota_{it}|=o_P(1),\quad |\frac{1}{nT}\widehat\eta'\widehat\eta-\frac{1}{nT}\sum_{it}E\iota
_{it}|=o_P(1),
\end{equation}
and $\frac{1}{nT}\widehat\eta'\widehat\eta$ is bounded away from zero.

Let $z_{n,i}=\frac{1}{\sqrt{T}}\sum_t( \eta_{it}-\bar \eta_{i\cdot})(\epsilon_{it}-\bar\epsilon_{i\cdot}) $, $b_{n}=\left[\Var(\frac{1}{\sqrt{n}}\sum_i  z_{n,i})\right]^{-1/2}$, and $x_{n,i}=b_{n}z_{n,i}$. In addition, let $s_n^2=\sum_i\Var(x_{n,i})=\sum_i \Var(z_{n,i})b_n^2=n.$
  
\textbf{Step 2:} Show $\frac{b_n}{\sqrt{nT}}\tilde\eta'\tilde\epsilon=\frac{1}{s_n}\sum_ix_{n,i}+o_P(1).$
  
Note that 
\begin{align}\label{eb.2add}
\begin{split}
E\left(\frac{1}{{T}}\sum_{t}\bar \eta_{\cdot t}\bar \epsilon_{\cdot t}\right)^2 &= \frac{1}{{T}}\sum_{t} \frac{1}{n}\sum_i\frac{1}{n}\sum_j \frac{1}{{T}}\sum_{s=1}^T \frac{1}{n}\sum_{l=1}^n\frac{1}{n}\sum_{v=1}^n E\eta_{ls}\epsilon_{vs}\eta_{it}\epsilon_{jt}\\
&= \sum_{t} \frac{1}{n^4}\sum_i  \frac{1}{{T^2}}\sum_{s=1}^T    E\eta_{is}\epsilon_{is}\eta_{it}\epsilon_{it} \\
& \qquad +\frac{1}{{T}}\sum_{t} \frac{1}{n^4}\sum_i \sum_{j\neq i}\frac{1}{{T}}\sum_{s=1}^T  E\eta_{js}\epsilon_{jt}E\epsilon_{is}\eta_{it} \\
& \qquad +\frac{1}{{T}}\sum_{t} \frac{1}{n^4}\sum_i\sum_{j\neq i}\frac{1}{{T}}\sum_{s=1}^T E\eta_{is}\eta_{it}E\epsilon_{js}\epsilon_{jt} \\
& \qquad +\frac{1}{{T}}\sum_{t} \frac{1}{n^4}\sum_i \frac{1}{{T}}\sum_{s=1}^T  \sum_{l\neq i}  E\eta_{ls}\epsilon_{ls}\eta_{it}\epsilon_{it} \\
& \qquad +\frac{1}{{T}}\sum_{t} \frac{1}{n^4}\sum_i\sum_{j\neq i}\frac{1}{{T}}\sum_{s=1}^T   E\eta_{js}\epsilon_{js}\eta_{it}\epsilon_{jt} \\
&= O\left(\frac{1}{n^2}\right).
\end{split}
\end{align}

We have $E[z_{n,i}]=0$. We also have, under our assumptions, $\Var(z_{n,i})$ and $b_n$ bounded away from both zero and infinity uniformly in $i$. 
Then
\begin{align*}
\frac{b_n}{\sqrt{nT}}\tilde\eta'\tilde\epsilon &= \frac{b_n}{\sqrt{nT}}\sum_{it}( \eta_{it}-\bar \eta_{i\cdot})(\epsilon_{it}-\bar\epsilon_{i\cdot})    
-\frac{b_n\sqrt{nT}}{{T}}\sum_{t}\bar \eta_{\cdot t}\bar \epsilon_{\cdot t}+b_n\sqrt{nT}\bar{\bar{\eta}}\bar{\bar{\epsilon}} \\
&= \frac{1}{s_n}\sum_ix_{n,i} + o_P(1).
\end{align*}

 \textbf{Step 3: } Apply the CLT.
 
We now verify the Lindeberg condition for the triangular array $\{x_{n,i}\}$. For any $\varepsilon>0$,
$$E\left(\frac{1}{n}\sum_ix_{n,i}^21\{|x_{n,i}|>\varepsilon \sqrt{n}\}\right)\leq E\left[\frac{1}{n}\sum_ix_{n,i}^2\right]=1.$$ Hence by the dominated convergence theorem,
$$
s_n^{-2}\sum_iE\left[x_{n,i}^21\{|x_{n,i}|>\varepsilon s_n\}\right]=E\left(\frac{1}{n}\sum_ix_{n,i}^21\{|x_{n,i}|>\varepsilon \sqrt{n}\}\right) \to 0.
$$
This implies, by the Lindeberg central limit theorem, 
$$
\frac{b_n}{\sqrt{nT}}\tilde\eta'\tilde\epsilon\to^d\mathcal{N}(0,1).
$$

In the previous subsections, we have proven $A_i=o_P(1)$ for $i=1,...,6$. Hence, it follows from (\ref{eb.1}) that 
$
 \sqrt{nT}(\frac{1}{nT}\widehat\eta'\widehat\eta)( \widehat\alpha-\alpha)= \frac{1}{\sqrt{nT}} \tilde\eta'  \tilde \epsilon+o_P(1).
$
In addition,
\begin{align*}
b_n|(\frac{1}{nT}&\widehat\eta'\widehat\eta-\frac{1}{nT}\sum_{i,t}E[\iota_{it}])\sqrt{nT}(\widehat\alpha-\alpha)| \\
&\leq b_n|\frac{1}{nT}\widehat\eta'\widehat\eta-\frac{1}{nT}\sum_{i,t}E[\iota_{it}]|(\frac{1}{nT}\widehat\eta'\widehat\eta)^{-1}| \frac{1}{\sqrt{nT}}\tilde\eta' \tilde  \epsilon+o_P(1)| \\
&=o_P(1).
\end{align*}
Therefore,  
\begin{align*}
\sigma_{\eta\epsilon}^{-1/2}\sigma_{\eta}^2\sqrt{nT}(\widehat\alpha-\alpha) &= b_n\frac{1}{nT}\sum_{i,t}E[\iota_{it}]\sqrt{nT}(\widehat\alpha-\alpha) \\
&=b_n\frac{1}{nT}\widehat\eta'\widehat\eta\sqrt{nT}(\widehat\alpha-\alpha)
+o_P(1) \\
&= \frac{b_n}{\sqrt{nT}} \tilde\eta'  \tilde \epsilon+o_P(1)
\to^d\mathcal{N}(0,1).
\end{align*}

(ii) To verify normality with the estimated  asymptotic variance, we need to prove consistency of $\widehat \sigma_{\eta\epsilon}$ and $\widehat \sigma_{\eta}^2$.  We have previously shown $|\frac{1}{nT}\widehat\eta'\widehat\eta-\frac{1}{nT}\sum_{i,t}E[\iota_{it}]|=o_P(1)$ which establishes consistency of $\sigma_{\eta}^2=\frac{1}{nT}\sum_{i,t}E[\iota_{it}]$. Hence, it remains to prove $\widehat \sigma_{\eta\epsilon}-\sigma_{\eta\epsilon}=o_P(1)$. 
Recall that
$$
\sigma_{\eta\epsilon}=\Var \left(\frac{1}{\sqrt{nT}}\sum_{i=1}^n\sum_{t=1}^T( \eta_{it}-\bar \eta_{i\cdot})(\epsilon_{it}-\bar\epsilon_{i\cdot}) \right).
$$

%Step 1: % We first prove $\sigma_{\eta\epsilon}=\Var (\frac{1}{\sqrt{nT}}\sum_{i=1}^p\sum_{t=1}^T\tilde \eta_{it}\tilde \epsilon_{it}) +o(1)$. In fact, 
%$$
%\frac{1}{\sqrt{nT}}\sum_{i=1}^p\sum_{t=1}^T\tilde \eta_{it}\tilde \epsilon_{it}=\underbrace{ \frac{1}{\sqrt{nT}}\sum_{it}( \eta_{it}-\bar \eta_{i\cdot})(\epsilon_{it}-\bar\epsilon_{i\cdot})    }_{x_1}-\underbrace{\frac{\sqrt{nT}}{{T}}\sum_{t}\bar \eta_{\cdot t}\bar \epsilon_{\cdot t}}_{x_2}.
%$$ It follows from (\ref{eb.2add}) that 
%$\Var(x_2)=nTE(\frac{1}{T}\sum_{t}\bar \eta_{\cdot t}\bar \epsilon_{\cdot t})^2=O(\frac{T}{n})=o(1).$
%Hence \\$\Var (\frac{1}{\sqrt{nT}}\sum_{i=1}^p\sum_{t=1}^T\tilde \eta_{it}\tilde \epsilon_{it}) -\sigma_{\eta\epsilon}=\Var(x_2)-2\Cov(x_1, x_2)\leq \Var(x_2)+2\sqrt{\sigma_{\eta\epsilon}\Var(x_2)}$. Since $\sigma_{\eta\epsilon}=O(1)$,  and  $\Var(x_2)=o(1)$, we have completed the proof of this step.

 Step 1: Bound $\Delta_1:=\widehat \sigma_{\eta\epsilon}-\frac{1}{nT}\sum_{i=1}^n(\sum_{t=1}^T\tilde\eta_{it}\tilde\epsilon_{it})^2
$. We have
\begin{align}
\nonumber
\Delta_1 &= \frac{1}{nT}\sum_{i=1}^n\left[\left(\sum_{t=1}^T\widehat\eta_{it}\widehat\epsilon_{it}\right)^2-\left(\sum_{t=1}^T\tilde \eta_{it}\tilde\epsilon_{it}\right)^2\right] \\
\nonumber
&=  \frac{1}{nT}\sum_{i=1}^n\left[\sum_{t=1}^T(\widehat\eta_{it}\widehat\epsilon_{it}+ \tilde \eta_{it}\tilde\epsilon_{it})\right]\left[\sum_{t=1}^T(\widehat\eta_{it}\widehat\epsilon_{it}- \tilde \eta_{it}\tilde\epsilon_{it})\right] \\
\label{Delta1.1}
&= \frac{1}{nT}\sum_{i=1}^n\left[\sum_{t=1}^T(\widehat\eta_{it}\widehat\epsilon_{it}- \tilde \eta_{it}\tilde\epsilon_{it})\right] ^2 \\
\label{Delta1.2}
& \qquad + \frac{2}{n}\sum_{i=1}^n\left(\frac{1}{T}\sum_{t=1}^T \tilde \eta_{it}\tilde\epsilon_{it}\right)\sum_{s=1}^T(\widehat\eta_{is}\widehat\epsilon_{is}- \tilde \eta_{is}\tilde\epsilon_{is}).
\end{align}
By Lemma \ref{lb.4}, term (\ref{Delta1.1}) is $o_P(1)$.
 
For term (\ref{Delta1.2}), we have
\begin{align*}
E\left[\frac{1}{n}\sum_i|\frac{1}{T}\sum_{t=1}^T\tilde\eta_{it}\tilde\epsilon_{it}|^2\right] &= E\left[ \frac{1}{n}\sum_i|    \frac{1}{T} \sum_{t=1}^T(\eta_{it} -\bar \eta_{\cdot t})  (\epsilon_{it}-\bar \epsilon_{\cdot t})-   (\bar\eta_{i \cdot}-\bar{\bar \eta})( \bar \epsilon_{i\cdot} -\bar{\bar \epsilon})|^2\right] \\
&= O(\frac{1}{T}).
\end{align*}
since the process $\{(\eta_t, \epsilon_t)\}_{t=-\infty}^{+\infty}$ satisfies the strong mixing condition with exponential tails. %employing the Bernstein inequality for weakly dependent data (e.g., \cite{merlevede2011bernstein}), we have (here $T$ can either grow or stay fixed )				
Thus by Cauchy-Schwarz,
\begin{align*}
\left[\frac{1}{n}\sum_{i=1}^n \right. & \left. (\frac{1}{T}\sum_{t=1}^T \tilde \eta_{it}\tilde\epsilon_{it})\sum_{s=1}^T(\widehat\eta_{is}\widehat\epsilon_{is}- \tilde \eta_{is}\tilde\epsilon_{is})\right]^2 \\
&\leq \left[\frac{1}{n}\sum_{i=1}^n (\frac{1}{T}\sum_{t=1}^T \tilde \eta_{it}\tilde\epsilon_{it})^2\right]  \left[\frac{1}{n}\sum_{i=1}^n(\sum_{s=1}^T(\widehat\eta_{is}\widehat\epsilon_{is}- \tilde \eta_{is}\tilde\epsilon_{is}))^2\right] \\
&\leq O_P\left(\frac{1}{T}\right)\frac{1}{n}\sum_{i=1}^n\left(\sum_{s=1}^T(\widehat\eta_{is}\widehat\epsilon_{is}- \tilde \eta_{is}\tilde\epsilon_{is})\right)^2 \\
&= o_P(1).
\end{align*}
This result then implies $\Delta_1=o_P(1)$.

Step 2:  Bound $\Delta_2:=\frac{1}{nT}\sum_{i=1}^n(\sum_{t=1}^T\tilde\eta_{it}\tilde\epsilon_{it})^2-\frac{1}{nT}\sum_{i=1}^n(\sum_{t=1}^T(\eta_{it}-\bar\eta_{i\cdot})(\epsilon_{it}-\bar \epsilon_{i\cdot})   )^2$.

Note that
$$
\sum_{t=1}^T\tilde\eta_{it}\tilde\epsilon_{it}
  =    \sum_{t=1}^T(\eta_{it}-\bar\eta_{i\cdot})(\epsilon_{it}-\bar \epsilon_{i\cdot})
\underbrace{-   \sum_{t=1}^T \bar\eta_{\cdot t}(\epsilon_{it}-\bar\epsilon_{\cdot t})
-  \sum_{t=1}^T\eta_{it} \bar \epsilon_{\cdot t}
+T \bar{\bar \eta}(\bar \epsilon_{i\cdot}- \bar{\bar\epsilon})
+T\bar\eta_{i\cdot}\bar{\bar\epsilon}}_{B_i}.
$$ 
and that $\frac{1}{T}  \sum_{t=1}^T(\eta_{it}-\bar\eta_{i\cdot})(\epsilon_{it}-\bar \epsilon_{i\cdot})=\frac{1}{T}  \sum_{t=1}^T\eta_{it}\epsilon_{it}-\bar\eta_{i\cdot}\bar \epsilon_{i\cdot}$.
    
Hence,
\begin{align*}
\frac{1}{nT}\sum_{i=1}^n(\sum_{t=1}^T\tilde\eta_{it}\tilde\epsilon_{it})^2 &= \frac{1}{nT}\sum_{i=1}^n(\sum_{t=1}^T(\eta_{it}-\bar\eta_{i\cdot})(\epsilon_{it}-\bar \epsilon_{i\cdot})  +B_i)^2 \\
&= \frac{1}{nT}\sum_{i=1}^n(\sum_{t=1}^T(\eta_{it}-\bar\eta_{i\cdot})(\epsilon_{it}-\bar \epsilon_{i\cdot}))^2+ \frac{1}{nT}\sum_{i=1}^nB_i^2 \\
& \qquad + \frac{2}{nT}\sum_{i=1}^nB_i\sum_{t=1}^T(\eta_{it}-\bar\eta_{i\cdot})(\epsilon_{it}-\bar \epsilon_{i\cdot}).
\end{align*}
Note that 
\begin{align*}
\left[ \frac{1}{nT}\sum_{i=1}^n \right. & \left. B_i\sum_{t=1}^T(\eta_{it}-\bar\eta_{i\cdot})(\epsilon_{it}-\bar \epsilon_{i\cdot})\right]^2 \\
&\leq \frac{1}{nT}\sum_iB_i^2\frac{1}{n}\sum_i\left(\frac{1}{\sqrt{T}}\sum_{t=1}^T(\eta_{it}-\bar\eta_{i\cdot})(\epsilon_{it}-\bar \epsilon_{i\cdot})\right)^2 \\
&= \frac{1}{nT}\sum_iB_i^2O_P\left(\frac{1}{n}\sum_i\Var(\frac{1}{\sqrt{T}}\sum_{t=1}^T(\eta_{it}-\bar\eta_{i\cdot})(\epsilon_{it}-\bar \epsilon_{i\cdot}))\right) \\
&= \frac{1}{nT}\sum_iB_i^2 O_P(\sigma_{\eta\epsilon}).
\end{align*}
Therefore, $|\Delta_2|\leq \frac{1}{nT}\sum_iB_i^2+(\frac{1}{nT}\sum_iB_i^2)^{1/2}O_P(1)$.  It suffices to prove $\frac{1}{nT}\sum_iB_i^2=o_P(1)$.  In fact, $\frac{1}{nT}\sum_iB_i^2\leq C\sum_{l=1}^4\bar A_l$ for a constant $C>0$ and
\begin{eqnarray*}
\bar A_1&=& \frac{1}{nT}\sum_{i=1}^n\left( \sum_{t=1}^T\eta_{it} \bar \epsilon_{\cdot t}\right)^2, \quad\bar  A_2  =\frac{1}{nT}\sum_{i=1}^n\left( \sum_{t=1}^T \bar\eta_{\cdot t}(\epsilon_{it}-\bar\epsilon_{\cdot t})\right)^2,\cr
 \bar A_3&=& \frac{T}{n}\bar{\bar\epsilon}^2\sum_{i=1}^n \bar\eta_{i\cdot}^2,\quad \bar  A_4 =  \frac{T}{n} \bar{\bar \eta}^2\sum_{i=1}^n(\bar \epsilon_{i\cdot}- \bar{\bar\epsilon})^2.
\end{eqnarray*}
where each $\bar A_l=O_P(E\bar A_l)$. We then have 
\begin{eqnarray*}
    E   \bar A_1   &=&\frac{1}{n^3T}\sum_{j=1}^n\sum_{i=1}^n \sum_{m=1}^n      \sum_{s=1}^T \sum_{t=1}^TE\eta_{it}\epsilon_{jt}
\eta_{is}  \epsilon_{m s}
=\frac{1}{n^3T}\sum_{i=1}^n      \sum_{s=1}^T \sum_{t=1}^TE\eta_{it}\epsilon_{it}
\eta_{is}  \epsilon_{i s}\cr
&&+\frac{1}{n^3T}\sum_{i=1}^n \sum_{j\neq i}      \sum_{s=1}^T \sum_{t=1}^TE\eta_{it}
\eta_{is} E \epsilon_{j s}\epsilon_{jt}=O\left(\frac{T}{n}\right)=o(1).
\end{eqnarray*}
Similarly, $E\bar A_2=o(1)$.  In addition, $\bar{\bar \epsilon}^2=O_P(n^{-1})$ and $\bar{\bar\eta}^2=O_P(n^{-1})$, so $\bar A_3$ and $\bar A_4$ are each $O_P(T/n)=o_P(1)$. Combining verifies that
\begin{equation}\label{eb.7a}
\Delta_2:=\frac{1}{nT}\sum_{i=1}^n\left(\sum_{t=1}^T\tilde\eta_{it}\tilde\epsilon_{it}\right)^2-\frac{1}{nT}\sum_{i=1}^n\left(\sum_{t=1}^T(\eta_{it}-\bar\eta_{i\cdot})(\epsilon_{it}-\bar \epsilon_{i\cdot})   \right)^2=o_P(1).
\end{equation}

Step 3:  Bound $\Delta_3:=\frac{1}{nT}\sum_{i=1}^n(\sum_{t=1}^T(\eta_{it}-\bar\eta_{i\cdot})(\epsilon_{it}-\bar \epsilon_{i\cdot})   )^2-\sigma_{\eta\epsilon}$.
             
Note that $\sigma_{\eta\epsilon}=E\left[\frac{1}{nT}\sum_{i=1}^n\left(\sum_{t=1}^T(\eta_{it}-\bar\eta_{i\cdot})(\epsilon_{it}-\bar \epsilon_{i\cdot})\right)^2\right]$, and let 
$$
z_{n,i}=\frac{1}{\sqrt{T}}\sum_t( \eta_{it}-\bar \eta_{i\cdot})(\epsilon_{it}-\bar\epsilon_{i\cdot}).
$$ 
Then $\Delta_3=\frac{1}{n}\sum_{i=1}^n(z_{n,i}^2-Ez_{n,i}^2)$. Because $\frac{1}{n}\sum_i\Var(z_{n,i}^2)=O(1)$, we have
\begin{equation}\label{eb.4}
E\Delta_3^2=\Var(\Delta_3)=\Var\left(\frac{1}{n}\sum_iz_{n,i}^2\right)=\frac{1}{n^2}\sum_i\Var(z_{n,i}^2)=o(1),
\end{equation}
which implies
\begin{equation}\label{eb.91}
\Delta_3:=\frac{1}{nT}\sum_{i=1}^n\left(\sum_{t=1}^T(\eta_{it}-\bar\eta_{i\cdot})(\epsilon_{it}-\bar \epsilon_{i\cdot}) \right)^2-\sigma_{\eta\epsilon}=o_P(1).
\end{equation}
 
Combining the above three steps, we reach $|\widehat \sigma_{\eta\epsilon}-\sigma_{\eta\epsilon}|=o_P(1)$. $\blacksquare$
          
  \textbf{Proof of Corollary \ref{co3.1}} Given Theorem \ref{t3.1}, the corollary follows from the same argument as that of Corollary 1 (i) of \cite{belloni2014inference}. We thus refer to \cite{belloni2014inference} for details.  $\blacksquare$

\section{Convergence of the \texorpdfstring{$k$}{k}-step bootstrap lasso}\label{sd}
 
 In this section, we obtain the  statistical convergence rate (in $O_{P^*}$) of the $k$-step bootstrap lasso estimators $\tilde\gamma_d^*$  $\tilde\gamma_y^*$. We focus on $\tilde\gamma_y^*$, as the proof of $\tilde\gamma_d^*$ is similar. Recall that
   \begin{eqnarray}\label{ed.0}
\mathcal{L}^*_y(\gamma)&=& \frac{1}{nT}\sum_{t=1}^T\sum_{i=1}^n( \tilde y_{it}^*-   \widehat\delta_{yt}^{*'}\widehat f_i^*-\widehat U_{it}^{*'}\gamma)^2,\cr
\tilde \gamma_{y,lasso}^*&=&\arg\min_{\gamma\in\mathbb{R}^{p}}\mathcal{L}^*_y(\gamma)+\kappa_n\|\widehat\Psi^y\gamma\|_1.
\end{eqnarray}
and that 
   \begin{eqnarray*}
   \widehat\gamma_y&=& \text{ the post-lasso estimator based on the original data}\cr
   \tilde\gamma_y^*&=& \text{ the k-step lasso estimator based on the bootstrap data}\cr
\tilde \gamma_{y,lasso}^*&=&  \text{ the  lasso estimator based on the bootstrap data }\cr
&&\text{ if a complete lasso program is carried out}.
\end{eqnarray*}
In particular, $\widehat\gamma_y$ is used as the coefficient when generating the bootstrap data.

 We divide the proof into three subsections. Section \ref{sd.1} proves the statistical convergence of  $\|\tilde \gamma_{y,lasso}^*-\widehat\gamma_y\|_1$ in the bootstrap sampling space.  Section \ref{sd.2} quantifies the computational error $\|\tilde\gamma_{y}^*-\tilde \gamma_{y,lasso}^*\|_1$ and shows that  the computational error of the k-step lasso  is negligible using the assumed high-level conditions on the iterative scheme $\mathcal{S}_y(\cdot)$. Section  \ref{sd.3} verifies the high-level conditions for the coordinate descent, or ``shooting'', method (\cite{fu1998penalized, kadkhodaie2014linear}).
 
We employ the usual definition of $o_{P^*}(1)$ and  $O_{P^*}(1)$.  We say that a sequence $X_n^*$ in the bootstrap sampling space is   $o_{P^*}(1)$  if for any $\varepsilon,\delta>0$, 
 $$
 P\{P^*(|X_n^*|>\varepsilon)>\delta\}\to0,
 $$
and that $X_n^*=O_{P^*}(1)$ if for any $\delta>0$, there is $M>0$, such that
$$
 P\{P^*(|X_n^*|>M)>\delta\}\to0.
$$

\subsection{The convergence of lasso on bootstrap data}\label{sd.1}
The main result in this subsection is the following proposition.
\begin{proposition}\label{pd.1}
$\|\tilde \gamma_{y,lasso}^*-\widehat\gamma_y\|_1=O_{P^*}( \kappa_n|J|_0).$
\end{proposition}
 
\proof   
Recall that $\widehat F$,  $\widehat\delta_{yt}$, and $\widehat \gamma_y$ respectively denote the estimated factors, the estimator of $\delta_{yt}$, and the post-lasso estimator of $\gamma_y$ using the original data.  We also have that $\tilde U_t^*$ denotes the wild bootstrapped idiosyncratic term in the factor equation and that the following relations hold:
\begin{equation}\label{ec.2}
\tilde Y_{t}^*=\widehat F\widehat \delta_{yt}+\tilde U_t^*\widehat\gamma_y+\tilde e_t^*,\quad \tilde e_t^*=\tilde \epsilon_t^*+\tilde \eta_t^*\widehat\alpha.
\end{equation}
In addition, recall that $\widehat\delta_{yt}^*$ and $\widehat U_t^*$ denote the estimates obtained from the bootstrap data.
Define
\begin{equation}\label{ec.3}
M_t^*=\widehat F\widehat\delta_{yt}-\widehat F^*\widehat\delta_{yt}^*+(\tilde U_t^*-\widehat U_t^*) \widehat\gamma_y,\quad \Delta_{\gamma}^*=\widehat\gamma_y-\tilde\gamma_{y,lasso}^*.
\end{equation} 

By definition, $\mathcal{L}^*_y(\tilde\gamma^*_{y,lasso})+\kappa_n\|\widehat\Psi^y\tilde\gamma^*_{y,lasso}\|_1\leq \mathcal{L}^*_y(\widehat \gamma_y)+\kappa_n\|\widehat\Psi^y\widehat\gamma_y\|_1$, which  implies 
\begin{equation}\label{ed.1}
\frac{1}{nT}\sum_{t=1}^T\left(\|\widehat U_t^* \Delta^*_{\gamma}\|_2^2+2 ( \tilde e_t^{*'}+M_t^{*'})\widehat U_t ^*\Delta_{\gamma} ^*\right)+\kappa_n\|\widehat\Psi^y\tilde\gamma^*_{y,lasso}\|_1
\leq  \kappa_n\|\widehat\Psi^y\widehat\gamma_y\|_1.
\end{equation}
By Lemma \ref{ld.5new} and $\kappa_n=\frac{2c_0}{\sqrt{nT}}\Phi^{-1}(1-q_n/(2p))$ for some $c_0>1$,
\begin{align*}
|\frac{1}{nT}\sum_{t=1}^T & 2 ( \tilde e_t^{*'}+M_t^{*'})\widehat U_t^* \Delta_{\gamma}^*  |  \\
&\leq \|\frac{1}{nT}\sum_{t=1}^T 2  \tilde e_t^{*'}\tilde U_t^* \widehat \Psi^{y-1}\|_{\infty}\|\widehat\Psi^y\Delta_{\gamma} ^* \|_1 + (\|\frac{1}{nT}\sum_{t=1}^T 2  \tilde e_t^{*'}(\widehat U_t^*-\tilde U_t^*)\|_{\infty} \\
& \qquad +\|\frac{1}{nT}\sum_{t=1}^T 2  M_t^{*'}\widehat U_t^* \Delta_{\gamma}^*  \|_{\infty})\| \widehat \Psi^{y}\Delta_{\gamma} ^* \|_1\max_{m}\widehat\Psi^y_m  \\
&\leq \left[\frac{2}{\sqrt{nT}}\Phi^{-1}(1-\frac{q_n}{2p})(1+o_{P^*}(1))+o_{P^*}\left(\sqrt{\frac{\log p}{nT}}\right)\right]\|\widehat\Psi^y\Delta_{\gamma} ^* \|_1 \\
&\leq \frac{(c_0+1)}{\sqrt{nT}}\Phi^{-1}\left(1-\frac{q_n}{2p}\right)\|\widehat\Psi^y\Delta_{\gamma} ^* \|_1 \\
&=\frac{c_0+1}{2c_0}\kappa_n\|\widehat\Psi^y\Delta_{\gamma} ^* \|_1.
\end{align*}
with $P^*$ approaching one. Equation (\ref{ed.1}) then implies, for the support set $\widehat J$ of $\widehat \gamma_y$,  
 \begin{equation}\label{ed.3ad}
 \frac{1}{nT}\sum_{t=1}^T\|\widehat U_t^* \Delta^*_{\gamma}\|_2^2+\frac{c_0-1}{2c_0}\kappa_n\|(\widehat\Psi^y\Delta_{\gamma} ^*)_{\widehat J^c}\|_1
\leq  \kappa_n\|(\widehat\Psi^y\Delta_{\gamma} ^*)_{\widehat J}\|_1\frac{3c_0+1}{2c_0}.
 \end{equation}
 Hence, $\|(\Delta_{\gamma} ^*)_{\widehat J^c}\|_1
\leq c\|(\Delta_{\gamma} ^*)_{\widehat J}\|_1$ for some $c>0$.  This also implies 
 for some generic $C>0$, $\|\Delta_{\gamma}^*\|_1^2\leq C\|\Delta_{\gamma, \widehat J}\|_1^2\leq C\|\Delta_{\gamma}^*\|_2^2O_P(|J|_0)$ as $|\widehat J|_0=O_P(|J|_0)$.

We can now apply Lemma \ref{ld.5new} to obtain 
$$
\frac{1}{nT}\sum_{t=1}^T \|\widehat U_t^* \Delta_{\gamma}^*\|_2^2\geq 
\frac{1}{nT}\sum_{t=1}^T \|\tilde U_t \Delta_{\gamma}^*\|_2^2-\|\Delta_{\gamma}^*\|_2^2o_{P^*}(1).
$$ 
In addition,  the sparse-eigenvalue condition implies the restricted eigenvalue condition: For any $m>0$, there is $\underline{\phi}>0$, so that, for $|\widehat J|=O_P(|J|_0)$, with probability arbitrarily close to one,  
$$
 \inf_{\delta\in\mathbb{R}^{p}: \|\delta_{\widehat J}\|_1\leq m\|\delta_{\widehat J^c}\|_1} 
 \frac{ \delta' \frac{1}{nT}\sum_{i=1}^n\sum_{t=1}^T\tilde U_{it}\tilde U_{it}'\delta}{\delta'\delta}\geq \underline{\phi}.
 $$
Hence  $\frac{1}{nT}\sum_{t=1}^T \|\widehat U_t^* \Delta_{\gamma}^*\|_2^2 
\geq \underline{\phi}\|\Delta_{\gamma}^*\|_2^2/2.
$
\begin{equation}\label{ed.4ad}
  \frac{1}{nT}\sum_{t=1}^T\|\widehat U_t ^*\Delta_{\gamma}^*\|_2^2= O_{P^*}(\kappa_n^2|J|_0),\quad \| \Delta_{\gamma}^*\|_1=O_{P^*}( \kappa_n|J|_0).
\end{equation}
 $\blacksquare$

\subsection{The computational error of the k-step lasso}\label{sd.2}

 The main result in this subsection is the following proposition.
 \begin{proposition}  \label{pd.2}
 (i)
$\|\tilde\gamma_{y}^*-\tilde \gamma_{y,lasso}^*\|_1\leq
c \| \widehat\gamma_y-\tilde\gamma_{y,lasso}^* \|_1
 +O_{P^*}\left(\frac{a_n}{\kappa_n}+ \sqrt{a_n| J|_0}\right)
$ for some $c>0$. \\
(ii)  $\|\widehat\gamma_y-\tilde\gamma_y^*\|_1=O_{P^*}\left(\kappa_n|J|_0+\frac{a_n}{\kappa_n}+ \sqrt{a_n| J|_0}\right)$,\\
(iii)  $ \frac{1}{nT}\sum_{t=1}^T\|\widehat U_t ^* (\widehat\gamma_y-\tilde\gamma_y^*)\|_2^2=O_{P^*}(\kappa_n^2|J|_0+a_n)$.

\end{proposition}
 
 \proof   (i)  We apply Lemma \ref{ec.3add} below. Note that condition (\ref{ec.13}) in this lemma is satisfied under   Assumption \ref{a6.1} with $b_n=O_{P^*}(a_n)$. Hence applying Lemma \ref{ec.3add} with $\gamma=\tilde\gamma_y^*$ immediately implies the result.
 
 (ii) The conclusion follows immediately from part (i) and Proposition \ref{pd.1}.
 
 (iii) By equation (\ref{ed.3new}) given below in the proof of Lemma \ref{ec.3add} with $b_n=O_{P^*}(a_n)$,
$$
\frac{2}{nT}\sum_{it}\|\widehat U_{it}^*(\tilde\gamma_y^*-\tilde\gamma_{y,lasso}^*)\|_2^2=O_{P^*}(a_n).
$$
Hence by  (\ref{ed.4ad}),
$
\frac{1}{nT}\sum_{t=1}^T\|\widehat U_t ^* (\widehat\gamma_y-\tilde\gamma_y^*)\|_2^2\leq O_{P^*}(\kappa_n^2|J|_0+a_n).
$
  $\blacksquare$

 \begin{lemma} \label{ec.3add}For each $\gamma$,
suppose   for some $b_n$ (either stochastic or deterministic),
\begin{equation}\label{ec.13}
\mathcal{L}^*(\gamma)+\kappa_n\|\widehat\Psi\gamma\|_1\leq\mathcal{L}^*(\tilde\gamma^*_{lasso})+\kappa_n\|\widehat\Psi\tilde\gamma^*_{lasso}\|_1 +b_n,
\end{equation}
  then 
\begin{eqnarray*}
&&\|\gamma-\tilde\gamma_{lasso}^*\|_1\leq  C \|( \widehat\gamma_y-\tilde\gamma_{y,lasso}^*)_{\widehat J}\|_1
 + \frac{b_n}{\kappa_n}+O_{P^*}( \sqrt{b_n| J|_0}),\cr
 &&\|\gamma-\tilde\gamma_{lasso}^*\|_2\leq b_n^{1/2}+   o_{P^*}\left(  |J|_0^{-1/2}  \right)
\left(C \|( \widehat\gamma_y-\tilde\gamma_{y,lasso}^*)_{\widehat J}\|_1
 +\frac{b_n}{\kappa_n}\right).
\end{eqnarray*}
 \end{lemma}

 \proof We prove for $ \mathcal{L}^*= \mathcal{L}^*_y$. The case with $ \mathcal{L}^*= \mathcal{L}^*_d$ follows by the same argument. 
 
 \textbf{Step 1: Show $ \| \Delta\|_2^2\leq O_{P^*}(b_n)+ \| \Delta\|_1^2 o_{P^*}(  |J|_0^{-1}  ).$} Here $ \Delta=\gamma-\tilde\gamma_{y,lasso}^*$.
 
 Since $\mathcal{L}^*_y(\gamma)$ is quadratic, for any $\gamma_1,\gamma_2$, 
 $$
 \mathcal{L}^*_y(\gamma_1)-\mathcal{L}^*_y(\gamma_2)=(\gamma_1-\gamma_2)'\nabla \mathcal{L}^*_y(\gamma_2)+(\gamma_1-\gamma_2)'\nabla^2 \mathcal{L}^*_y(\gamma_2) (\gamma_1-\gamma_2),
 $$
 where 
\begin{align*}
\nabla\mathcal{L}^*_y(\gamma_2) &=- \frac{2}{nT}\sum_{t=1}^T\sum_{i=1}^n\widehat U_{it}^*( \tilde y_{it}^*-   \widehat\delta_{yt}^{*'}\widehat f_i^*-\widehat U_{it}^{*'}\gamma_2), \\
\nabla^2\mathcal{L}^*_y(\gamma_2) &=  \frac{2}{nT}\sum_{t=1}^T\sum_{i=1}^n\widehat U_{it}^*\widehat U_{it}^{*'}.
\end{align*}
Now let $\gamma_1=\gamma, $ and $ \gamma_2=\tilde\gamma_{y,lasso}^*$.
Condition (\ref{ec.13}) then implies 
\begin{eqnarray}\label{ed.3new}
\Delta' \frac{2}{nT}\sum_{t=1}^T\sum_{i=1}^n\widehat U_{it}^*\widehat U_{it}^{*'} \Delta
  &\leq& b_n+ \kappa_n\|\widehat\Psi^y\tilde\gamma_{y,lasso}^*\|_1-\kappa_n\|\widehat\Psi^y\gamma\|_1-\Delta'\nabla \mathcal{L}^*_y(\tilde\gamma_{y,lasso}^*)\cr
  &\leq& b_n
\end{eqnarray}
where, to establish the last inequality, we used $\kappa_n\|\widehat\Psi^y\tilde\gamma_{y,lasso}^*\|_1-\kappa_n\|\widehat\Psi^y\gamma\|_1-\Delta'\nabla \mathcal{L}^*_y(\tilde\gamma_{y,lasso}^*)\leq 0$ which follows due
to the  first order condition of (\ref{ed.0}) and the convexity of $\|.\|_1$. (See the proof of Lemma 11
of \cite{agarwal2012fast}.)  

We now establish a lower bound for the left hand side of (\ref{ed.3new}). 
\begin{align*}
\Delta' \frac{2}{nT}\sum_{t=1}^T\sum_{i=1}^n\widehat U_{it}^*\widehat U_{it}^{*'} \Delta
&= \frac{2}{nT}\sum_{t=1}^T \|\widehat U_t^* \Delta\|_2^2 \\
& \geq^{(a)}  \frac{2}{nT}\sum_{t=1}^T \|\tilde U_t  \Delta\|_2^2
-\| \Delta\|_1^2 o_{P^*}(  |J|_0^{-1}  ) \\
&\geq^{(b)} c\| \Delta\|_2^2-\| \Delta\|_1^2 o_{P^*}(  |J|_0^{-1}  )
  \end{align*}
  where (a) follows from (\ref{ed.18}) and (b) follows from Assumption \ref{a6.2new}. Substituting this lower bound in for the left-hand-side of (\ref{ed.3new}) then yields
\begin{equation}\label{ed.5add}
   \| \Delta\|_2^2\leq b_n+ \| \Delta\|_1^2 o_{P^*}(  |J|_0^{-1}  ).
\end{equation}

\textbf{Step 2: Show $\| \Delta\|_1\leq O_{P^*}(b_n/\kappa_n)+\|(\widehat\gamma_y-\widehat\gamma_{y,lasso}^*)_{\widehat J}\|_1+\| \Delta\|_2O_P(\sqrt{|J|_0}).$}

We re-visit the proof of Proposition \ref{pd.1}. Note that (\ref{ed.3ad}) implies, for some $c>0$,
$$
 \kappa_n\|(\widehat\gamma_y-\tilde\gamma_{y,lasso}^*)_{\widehat J^c}\|_1
\leq  \kappa_n\|( \widehat\gamma_y-\tilde\gamma_{y,lasso}^*)_{\widehat J}\|_1c.
$$
The same argument also applies using $\gamma$ in place of $\tilde\gamma_{y,lasso}^*$ due to Condition (\ref{ec.13}), yielding
$$
 \kappa_n\|(\widehat\gamma_y-\gamma)_{\widehat J^c}\|_1
\leq  \kappa_n\|( \widehat\gamma_y-\gamma)_{\widehat J}\|_1c+b_n.
$$
Adding these two inequalities and using the triangle inequality, we have
\begin{align*}
\|( \Delta)_{\widehat J^c}\|_1 &\leq  \|(\widehat\gamma_y-\tilde\gamma_{y,lasso}^*)_{\widehat J^c}\|_1 +  \|(\widehat\gamma_y-\gamma)_{\widehat J^c}\|_1 \\
&\leq \|( \widehat\gamma_y-\tilde\gamma_{y,lasso}^*)_{\widehat J}\|_1c+\|( \widehat\gamma_y-\gamma)_{\widehat J}\|_1c+\frac{b_n}{\kappa_n} \\
&\leq 2c \|( \widehat\gamma_y-\tilde\gamma_{y,lasso}^*)_{\widehat J}\|_1
 +\|( \Delta)_{\widehat J}\|_1c
 +\frac{b_n}{\kappa_n} \\
&\leq  2c \|( \widehat\gamma_y-\tilde\gamma_{y,lasso}^*)_{\widehat J}\|_1
 +\|( \Delta)_{\widehat J}\|_2c\sqrt{|\widehat J|_0}
 +\frac{b_n}{\kappa_n}.
\end{align*}
We then obtain
\begin{align}\label{ed.6a}
\begin{split}
 \| \Delta\|_1 &\leq  \|( \Delta)_{\widehat J^c}\|_1 + \|( \Delta)_{\widehat J}\|_1 \\
 &\leq  2c \|( \widehat\gamma_y-\tilde\gamma_{y,lasso}^*)_{\widehat J}\|_1
 +\| \Delta\|_2O_{P^*}(\sqrt{| J|_0})
 +\frac{b_n}{\kappa_n}.
\end{split}
\end{align}

 \textbf{Step 3: Complete the proof.} 
Substituting (\ref{ed.6a}) in for the right-hand-side of (\ref{ed.5add}) gives
$$  \| \Delta\|_2^2\leq b_n+   o_{P^*}(  |J|_0^{-1}  )
\left(C \|( \widehat\gamma_y-\tilde\gamma_{y,lasso}^*)_{\widehat J}\|_1
 +\frac{b_n}{\kappa_n}\right)^2 +\| \Delta\|_2^2o_{P^*}(1),
$$
 yielding
 $  \| \Delta\|_2^2\leq b_n+   o_{P^*}(  |J|_0^{-1}  )
\left(C \|( \widehat\gamma_y-\tilde\gamma_{y,lasso}^*)_{\widehat J}\|_1
 +\frac{b_n}{\kappa_n}\right)^2,
$
and  thus
$$
 \| \Delta\|_2\leq  b_n^{1/2}+   o_{P^*}(  |J|_0^{-1/2}  )
\left(C \|( \widehat\gamma_y-\tilde\gamma_{y,lasso}^*)_{\widehat J}\|_1
 +\frac{b_n}{\kappa_n}\right).
$$
Substituting this bound back in for $\| \Delta\|_2$ in (\ref{ed.6a}) then yields
\begin{equation}\label{ed.7}
 \| \Delta\|_1\leq   C \|( \widehat\gamma_y-\tilde\gamma_{y,lasso}^*)_{\widehat J}\|_1
 + \frac{b_n}{\kappa_n}+O_{P^*}( \sqrt{b_n| J|_0}).
\end{equation}
$\blacksquare$

 \subsection{Verifying Assumption \ref{a6.1}}\label{sd.3}
 
 We now prove Proposition \ref{p6.1}, which states that the shooting method of \cite{fu1998penalized} satisfies 
 Assumption \ref{a6.1}.

We make use of the following lemma in proving Lemma \ref{lemma: shooting}.

\begin{lemma}\label{lc.1a} Recall that $\widehat\gamma$ denotes the post-lasso estimator using the original data     and $\gamma^*_{lasso}$  denotes the lasso estimator using the   bootstrap data.  We have that
\begin{equation}\label{ec.9a}
 0\leq\mathcal{L}^*(\widehat\gamma)+\kappa_n\|\widehat\Psi  \widehat \gamma\|_1- (\mathcal{L}^*(\widehat\gamma^*_{lasso})+\kappa_n\|\widehat\Psi  \widehat \gamma^*_{lasso}\|_1)=O_{P^*}(\kappa_n^2|J|_0).
\end{equation}
\end{lemma}

\proof The first inequality follows from the definition of $ \widehat \gamma^*_{lasso}$. 

We now show the equality.  Note that for each $\gamma$,
 $$
 \mathcal{L}^*(\gamma)=\frac{1}{nT}\sum_{t=1}^T\left(\|\widehat U_t^* (\widehat\gamma-\gamma)\|_2^2+\|M_t^*+\tilde e_t^*\|_2^2+2 ( \tilde e_t^{*'}+M_t^{*'})\widehat U_t ^*(\widehat\gamma-\gamma)\right)+\kappa_n\|\widehat\Psi^y\gamma\|_1
 $$
where $M_t^*$ and $\tilde e_t^{*}$ are defined in the proof of Proposition \ref{pd.1}.
Hence by Proposition \ref{pd.1} and  Lemma \ref{ld.5new},
\begin{align*} 
\mathcal{L}^*(\widehat\gamma)&+\kappa_n\|\widehat\Psi  \widehat \gamma\|_1- (\mathcal{L}^*(\widehat\gamma^*_{lasso})+\kappa_n\|\widehat\Psi  \widehat \gamma^*_{lasso}\|_1) \\
&= \kappa_n\|\widehat\Psi^y\widehat\gamma\|_1-\kappa_n\|\widehat\Psi  \widehat \gamma^*_{lasso}\|_1 \\
& \qquad - \frac{1}{nT}\sum_{t=1}^T\left(\|\widehat U_t^* (\widehat\gamma-\widehat\gamma^*_{lasso})\|_2^2+ 2 ( \tilde e_t^{*'}+M_t^{*'})\widehat U_t ^*(\widehat\gamma-\widehat\gamma^*_{lasso})\right) \\
&\leq  \kappa_n\|\widehat\Psi^y(\widehat\gamma- \widehat \gamma^*_{lasso})\|_1
+\| \frac{2}{nT}\sum_{t=1}^T   ( \tilde e_t^{*'}+M_t^{*'})\widehat U_t ^*\|_{\infty}\|\widehat\gamma-\widehat\gamma^*_{lasso}\|_1 \\
&\leq O_{P^*}(\kappa_n)\|\widehat\gamma-\widehat\gamma^*_{lasso}\|_1 = O_{P^*}(\kappa_n^2|J|_0).
\end{align*}
 $\blacksquare$
  
  \begin{lemma}\label{lemma: shooting} For the shooting method, 
 (i) $\mathcal{L}^*(\gamma_{l})+\kappa_n\|\widehat\Psi  \gamma_l\|_1\leq \mathcal{L}^*(\gamma_{l-1})+\kappa_n\|\widehat\Psi  \gamma_{l-1}\|_1.$\\
 (ii) $\mathcal{L}^*(\tilde\gamma^*)+\kappa_n\|\widehat\Psi  \tilde\gamma^*\|_1
 \leq    \mathcal{L}^*(\widehat\gamma^*_{lasso})+\kappa_n\|\widehat\Psi  \widehat \gamma^*_{lasso}\|_1+O_{P^*}(\kappa_n^2|J|_0).$\\
 (iii) $|\widehat J^*|_0=O_{P^*}(|J|_0)$.
 \end{lemma}
 
 \proof Write $\gamma_{l}=(\gamma_{l,1},...,\gamma_{l,p})'$, where $\gamma_l$ can be either $\gamma_{d,l}$ or $\gamma_{y,l}$, to denote the solution after the $l^{\text{th}}$ iteration.   Note that $\gamma_k=\tilde\gamma^*$ is the $k$-step lasso estimator. 
 
 (i) For the shooting method, each $\gamma_{l,m}$ for $m\leq p$ is defined as
 $$
 \gamma_{l,m}=\arg\min_{g}\frac{1}{nT}\sum_{i,t}(\tilde y_{it}^*-\widehat\delta_{yt}^{*'}\widehat f_i^*-\widehat U_{it,m^-}^{*'}\gamma_{l,m^-}-\widehat U_{it,m^+}^{*'}\gamma_{l-1,m^+}-\widehat U_{it,m}^* g)^2+\kappa_n|\widehat\Psi_mg|.
 $$
 As is discussed in Section \ref{sks}, after the $m^{\text{th}}$ element is updated in the $l^{\text{th}}$ iteration, the vector becomes $\gamma_{l}^{(m)}:=(\gamma_{l, m^-}, \gamma_{l,m}, \gamma_{l-1, m^+})'$, where $m^-=\{j: j< m\}$, and $m^+=\{j: j>m\}$;   $\gamma_{l,m^-}$   represents the subvector of $ \gamma_{l}$ whose indices are in $m^-$  and $ \gamma_{l-1,m^+}$  represents  the subvector of  $\gamma_{l-1}$ whose indices are in $m^+$.
% Here $m^-=\{j: j< m\}$, and $m^+=\{j: j>m\}$; $U_{it,m^-}^{*} , \gamma_{l,m^-}$ represent the subvectors of $U_{it}^{*} , \gamma_{l}$ whose indices are in $m^-$ (that is, the first $m-1$ elements). When $m=1$, $m^-$ is empty and the subvectors are defined as zero. Similarly, $U_{it,m^+}^{*} , \gamma_{l-1,m^+}$ represent the subvectors of $U_{it}^{*} , \gamma_{l-1}$ whose indices are in $m^+$ (that is, the   $m+1,...,p$ th elements). When $m=p$, $m^+$ is empty and the subvectors are defined as  zero.  Note that in the $l^{\text{th}}$ iteration,  the previous $m-1$ elements have been updated when   $\gamma_{l,m}$ is being updated, but the remaining $p-m$ elements are yet to be updated. Thus $\gamma_{l,m^-}$ is a subvector of $\gamma_{l}$, but $\gamma_{l-1,m^+}$ is a subvector of $\gamma_{l-1}$. After the $m^{\text{th}}$ element is updated in the $l^{\text{th}}$ iteration, the vector becomes $\gamma_{l}^{(m)}:=(\gamma_{l, m^-}, \gamma_{l,m}, \gamma_{l-1, m^+})'$.
 With  this  notation, after the $(m-1)^{\text{th}}$ element is updated in the $l^{\text{th}}$ iteration, the current solution vector is $\gamma_{l}^{(m-1)}=(\gamma_{l, (m-1)^-}, \gamma_{l,m-1}, \gamma_{l-1, (m-1)^+})'$. This vector can be rearranged  as
$$
(\gamma_{l, (m-1)^-}, \gamma_{l,m-1}, \gamma_{l-1, (m-1)^+})' =(\gamma_{l, m^-}, \gamma_{l-1,m}, \gamma_{l-1, m^+})'.
$$
It can be seen that the loss function is non-increasing after the $m^{\text{th}}$ element is updated:
\begin{align*}
\mathcal{L}^*(\gamma_{l}^{(m)})&+\kappa_n\|\widehat\Psi  \gamma_l^{(m)}\|_1 \\
&= \mathcal{L}^*((\gamma_{l, m^-}, \gamma_{l,m}, \gamma_{l-1, m^+})) \\
& \qquad +\kappa_n\|\widehat\Psi_{m^-}\gamma_{l, m^-}\|_1+\kappa_n|\widehat\Psi_m\gamma_{lm}|+\kappa_n\|\widehat\Psi_{m^+}\gamma_{l-1,m^+}\|_1 \\
&\leq  \mathcal{L}^*((\gamma_{l, m^-}, \gamma_{l-1,m}, \gamma_{l-1, m^+})) \\
& \qquad +\kappa_n\|\widehat\Psi_{m^-}\gamma_{l, m^-}\|_1+\kappa_n|\widehat\Psi_m\gamma_{l-1,m}|+\kappa_n\|\widehat\Psi_{m^+}\gamma_{l-1,m^+}\|_1 \\
&=\mathcal{L}^*((\gamma_{l, (m-1)^-}, \gamma_{l,m-1}, \gamma_{l-1, (m-1)^+})) \\
& \qquad +\kappa_n\|\widehat\Psi_{(m-1)^-}\gamma_{l, (m-1)^-}\|_1
+\kappa_n|\widehat\Psi_{m-1}\gamma_{lm-1}|+\kappa_n\|\widehat\Psi_{(m-1)^+}\gamma_{l-1,(m-1)^+}\|_1 \\
&=\mathcal{L}^*(\gamma_{l}^{(m-1)})+\kappa_n\|\widehat\Psi  \gamma_l^{(m-1)}\|_1.
\end{align*}
Note that $\gamma_{l}^{(p)}=\gamma_l$. Hence by (\ref{ec.9a}) in Lemma \ref{lc.1a}, 
\begin{align*}
\mathcal{L}^*(\gamma_{l})+\kappa_n\|\widehat\Psi  \gamma_l\|_1 &\leq \mathcal{L}^*(\gamma_{l}^{(1)})+\kappa_n\|\widehat\Psi  \gamma_l^{(1)}\|_1\leq \mathcal{L}^*(\gamma_{l-1}^{(p)})+\kappa_n\|\widehat\Psi  \gamma_{l-1}^{(p)}\|_1 \\
&=\mathcal{L}^*(\gamma_{l-1})+\kappa_n\|\widehat\Psi  \gamma_{l-1}\|_1.
\end{align*}
 
 (ii) From (i), $\mathcal{L}^*(\gamma_{k})+\kappa_n\|\widehat\Psi  \gamma_k\|_1
 \leq \mathcal{L}^*(\gamma_{0})+\kappa_n\|\widehat\Psi  \gamma_0\|_1
 =\mathcal{L}^*(\widehat\gamma)+\kappa_n\|\widehat\Psi  \widehat \gamma\|_1.
 $
 In addition, by (\ref{ec.9a}) in Lemma \ref{lc.1a},  
 $$
 \mathcal{L}^*(\widehat\gamma)+\kappa_n\|\widehat\Psi  \widehat \gamma\|_1- (\mathcal{L}^*(\widehat\gamma^*_{lasso})+\kappa_n\|\widehat\Psi  \widehat \gamma^*_{lasso}\|_1)=O_{P^*}(\kappa_n^2|J|_0)
 $$
for $\widehat\gamma$ and $\gamma^*_{lasso}$ respectively denoting the completed lasso estimator (as opposed to the k-step lasso solution) using the original data and the bootstrap data.  
 Note that $\kappa_n^2|J|_0\sqrt{nT}=o(1)$ and  $\gamma_k=\tilde\gamma^*$, so
\begin{align*}
 \mathcal{L}^*(\tilde\gamma^*)+\kappa_n\|\widehat\Psi  \tilde\gamma^*\|_1
 &\leq  \mathcal{L}^*(\widehat\gamma)+\kappa_n\|\widehat\Psi  \widehat \gamma\|_1 \\
&\leq \mathcal{L}^*(\widehat\gamma^*_{lasso})+\kappa_n\|\widehat\Psi  \widehat \gamma^*_{lasso}\|_1+o_{P^*}((nT)^{-1/2})
\end{align*}
which verifies  Assumption \ref{a6.1}(i).

(iii)  We now focus on the $k$-step lasso estimator $\gamma_k=\tilde\gamma^*$ and let $\gamma_{k,m}$ denote its $m^{\text{th}}$ element.   By the KKT condition, if $\gamma_{k,m}\neq 0$, then 
 \begin{eqnarray}\label{ec.9}
-\kappa_n\widehat\Psi_msgn(\gamma_{k,m})&=&\frac{2}{nT}\sum_{it}\widehat U_{it,m}^{*}(\tilde y_{it}^*-\widehat\delta_{yt}^{*'}\widehat f_i^*-\widehat U_{it,m^-}^{*'}\gamma_{k,m^-}-\widehat U_{it,m^+}^{*'}\gamma_{k-1,m^+}-\widehat U_{it,m}^* \gamma_{k,m})\cr
&=&\frac{2}{nT}\sum_{it}\widehat U_{it,m}^{*}(\tilde y_{it}^*-\widehat\delta_{yt}^{*'}\widehat f_i^*-\widehat U_{it}^{*'}\gamma_{k}^{(m)})\cr
&=&\frac{2}{nT}\sum_{it}\widehat U_{it,m}^{*}(
 M_{it}^*+\tilde e_{it}^*+\widehat U_{it}^{*'}(\widehat\gamma
-\gamma_{k}^{(m)}))
 \end{eqnarray}
where $\gamma_{k}^{(m)}:=(\gamma_{k, m^-}, \gamma_{k,m}, \gamma_{k-1, m^+})'$, and  $M_{it}^*$, $\tilde e_{it}^*$ are respectively defined in (\ref{ec.2})(\ref{ec.3}).  Let $\widehat U_{it,\widehat J^*}^*$ denote the subvector of $\widehat U_{it}^*$, consisting of $\{\widehat U_{it,m }^{*}: \gamma_{k,m}\neq 0, m\leq p\}=\{\widehat U_{it,m}^{*}: m\in\widehat J^*\}$.
  Then  the vector form of (\ref{ec.9}) is
$$
-\kappa_n\widehat\Psi(\widehat J^*)sgn(\gamma_{k,m}: m\in\widehat J^*)=\frac{2}{nT}\sum_{it}\widehat U_{it,\widehat J^*}^{*}(M_{it}^*+\tilde e_{it}^*) +A,
$$
 where  (without loss of generality, we assume  $\{m\leq p: \gamma_{k,m}\neq 0\}=\{1,...,|\widehat J^*|_0\}$)
 \begin{eqnarray*}
 A&=&\begin{pmatrix}
\frac{2}{nT}\sum_{it}\widehat U_{it,1}^{*}\widehat U_{it}^{*'}(\widehat\gamma
-\gamma_{k}^{(1)})\\
\vdots\\
\frac{2}{nT}\sum_{it}\widehat U_{it,|\widehat J^*|_0}^{*}\widehat U_{it}^{*'}(\widehat\gamma
-\gamma_{k}^{(|\widehat J^*|_0)})
 \end{pmatrix}=
 \begin{pmatrix}
\frac{2}{nT}\widehat U_{1}^{*'}\widehat U^*(\widehat\gamma
-\gamma_{k}^{(1)})\\
\vdots\\
\frac{2}{nT}\widehat U_{ |\widehat J^*|_0}^{*'}\widehat U^*(\widehat\gamma
-\gamma_{k}^{( |\widehat J^*|_0)})
 \end{pmatrix}\cr
 &=&
  \begin{pmatrix}
  (\widehat\gamma
-\gamma_{k}^{(1)})'\frac{2}{nT}\widehat U^{*'}& 0\cdots&0\\
0 &\ddots&\\
0&\cdots&  (\widehat\gamma
-\gamma_{k}^{(|\widehat J^*|_0)})'\frac{2}{nT}\widehat U^{*'}
   \end{pmatrix}\begin{pmatrix}
   \widehat U_1^*\\
   \vdots\\
      \widehat U_{|\widehat J^*|_0}^*
   \end{pmatrix}:=\frac{2}{nT}B\widehat U_{\widehat J^*}.
 \end{eqnarray*}
 Therefore 
\begin{equation}\label{ec.11}
\kappa_n\|\widehat \Psi(\widehat J^*)\|_2\leq \max_j|\frac{2}{nT}\sum_{it}\widehat U_{it, j}^{*}(M_{it}^*+\tilde e_{it}^*)|\sqrt{|\widehat J^*|_0}+\|\frac{2}{\sqrt{nT}}B\|\|\frac{1}{\sqrt{nT}}\widehat U_{\widehat J^*}\|.
 \end{equation} 
 Note that here the norm in both $\|\frac{2}{\sqrt{nT}}B\|$ and $\|\frac{1}{\sqrt{nT}}\widehat U_{\widehat J^*}\|$ is the operator norm and we have used the inequality $\|B\widehat U_{\widehat J^*}\|_2\leq \|B\|\|\widehat U_{\widehat J^*}\|$.
 
 Now  by (\ref{ed.13}),
\begin{align}\label{ec.12}
\begin{split}
\frac{1}{nT}\|\widehat U^*_{\widehat J^*}\|^2 &\leq \frac{2}{nT}\|\tilde U^*_{\widehat J^*}\|^2+\frac{2}{nT}\|\widehat U^*_{\widehat J^*}-\tilde U^*_{\widehat J^*} \|^2 \\
&\leq 2\phi_{\max}(|\widehat J^*|_0)+\frac{2}{nT}\sum_{t=1}^T\sum_{i=1}^n\sum_{m\in\widehat J^*}(\widehat U^*_{it, m}-\tilde  U^*_{it, m})^2 \\
&\leq 2\phi_{\max}(|\widehat J^*|_0)+ O_{P^*}(\Delta_F^{*2}+\frac{\log (pT)}{n})|\widehat J^*|_0, \ \text{and} \\
\frac{1}{{nT}}\|B\|^2 &=\max_{m\in\widehat J^*}\frac{4}{nT}\|\widehat U^*(\widehat\gamma- \gamma_k^{(m)})\|^2 \\
&\leq  \frac{8}{nT}\|\widehat U^*(\widehat\gamma-\tilde\gamma_{lasso}^{*})\|_2^2+\max_{m\in\widehat J^*}\frac{8}{nT}\|\widehat U^*(\tilde\gamma_{lasso}^*-\gamma_k^{(m)})\|^2_2 \\
&\leq O_{P^*}(\kappa_n^2|J|_0).
\end{split}
\end{align}
where $\frac{1}{nT}\|\widehat U^*(\widehat\gamma-\tilde\gamma_{lasso}^{*})\|_2^2=O_{P^*}(\kappa_n^2|J|_0)$ follows from (\ref{ed.4ad}). To show 
$$
\max_{m\in\widehat J^*}\frac{8}{nT}\|\widehat U^*(\tilde\gamma_{lasso}^*-\gamma_k^{(m)})\|^2_2=O_{P^*}(\kappa_n^2|J|_0),
$$ we note that part (i) and (\ref{ec.9a}) demonstrate
\begin{align*}
\mathcal{L}^*(\gamma_k^{(m)})+\kappa_n\|\widehat\Psi  \gamma_k^{(m)}\|_1 &\leq \mathcal{L}^*(\gamma_{0})+\kappa_n\|\widehat\Psi  \gamma_{0}\|_1 \\
&= \mathcal{L}^*(\widehat\gamma)+\kappa_n\|\widehat\Psi  \widehat\gamma\|_1
&\leq \mathcal{L}^*(\widehat\gamma^*_{lasso})+\kappa_n\|\widehat\Psi  \widehat \gamma^*_{lasso}\|_1+O_{P^*}(\kappa_n^2|J|_0).
\end{align*}
Thus, the same argument as used in equation (\ref{ed.3new}) leads to 
 $$
 \max_{m\in\widehat J^*}\frac{8}{nT}\|\widehat U^*(\tilde\gamma_{lasso}^*-\gamma_k^{(m)})\|^2_2\leq O_{P^*}(\kappa_n^2|J|_0).
 $$

By Lemma \ref{ld.5new}, $ \max_j|\frac{2}{nT}\sum_{it}\widehat U_{it, j}^{*}(M_{it}^*+\tilde e_{it}^*)|=o_{P^*}(\kappa_n)$ and  $\kappa_n\|\widehat \Psi(\widehat J^*)\|_2\geq c\kappa_n$. Hence, (\ref{ec.11}) and (\ref{ec.12}) imply  
 $$
 \kappa_n^2{|\widehat J^*|_0}\leq   (C\phi_{\max}(|\widehat J^*|_0)+ O_{P^*}(\Delta_F^{*2}+\frac{\log (pT)}{n})|\widehat J^*|_0)\kappa_n^2|J|_0.
 $$
We thus obtain exactly the same inequality as given (\ref{ea.3add}). The rest of the proof then follows from the same argument as used to show Proposition \ref{p2.1} (ii). We conclude $|\widehat J^*|_0=O_{P^*}(|J|_0)$ which verifies  Assumption \ref{a6.1}(ii).
$\blacksquare$

  The following lemma is useful to uniformly bound terms in the bootstrap sampling space. 
  \begin{lemma}\label{ld.1ad}
  Suppose the following conditions hold:\\
  (i)   $\{Z_{ijm}\}$ is a sequence of random variables in the original sampling space, satisfying 
$$
\max_{m\leq p, i\leq n}\frac{1}{n}\sum_{j=1}^nZ_{ijm}^2=O_P(a_n^2)
$$ 
for some deterministic sequence $a_n>0.$ \\
  (ii)  $\{X_i^*, Y_i^*\}_{i\leq n}$ is an i.i.d. sequence in the bootstrap sampling space such that $\{X_i^*\} $ is independent of  $\{Y_i^*\}$, $EX_i^*=EY_i^*=0$, and $\Var^*(X_i)<C$ and $\Var^*(Y_i)<C$ for a constant $C>0$ where $C$ is non-random in both the original and bootstrap sampling space.\\
  (iii) Both $X_i^*$ and $Y_i^*$ are sub-exponential random variables satisfying Assumption \ref{a3.2} (iv).  
  
  Then for any $\varepsilon_1, \varepsilon_2>0$, there is a $C_{\varepsilon_1,\varepsilon_2}>0$ such that
  $$
  P\left(P^*\left(\max_{m\leq p}\left|\frac{1}{n^2}\sum_{i,j\leq n}X_i^*Y_j^*Z_{ijm}\right|>\frac{2a_nC\sqrt{C_{\varepsilon_1, \varepsilon_2}\log p\log(pn)}}{n}\right)>\varepsilon_1\right)<\varepsilon_2.
  $$
  Thus $\max_{m\leq p}|\frac{1}{n^2}\sum_{i,j\leq n}X_i^*Y_j^*Z_{ijm}|=O_{P^*}\left(\frac{a_n\sqrt{\log p\log(np)}}{n}\right)$. 
  \end{lemma}
	
  \proof By condition (i), for any $\delta>0$, there is $C_{\delta}>0$ such that with probability at least $1-\delta$ the event $A_{\delta}:=\{\max_{m\leq p, i\leq n}\frac{1}{n}\sum_{j=1}^nZ_{ijm}^2<a_n^2C_{\delta}\}$ holds. 
  
Let $V^*= \max_{mi}|\frac{1}{n}\sum_jY_j^*Z_{ijm}|$. Define $W_{im}^*=X_i^*\frac{1}{n}\sum_jY_j^*Z_{ijm}$ and $Y^*=\{Y_i^*\}_{i\leq n}$.  Since $\{X_i^*\}$ and $\{Y_i^*\}$ are independent,   then on the event $A_{\delta}$,
\begin{align}\label{ed.2a}
\begin{split}
\max_{m,i}\frac{1}{n}\sum_j\Var^*(Y_j^*Z_{ijm}) &=\max_{m,i}\frac{1}{n}\sum_jZ_{ijm}^2\Var(Y_j^*) \\
&<a_n^2CC_{\delta} \\
\max_m\frac{1}{n}\sum_i\Var^*(W_{im}^*|Y^*)
&=\max_m\frac{1}{n}\sum_i\left(\frac{1}{n}\sum_jY_j^*Z_{ijm}\right)^2\Var^*(X_i^*) \\
&\leq   CV^{*2}.
\end{split}
\end{align}  
In the bootstrap sampling space (BSS), {$\{Y_j^*Z_{ijm}\}_{j \leq n}$ is independent across $j$} and $E^*Y_j^*Z_{ijm}=0$. By the Bernstein inequality,  for $y=(2a_n^2CC_{\delta}\log(pn)/n)^{1/2}$,
\begin{align}\label{ed.3a}
\begin{split}
P^*(V^*>y)1\{A_{\delta}\} &\leq pn\max_{m,i}P^*\left(|\frac{1}{n}\sum_jY_j^*Z_{ijm}|>y\right)1\{A_{\delta}\} \\
&\leq \exp\left(\log(pn)-\frac{ny^2}{\max_{m,i}\frac{1}{n}\sum_j\Var^*(Y_j^*Z_{ijm})}\right)1\{A_{\delta}\} \\
&\leq\exp\left(\log(pn)-\frac{ny^2}{ a_n^2CC_{\delta}}\right) = (pn)^{-1}.
\end{split}
\end{align}
In the BSS, $\{W_{im}^*\}_{i\leq n}$ is independent across $i$ conditional on $Y^*$.
 By (\ref{ed.2a}) and the Bernstein inequality, for $x=y\sqrt{\frac{2C\log p}{n}}=\frac{2a_nC\sqrt{C_{\delta}\log p\log(pn)}}{n}$,
\begin{align}\label{ed.4a}
\begin{split}
P^*(\max_{m\leq p}|\frac{1}{n}&\sum_{i}W_{im}^*|>x|Y^*)1\{V^*<y\} \\
&\leq p\max_mP^*\left(|\frac{1}{n}\sum_{i}W_{im}^*|>x |Y^*\right)1\{V^*<y\} \\
&\leq \exp\left(\log p-\frac{nx^2}{\max_m\frac{1}{n}\sum_i\Var^*(W_{im}^*|Y^*)}\right)1\{V^*<y\} \\
&\leq \exp\left(\log p-\frac{nx^2}{CV^{*2}}\right)1\{V^*<y\} \\
&\leq \exp\left(\log p-\frac{nx^2}{Cy^2}\right)=p^{-1}.
\end{split}
\end{align}

Let $E_{Y^*}$ denote the expectation operator with respect to the marginal distribution of $Y^*$ in the bootstrap sampling space; i.e., $E_{Y^*}$ is the conditional distribution of $Y^*$ given the original data. By the law of iterated expectations, $E_{Y^*}\left[P^*(\cdot|Y^*)\right]=P^*(\cdot)$. We then have
\begin{align*}
P^*&\left(\max_{m\leq p}|\frac{1}{n^2}\sum_{i,j\leq n}X_i^*Y_j^*Z_{ijm}|>x\right) \\
&\leq P^*\left(\max_{m\leq p}|\frac{1}{n^2}\sum_{i,j\leq n}X_i^*Y_j^*Z_{ijm}|>x\right)1\{A_{\delta}\}+1\{A_{\delta}^c\} \\
&= P^*\left(\max_{m\leq p}|\frac{1}{n}\sum_{i}W_{im}^*|>x\right)1\{A_{\delta}\}+1\{A_{\delta}^c\} \\
&= E_{Y^*}P^*\left(\max_{m\leq p}|\frac{1}{n}\sum_{i}W_{im}^*|>x|Y^*\right)1\{A_{\delta}\}+1\{A_{\delta}^c\} \\
&\leq^{(a)} E_{Y^*}P^*\left(\max_{m\leq p}|\frac{1}{n}\sum_{i}W_{im}^*|>x|Y^*\right)1\{V^*<y\}1\{A_{\delta}\}+E_{Y^*} 1\{V^*\geq y\}1\{A_{\delta}\}+1\{A_{\delta}^c\} \\
&\leq^{(b)} p^{-1}+P^*(V^*\geq y)1\{A_{\delta}\}+1\{A_{\delta}^c\} \\
&\leq^{(c)} p^{-1}+(pn)^{-1}+1\{A_{\delta}^c\},
\end{align*}
where we used $P^*(\cdot |Y^*)\leq P^*(\cdot |Y^*)1\{V^*<y\}+1\{V^*\geq y\}$ in (a), (\ref{ed.4a}) in (b), and (\ref{ed.3a}) in (c). Because $P(A_{\delta}^c)\leq \delta$, taking the expectation with respect to the distribution of the original data on both sides yields
    $$
  EP^*(\max_{m\leq p}|\frac{1}{n^2}\sum_{i,j\leq n}X_i^*Y_j^*Z_{ijm}|>x)\leq  p^{-1}+(pn)^{-1}+\delta.
  $$
For any $\varepsilon_1, \varepsilon_2>0$, let $\delta=\varepsilon_1\varepsilon_2/2$, and call $C_{\delta}$ in $x$ to  be $C_{\varepsilon_1,\varepsilon_2}$. By the Markov Inequality, we then have 
\begin{align*}
 P\left(P^*(\max_{m\leq p}|\frac{1}{n^2}\sum_{i,j\leq n}X_i^*Y_j^*Z_{ijm}|>x)>\varepsilon_1\right)
 &\leq \frac{1}{\varepsilon_1}(p^{-1}+(pn)^{-1}+\delta) \\
&\leq\frac{\varepsilon_1\varepsilon_2/2+\delta}{\varepsilon_1}=\varepsilon_2.
\end{align*}
   $\blacksquare$

\section{Verifying Conditions for Estimating the Factors } \label{sf}

This section verifies Assumptions \ref{ca.1} and \ref{a6.2} when factors are estimated using PCA.

\subsection{Proof of Proposition \ref{p4.1} (for \texorpdfstring{$\widehat F$}{F-hat} using the original data)} 

%We first prove a less sharp rate: $ \frac{1}{n}\sum_{i=1}^n\|\widehat f_i-H' f_i\|_2^2=O_P(\frac{1}{pT}+\frac{1}{n})$, that is, $n^{-2}$ is replaced with $n^{-1}$. Then we derive the shaper rate.
 
 %\begin{lemma}
 %$ \frac{1}{n}\sum_{i=1}^n\|\widehat f_i-H' \tilde f_i\|_2^2=O_P(\frac{1}{pT}+\frac{1}{n})$
 %\end{lemma}
 
(i) By Assumption \ref{pervasive},     %the pervasive condition, 
it can be shown that $\|H\|=O_P(1)=\|V^{-1}\|$.
In addition, we have the following identity:
 $$
 \widehat f_i-H'\tilde f_i=V^{-1} \sum_{l=1}^5A_{il},
 $$
 where 
\begin{align}\label{ec.1}
\begin{split}
 A_{i1} &= \frac{1}{pTn}\sum_{j=1}^n\widehat f_j\sum_{t=1}^T (\tilde U_{it}'\tilde U_{jt}-U_{it}' U_{jt}), \\ 
 A_{i2} &= \frac{1}{pTn}\sum_{j=1}^n\widehat f_j\sum_{t=1}^T (U_{it}' U_{jt}-EU_{it}' U_{jt}), \\
 A_{i3} &= \frac{1}{pTn}\sum_{j=1}^n\widehat f_j\sum_{t=1}^T EU_{it}'U_{jt}, \\
 A_{i4} &= \frac{1}{pTn}\sum_{j=1}^n\widehat f_j\sum_{t=1}^T\tilde f_j'\tilde \Lambda_t'\tilde U_{it}, \\
 A_{i5} &= \frac{1}{pTn}\sum_{j=1}^n\widehat f_j\sum_{t=1}^T\tilde f_i'\tilde \Lambda_t'\tilde U_{jt}.
\end{split}
\end{align}
Each term can be written in the form $A_{il}=\frac{1}{pTn}\sum_{j=1}^n\widehat f_j\sum_{t=1}^TB_{ijt, l}$. By Cauchy-Schwarz,
\begin{align}\label{ee.2}
\begin{split}
 \frac{1}{n}\sum_{i=1}^n\| \widehat f_i-H'\tilde f_i\|_2^2 &= O_P(1)  \sum_{l=1}^5\frac{1}{n}\sum_{i=1}^n\|A_{il}\|_2^2  \\
&\leq  O_P(1) \frac{1}{n^2}\sum_{i=1}^n \sum_{j=1}^n\left(\frac{1}{pT}\sum_{t=1}^TB_{ijt,l}\right)^2.
\end{split}
\end{align}
We bound the terms in (\ref{ee.2}) in Lemmas \ref{lc.1} and \ref{lc.2} below.  Then, applying the bounds in Lemmas \ref{lc.1} and \ref{lc.2} and using $T=o(n)$, we have 
$$
\frac{1}{n}\sum_{i=1}^n\|\widehat f_i-H'\tilde f_i\|_2^2=O_P(\Delta_F^2),\quad \Delta_F^2=\frac{1}{n^2}+ 
\frac{1}{nT^2}+\frac{1}{pT}.
$$
It  is then straightforward to verify that $|J|^2_0\Delta_F^2=o\left(\sqrt{\frac{1}{nT}}\right)$
  holds when $|J|_0^4=o(nT^3)$, $|J|_0^4n=o(p^2T)$, $|J|_0^2 \sqrt{\log p }  \log(pT)=o(n)$, and  $|J|_0^2T=o(n)$.  For example, to show $|J|_0^2 \sqrt{\log p }\log(T)=o(n)$, note that $|J|_0 =o(\sqrt{n/(\log p)})$ and $|J|_0=o(\sqrt{n/T})$ implies 
$$
|J|_0^2 \sqrt{\log p }  \log(T)=o(n\log T\sqrt{\log p}/\sqrt{T\log p})=o(n).
$$ 
  
%$|J|_0^2=o(n)$,  $|J|_0^2\log p=o(nT)$,   

(ii) We now verify that we can produce sequences $\Delta_{eg}$ so that $\Delta_{eg}=o\left(\frac{1}{\sqrt{nT}}\right)$. First, note that we can set $g_{tm} \in \{\gamma_d'\tilde\Lambda_t, \tilde\lambda_{tm}, \tilde\delta_t\}$ in applying Lemma \ref{lc.4}, each of which yields $\omega_n=O(|J|_0^2)$ for $\omega_n$ defined in Lemma \ref{lc.4}.  It then follows from Lemma \ref{lc.4} that we can take $\Delta_{eg}=(\frac{1}{\sqrt{npT}}+\frac{1}{n}) |J|_0$ so that $\Delta_{eg}=o\left(\sqrt{\frac{1}{nT}}\right),$ given $T|J|_0^2=o(n)$ and $|J|_0^2=o(p)$.  Note that $|J|_0^2=o(p)$ is implied by the assumption that $|J|_0^4n=o(p^2T)$.

% Then $\omega_n$ in Lemma \ref{lc.4} is $\omega_n=O(1)$, and hence $\Delta_{fel}=(\frac{1}{\sqrt{npT}}+\frac{1}{n})$ satisfies $\Delta_{fel}=o(\sqrt{\frac{\log p}{nT}})$.

(iii)
By Lemma \ref{lc.6}, we take $\Delta_{fum}=\frac{1}{n} + \frac{1}{T\sqrt{n}}+\frac{1}{T\sqrt{p}} +\sqrt{\frac{\log (pT)}{npT}}$, and $\Delta_{fe}^2=\frac{1}{n^2}+ \frac{1}{Tpn}$. We then have 
$$ 
\max_{s\leq T, m\leq p}|\frac{1}{n}  \sum_{i=1}^n (H^{'}\tilde f_i-\widehat f_i)\tilde U_{is,m} |=
O_P(  \Delta_{fum})
$$ 
and 
$$
\frac{1}{T} \sum_{t=1}^T\|\frac{1}{n}\sum_{i=1}^n \tilde e_{it}(\widehat  f_i-H'\tilde f_i)\|^2_2
 =O_P(\Delta_{fe}^2).  
$$    
Then, it is straightforward to check    $\Delta_{fum}^2=o\left(\frac{\log p}{T|J|^2\log(pT)}\right)$ and $\Delta_{fe}^2=o\left(\frac{\log p}{T\log(pT)}\right).$ 
 
(iv) By Lemma \ref{lc.7}, we can define $\Delta_{ud}=   \frac{1}{\sqrt{n}} \sqrt{\frac{\log(pT)}{nT}}  
 +\sqrt{\frac{\log p}{nT}}(\frac{1}{\sqrt{pT}}+   \frac{1}{T\sqrt{n}} )+\frac{1}{pT}$.  Given $|J|_0^4n=o(p^2T)$ and $|J|_0^4=o(nT^3)$, it is straightforward to verify that  $\Delta_{ud}=o\left(\sqrt{\frac{\log p}{nT}}\right) $ and 
 $|J|_0^2 \sqrt{\log p }      \Delta_{ud}=o(1)$.  This result follows by verifying $\sqrt{\frac{\log p}{nT}}\frac{1}{\sqrt{pT}}|J|_0^2\sqrt{\log p}=o(1)$ which can be shown by noting that $|J|_0^4n=o(p^2T)$ implies $|J_0|^2=o(p\sqrt{T/n})$ and that $|J|_0^2\log (p)=o(n)$. Thus, because $\log^2 p=o(n)$,
 $$
\left( \sqrt{\frac{\log p}{nT}}\frac{|J|_0^2\sqrt{\log p}}{\sqrt{pT}}\right)^2=\frac{\log^2 p|J|_0^4}{npT^2}=o\left(\frac{\log(p)pn\sqrt{T}}{npT^2\sqrt{n}}\right)=o\left(\frac{\log(p)\sqrt{T}}{T^2\sqrt{n}}\right)=o(1).
 $$

(v) First, by Lemma \ref{lc.8}, we can take $\Delta_{\max}=\frac{1}{\sqrt{n}}  +\sqrt{\frac{\log n}{Tp}}$.
 %Secondly, by Lemma \ref{lc.6}, we have
    % $ \frac{1}{T}\sum_t\|\frac{1}{n}\sum_i (\widehat f_i-H'\tilde f_i)\tilde U_{it}'\gamma\|_2^2\leq \max_{ms}|\frac{1}{n}  \sum_{i=1}^n (H^{'}\tilde f_i-\widehat f_i)\tilde U_{is,m} |^2\|\gamma\|_1^2= O_P(\Delta_{fu}^2),$where  $\Delta_{fu}^2=(\frac{1}{n^2}+   \frac{1}{T^2{n}}+\frac{1}{T^2{p}} +{\frac{\log (pT)}{npT}}  )|J|_0^2$.
%Also by parts (i)(iii)$\Delta_{fe}^2=\frac{1}{n^2}+ \frac{1}{Tpn}$ and 
Also $\Delta_F^2=\frac{1}{n^2}+ 
 \frac{1}{nT^2}+\frac{1}{pT}$. This implies 
 $
 \Delta_F^2|J|^2_0+ {\frac{|J|_0}{n}}=O\left( \frac{|J|_0}{n}+  \frac{|J|_0^2}{nT^2}+\frac{|J|_0^2}{pT}\right).
 $
 In addition,  $ \kappa_n^2|J|_0\sqrt{nT}=o(1)$  and $\|R_y\|_1^2=o\left(\frac{\log p}{nT}\right)$ imply 
  $\lambda^2_n|J|_0+\|R_y\|_1^2=O\left(\frac{1}{\sqrt{nT}}\right)$. Hence with the conditions  $|J|_0^4=o(nT^3)$, $|J|_0^4n=o(p^2T)$, we have
  $$
  \lambda^2_n|J|_0+\|R_y\|_1^2+\Delta_F^2|J|^2_0+ {\frac{|J|_0}{n}}=O\left(\frac{1}{\sqrt{nT}}\right).
  $$
Thus, in order to verify 
$\Delta_{\max}^2|J|_0^2T(\lambda^2_n|J|_0+\|R_y\|_1^2+\Delta_F^2|J|^2_0+ {\frac{|J|_0}{n}})=o(1),$
it suffices to verify 
$$
\left(\frac{1}{n}+\frac{\log n}{Tp}\right)|J|_0^2T \frac{1}{\sqrt{nT}}=o(1),
$$
which holds given the conditions $|J|_0^2T=o(n)$ and $|J|_0^2=o(p)$.  Note that $|J|_0^2=o(p)$ is implied by $|J|_0^4n=o(p^2T)$.
 $\blacksquare$

\begin{lemma}\label{lc.8}
$\max_i\|\widehat f_i-H'\tilde f_i\|_2=O_P \left(  \frac{1}{\sqrt{n}}  +\sqrt{\frac{\log n}{Tp}}\right):=O_P(\Delta_{\max}^2).$
\end{lemma}

\proof  We first bound  $\max_i\frac{1}{pT}\sum_{tm}  U_{it,m}^2$ and $\max_i\| \frac{1}{pT}\sum_{t=1}^T\tilde \Lambda_t' U_{it}\|_2$.
Since 
$$
\max_i\Var\left(\frac{1}{p}\sum_{m}  U_{it,m}^2\right)=O(p^{-1}),
$$ 
we have
\begin{align*}
\max_i\frac{1}{pT}\sum_{tm}  U_{it,m}^2 &\leq \max_i|\frac{1}{T}\sum_t(\frac{1}{p}\sum_{m}  U_{it,m}^2-E\frac{1}{p}\sum_{m}  U_{it,m}^2)|+\max_{itm}EU_{it,m}^2 \\
&=O_P\left(\sqrt{\frac{\log n}{pT}}+1\right)
\end{align*}
by the Bernstein inequality for strong mixing sequences.

For the $k^{\text{th}}$ row $\tilde\lambda_{tk}'$ of $\tilde\Lambda_t'$, 
\begin{align*}
\max_{itk} \Var\left(\frac{1}{p}\tilde\lambda_{tk}'U_{it}\right)
&= \frac{1}{p^2}\tilde\lambda_{tk,m}\tilde\lambda_{tk,l}\sum_{m,l}EU_{it,m}U_{it,l} \\
&\leq O\left(\frac{1}{p}\right)\max_{itm}\sum_l|EU_{it,m}U_{it,l}| \\
&=O\left(\frac{1}{p}\right).
\end{align*}  
Employing the Bernstein inequality for weakly dependent data (e.g., \cite{merlevede2011bernstein}), we then have
\begin{align}\label{ec.6}
\begin{split}
\max_i\| \frac{1}{pT}\sum_{t=1}^T\tilde\Lambda_t' U_{it}\|_2 &\leq O_P\left(\sqrt{\frac{\log n}{T}}\right)\sqrt{\max_{i\leq n, k\leq K, t\leq T}\Var\left(\frac{1}{p}\tilde\lambda_{tk}'U_{it}\right)} \\
&=O_P\left(\sqrt{\frac{\log n}{Tp}}\right).
\end{split}
\end{align}

Now, $\max_i\|\widehat f_i-H'\tilde f_i\|_2\leq\sum_{l=1}^3G_l$ where each $G_l$ is defined and bounded below.  Specifically, 
\begin{align*}
G_1 &= \max_i\|\frac{1}{pTn}\sum_{t=1}^T \tilde U_{it}'\sum_{j=1}^n\tilde U_{jt}\widehat f_j \|_2 \\
    &\leq \max_i\|\frac{1}{pTn}\sum_{t=1}^T   U_{it}'\sum_{j=1}^n\tilde U_{jt}\widehat f_j \|_2+\max_i \|\frac{1}{pTn}\sum_{t=1}^T   \bar U_{\cdot t}'\sum_{j=1}^n\tilde U_{jt}\widehat f_j \|_2 \\
%\leq\max_i\|\frac{1}{pTn}\sum_{t}\sum_j\sum_m\tilde U_{it,m}\tilde U_{jt,m}\widehat f_j\|_2\cr
    &\leq \left(2\max_i\frac{1}{pT}\sum_{tm}  U_{it,m}^2+  2\frac{1}{pT}\sum_{tm}  \bar U_{\cdot t,m}^2\right)^{1/2}\left(\frac{1}{pT}\sum_{tm} \| \frac{1}{n}\sum_j\tilde U_{jt,m}\widehat f_j\|_2^2\right)^{1/2} \\
    &= O_P\left(\left(\sqrt{\frac{\log n}{pT}}+1\right)\left( \frac{1}{T\sqrt{p}}  + \frac{1}{\sqrt{n}}\right)\right)  
\end{align*}
by equation (\ref{ec.4}) given below in Lemma \ref{lc.4}.  
\begin{align*}
G_2 &= \max_i\| \frac{1}{pTn}\sum_{j=1}^n\widehat f_j\sum_{t=1}^T\tilde f_j'\tilde\Lambda_t'\tilde U_{it}\|_2 \\
    &\leq O_P(1)\max_i\| \frac{1}{pT}\sum_{t=1}^T\tilde \Lambda_t'U_{it}\|_2+O_P(1)\| \frac{1}{pT}\sum_{t=1}^T\tilde \Lambda_t' \bar U_{\cdot t}\|_2 \\
    &= O_P\left(\sqrt{\frac{\log n}{Tp}}\right)
\end{align*}
by (\ref{ec.6}).  Finally,
\begin{align*}
G_3 &= \max_i\|\frac{1}{pTn}\sum_{t=1}^T\tilde f_i'\tilde\Lambda_t'\sum_{j=1}^n\tilde U_{jt}\widehat f_j'\|_2 \\
    &\leq \max_i\|\tilde f_i\|_2\|  \frac{1}{pTn}\sum_{t=1}^T\tilde\Lambda_t'\sum_{j=1}^n\tilde U_{jt}\widehat f_j'\|_F \\
    &\leq O_P(\sqrt{\log n})\| \frac{1}{pTn}\sum_{t=1}^T\tilde\Lambda_t'\sum_{j=1}^n\tilde U_{jt} \tilde f_j'\|_F
\\
	  & \qquad + O_P(\sqrt{\log n})\| \frac{1}{pTn}\sum_{t=1}^T\tilde\Lambda_t'\sum_{j=1}^n\tilde U_{jt}(\widehat f_j-H'\tilde f_j)'\|_F \\
    &= O_P\left(\sqrt{\frac{\log n}{pTn}} \right) + O_P(\sqrt{\log n}\Delta_F)\left(\frac{1}{n}\sum_j\|\frac{1}{pT}\sum_t\tilde\Lambda_t'\tilde U_{jt}\|_2^2\right)^{1/2} \\
		&= O_P\left(\sqrt{\frac{\log n}{pTn}} +\Delta_F\sqrt{\frac{\log n}{pT}}\right).
\end{align*}
Hence, $\max_i\|\widehat f_i-H'\tilde f_i\|_2=O_P \left(  \frac{1}{\sqrt{n}}  +\sqrt{\frac{\log n}{Tp}}\right).$
 $\blacksquare$

\begin{lemma}\label{lc.4}
Let $\{z_{it}\}$ be a random sequence with $E(z_{it}|f_t, U_t)=0$ and $\Var(z_{it})>0$.  In addition, let $\{g_{tm}\}$ be a deterministic sequence of vectors with a  fixed dimension, $m\leq p$.  Then for $\tilde z_{it}=z_{it}-\bar z_{i\cdot}-\bar z_{\cdot t}+\bar{\bar z}$,  and $\omega_n=\max_{m\leq p}\frac{1}{T}\sum_{t=1}^T \|g_{tm}\|_2^2,$
$$
\max_{m\leq p}\|  \frac{1}{nT}\sum_{i=1}^n\sum_{t=1}^Tg_{tm}\tilde z_{it}(\widehat f_i-H'\tilde f_i)'\|_F=
O_P\left(\frac{1}{\sqrt{npT}}+\frac{1}{n}\right)\omega_n^{1/2}.
$$
\end{lemma}
  
\proof It follows from equation (\ref{ec.1}) that 
$$
\max_{m\leq p}\|  \frac{1}{nT}\sum_{i=1}^n\sum_{t=1}^Tg_{tm}\tilde z_{it}(\widehat f_i-H'\tilde f_i)'\|_F\leq \sum_{l=1}^3\bar C_lO_P(1),
$$
where each term $\bar C_l$ is defined and bounded in  below.
\begin{align*}
\bar C_1 &= \max_{m\leq p}\|  \frac{1}{nT}\sum_{i=1}^n\sum_{t=1}^T \frac{1}{pTn}\sum_{j=1}^n\widehat f_j\sum_{s=1}^T \tilde U_{is}'\tilde U_{js}\tilde z_{it}g_{tm}'\|_F \\
&\leq \max_{m\leq p}\|  \frac{1}{nT}\sum_{i=1}^n\sum_{t=1}^T \frac{1}{pTn}\sum_{j=1}^n(\widehat f_j-H'\tilde f_j)\sum_{s=1}^T \tilde U_{is}'\tilde U_{js}\tilde z_{it}g_{tm}'\|_F \\
& \qquad + O_P(1)\max_{m\leq p}\|  \frac{1}{nT}\sum_{i=1}^n\sum_{t=1}^T \frac{1}{pTn}\sum_{j=1}^n\tilde f_j\sum_{s=1}^T \tilde U_{is}'\tilde U_{js}\tilde z_{it}g_{tm}'\|_F \\
 %&\leq&\left(  \frac{1}{pT^2}\sum_{t=1}^T \sum_{s=1}^T\|\frac{1}{n}\sum_{i=1}^n\tilde z_{it}\tilde U_{is}\|_2^2 \right)^{1/2}\left( \max_{m\leq p}\frac{1}{T}\sum_{t=1}^T \|g_{tm}\|_2^2\right)^{1/2}\left( \frac{1}{npT}\|\tilde U\|_F^2   \right)^{1/2}O_P(\Delta_F) \cr
% &&+\left(\frac{1}{pT^2}\sum_s  \sum_{t=1}^T   \|\frac{1}{n}\sum_{i=1}^n     \tilde z_{it}  \tilde  U_{is}  \|_2 ^2\right)^{1/2}  \left( \max_{m\leq p} \frac{1}{T} \sum_{t=1}^T\| g_{tm}\|_2^2\right)^{1/2}  O_P(1)(\frac{1}{pT}\sum_s  \| \frac{1}{n} \sum_{j=1}^n\tilde f_j \tilde U_{js}' \|_F^2  )^{1/2} \cr
&\leq (x_n\omega_n)^{1/2}\left( \left( \frac{1}{npT}\|\tilde U\|_F^2   \right)^{1/2}O_P(\Delta_F)  
 +O_P(1)(\frac{1}{pT}\sum_s  \| \frac{1}{n} \sum_{j=1}^n\tilde f_j \tilde U_{js}' \|_F^2  )^{1/2}
 \right) \\
&= (x_n\omega_n)^{1/2}O_P(\Delta_F+\frac{1}{\sqrt{n}}) \\
&=\omega_n^{1/2} O_P\left(\frac{1}{\sqrt{npT}}+\frac{1}{n}\right).
\end{align*}
where $x_n= \frac{1}{pT^2}\sum_{t=1}^T \sum_{s=1}^T\|\frac{1}{n}\sum_{i=1}^n\tilde z_{it}\tilde U_{is}\|_2^2=O_P(\frac{1}{{n}})$ and $\omega_n=\max_{m\leq p}\frac{1}{T}\sum_{t=1}^T \|g_{tm}\|_2^2$.  Next,
\begin{align*}
\bar C_2 &= \max_{m\leq p}\|  \frac{1}{nT}\sum_{i=1}^n\sum_{t=1}^T  \frac{1}{pTn}\sum_{j=1}^n\widehat f_j\sum_{s=1}^T\tilde f_j'\tilde \Lambda_s'\tilde U_{is} \tilde z_{it}g_{tm}'\|_F \\
&\leq O_P(\omega_n^{1/2})  \left( \frac{1}{T}\sum_{t=1}^T \| \frac{1}{pT} \frac{1}{n}  \sum_{i=1}^n\sum_{s=1}^T \tilde \Lambda_s'\tilde U_{is} \tilde z_{it}\|_2^2\right)^{1/2} \\
&= O_P\left(\omega_n^{1/2}\frac{1}{\sqrt{npT}}\right) 
\end{align*}
by Lemma \ref{lc.3}).  Finally,
\begin{align*}
\bar C_3 &= \max_{m\leq p}\|  \frac{1}{nT}\sum_{i=1}^n\sum_{t=1}^T\frac{1}{pTn}\sum_{j=1}^n\widehat f_j\sum_{s=1}^T\tilde f_i'\tilde \Lambda_s'\tilde U_{js} \tilde z_{it}g_{tm}'\|_F \cr
&\leq  \max_{m\leq p}\|  \frac{1}{nT}\sum_{i=1}^n\sum_{t=1}^T\frac{1}{pTn}\sum_{j=1}^n(\widehat f_j-H'\tilde f_j)\sum_{s=1}^T\tilde f_i'\tilde \Lambda_s'\tilde U_{js} \tilde z_{it}g_{tm}'\|_F\cr
& \qquad + O_P(1) \max_{m\leq p}\|  \frac{1}{nT}\sum_{i=1}^n\sum_{t=1}^T\frac{1}{pTn}\sum_{j=1}^n \tilde f_j\sum_{s=1}^T\tilde f_i'\tilde \Lambda_s'\tilde U_{js} \tilde z_{it}g_{tm}'\|_F\cr
&\leq  (\omega_nc_n)^{1/2}\left( O_P(\Delta_F)( \frac{1}{n}\sum_j \|\frac{1}{pT}  \sum_{s=1}^T\tilde \Lambda_s'\tilde U_{js}\|_2^2 )^{1/2}   +O_P(1)
 \max_{m\leq p}\| \frac{1}{pTn}  \sum_{j=1}^n\sum_{s=1}^T    \tilde \Lambda_s'\tilde U_{js}\tilde f_j '  \|_F\right) \\
 &= \omega_n^{1/2} O_P\left(\frac{1}{\sqrt{npT}}+\frac{1}{n}\right)
\end{align*}
where $ c_n=\frac{1}{T}\sum_{t=1}^T\|\frac{1}{n}\sum_{i=1}^n\tilde f_i\tilde z_{it}\|_2^2 =O_P(\frac{1}{n})$.
Therefore, 
$$
\max_{m\leq p}\|  \frac{1}{nT}\sum_{i=1}^n\sum_{t=1}^Tg_{tm}\tilde z_{it}(\widehat f_i-H'\tilde f_i)'\|_F=O_P\left(\frac{1}{\sqrt{npT}}+\frac{1}{n}\right)\omega_n^{1/2}.
$$
  $\blacksquare$
 
\begin{lemma}\label{lc.7}
Let $\{g_{tk}: k\leq p\}$ be a deterministic sequence of vectors of fixed dimension with $\max_{tk}\|g_{tk}\|_2=O(1)$, and let $\tilde g_{tk}=g_{tk}-\frac{1}{T}\sum_t g_{tk}$. Then
\begin{align*}
\max_{m,k\leq p}\|\frac{1}{nT} \sum_{i=1}^n  \sum_{t=1}^T(\widehat  f_i -H'\tilde f_i) \tilde U_{it,m} \tilde g_{tk}'  \|_F &= O_P\left(   \frac{1}{\sqrt{n}} \sqrt{\frac{\log(pT)}{nT}}  
 +\sqrt{\frac{\log p}{nT}}(\frac{1}{\sqrt{pT}}+   \frac{1}{T\sqrt{n}} )+\frac{1}{pT}\right) \\
&:=O_P(\Delta_{ud}).
\end{align*}
\end{lemma}
\proof

First, note that $\sum_t\tilde g_{tk}=0$. Hence,
\begin{align}
\nonumber 
\max_{m,k\leq p}\|\frac{1}{nT} &\sum_{i=1}^n  \sum_{t=1}^T(\widehat  f_i -H'\tilde f_i) \tilde U_{it,m} \tilde g_{tk}'  \|_F \\
\label{lemmaE3.i}
&\leq \max_{m,k\leq p}\|\frac{1}{nT} \sum_{i=1}^n  \sum_{t=1}^T(\widehat  f_i -H'\tilde f_i)  U_{it,m} \tilde g_{tk}'  \|_F \\
\label{lemmaE3.ii}
& \qquad +\|\frac{1}{n} \sum_{i=1}^n (\widehat  f_i -H'\tilde f_i) \|_2\max_{m,k\leq p}\| \frac{1}{T} \sum_{t=1}^T\bar U_{\cdot t,m} \tilde g_{tk}'  \|_F.
\end{align}
Term (\ref{lemmaE3.ii}) is $O_P\left(\Delta_F\sqrt{\frac{\log p}{nT}}\right)$. Term (\ref{lemmaE3.i}) is bounded by $\sum_{l=1}^7C_l$, where each $C_l$ is defined and bounded blow.
 
First, note that applying Lemma \ref{lc.6} gives     
\begin{align}\label{ec.4}
\begin{split}
\frac{1}{\sqrt{p}}\left(\frac{1}{T}\sum_t  \|   \frac{1}{n}\sum_j\widehat f_j\tilde U_{jt}'\|_F^2  \right)^{1/2}&\leq \frac{2}{\sqrt{p}}\left(\frac{1}{T}\sum_t  \|   \frac{1}{n}\sum_j(\widehat f_j-H'\tilde f_j)\tilde U_{jt}'\|_F^2  \right)^{1/2} \\
& \qquad +\frac{2}{\sqrt{p}}\left(\frac{1}{T}\sum_t  \|   \frac{1}{n}\sum_jH'\tilde f_j\tilde U_{jt}'\|_F^2  \right)^{1/2} \\
&= O_P\left(\frac{1}{n}+   \frac{1}{T\sqrt{n}}+\frac{1}{T\sqrt{p}} +\sqrt{\frac{\log (pT)}{npT}} + \frac{1}{\sqrt{n}} \right) \\
&= O_P( \frac{1}{T\sqrt{p}}  + \frac{1}{\sqrt{n}} ) \ \text{and} \\
\max_{tl}\|\frac{1}{n}\sum_{j=1}^n\widehat f_j \tilde U_{jt,l}\|_2 &= O_P\left(\sqrt{\frac{\log(pT)}{n}}+ \frac{1}{T\sqrt{p}} \right).
\end{split}
\end{align}
We then have, up to an $\|V^{-1}\|=O_P(1)$ term, 
\begin{align*}
C_1 &= \max_{m,k\leq p}\|\frac{1}{nT} \sum_{i=1}^n  \sum_{s=1}^T\frac{1}{pTn}\sum_{j=1}^n\widehat f_j\sum_{t=1}^T \tilde U_{jt}'(\tilde U_{it}   -U_{it}) U_{is,m} \tilde g_{sk}'  \|_F \\
%&\leq&\max_{m,k\leq p}\|\frac{1}{nT} \sum_{i=1}^n  \sum_{s=1}^T\frac{1}{pTn}\sum_{j=1}^n\widehat f_j\sum_{t=1}^T \tilde U_{jt}'\bar U_{i\cdot }    U_{is,m} \tilde g_{sk}'  \|_F\cr
&= \max_{m,k\leq p}\|\frac{1}{nT} \sum_{i=1}^n  \sum_{s=1}^T\frac{1}{pTn}\sum_{j=1}^n\widehat f_j\sum_{t=1}^T \tilde U_{jt}'\bar U_{\cdot t }    U_{is,m} \tilde g_{sk}'  \|_F  \text{ (because $\frac{1}{T}\sum_t\tilde U_{jt}=0$)} \\
&\leq \left(\frac{1}{T}\sum_t  \|   \frac{1}{n}\sum_j\widehat f_j\tilde U_{jt}'\|_F^2  \right)^{1/2} \left(\frac{1}{T}\sum_t\|\bar U_{\cdot t }  \|_2^2\right)^{1/2}\max_{mk} \|\frac{1}{nTp}\sum_{i=1}^n\sum_{s=1}^T  U_{is,m} \tilde g_{sk}'  \|_2 \\
&= O_P\left( \frac{1}{T\sqrt{p}}  + \frac{1}{\sqrt{n}} \right)
\frac{1}{\sqrt{n}} \sqrt{\frac{\log p}{nT}} \\
&= O_P\left( \frac{1}{T\sqrt{pn}}  + \frac{1}{n} \right)
 \sqrt{\frac{\log p}{nT}}.
\end{align*}
%&&+\max_{m,k\leq p}\|\frac{1}{nT} \sum_{i=1}^n  \sum_{s=1}^T\frac{1}{pTn}\sum_{j=1}^n\widehat f_j\sum_{t=1}^T \tilde U_{jt}'\bar {\bar U}    U_{is,m} \tilde g_{sk}'  \|_F\cr
\begin{align}
\label{ee.9a2}
\begin{split}
C_2 &= \max_{m,k\leq p}\left\|\frac{1}{nT} \sum_{i=1}^n  \sum_{s=1}^T\frac{1}{pTn}\sum_{j=1}^n\widehat f_j\sum_{t=1}^T \tilde U_{jt}'(U_{it}    U_{is,m}-EU_{it}U_{is,m}) \tilde g_{sk}'  \right\|_F \\
%&\leq &(\frac{1}{T}\sum_t  \|   \frac{1}{n}\sum_j\widehat f_j\tilde U_{jt}'\|_F^2  )^{1/2}(\max_{mkt}\frac{1}{T}\sum_t  \|   \frac{1}{nTp}\sum_{i=1}^n\sum_{s=1}^T  (U_{it}    U_{is,m}-EU_{it}U_{is,m}) \tilde g_{sk}'   \|_F^2  )^{1/2}
&\leq \left(\frac{1}{T}\sum_t  \|   \frac{1}{n}\sum_j\widehat f_j\tilde U_{jt}'\|_F^2  \right)^{1/2} \max_{mkt}  \|   \frac{1}{nTp}\sum_{i=1}^n\sum_{s=1}^T  (U_{it}    U_{is,m}-EU_{it}U_{is,m}) \tilde g_{sk}'   \|_F
\\
&= O_P\left( \frac{1}{T\sqrt{p}}  + \frac{1}{\sqrt{n}} \right) \left(\sqrt{\frac{\log(pT)}{nT}}\right)
\end{split}
\end{align}
using Lemma \ref{lc.5}.
\begin{align*}
C_3 &= \max_{m,k\leq p}\left\|\frac{1}{nT} \sum_{i=1}^n  \sum_{s=1}^T\frac{1}{pTn}\sum_{j=1}^n\widehat f_j\sum_{t=1}^T \tilde U_{jt}'(EU_{it}    U_{is,m}) \tilde g_{sk}'  \right\|_F \\
&\leq \max_{itm}\sum_{s=1}^T\sum_{l=1}^p |(EU_{it,l}    U_{is,m})   |   \max_{sk}\|\tilde g_{sk}\|_2   \frac{1}{pT}  \max_{tl}\|\frac{1}{n}\sum_{j=1}^n\widehat f_j \tilde U_{jt,l}\|_2 \\
&= O_P\left(\frac{1}{pT} \right) O_P\left(\sqrt{\frac{\log(pT)}{n}}+ \frac{1}{T\sqrt{p}} \right).
\end{align*}

\begin{align*}
C_4 &= \max_{m,k\leq p}\left\|\frac{1}{nT} \sum_{i=1}^n  \sum_{s=1}^T \frac{1}{pTn}\sum_{j=1}^n\widehat f_j\sum_{t=1}^T\tilde f_j'\tilde \Lambda_t'(\tilde U_{it} -U_{it}) U_{is,m} \tilde g_{sk}'  \right\|_F \\
&= \max_{m,k\leq p}\left\|\frac{1}{nT} \sum_{i=1}^n  \sum_{s=1}^T \frac{1}{pTn}\sum_{j=1}^n\widehat f_j\sum_{t=1}^T\tilde f_j'\tilde \Lambda_t' \bar U_{\cdot t} U_{is,m} \tilde g_{sk}'  \right\|_F \text{ (because $\sum_t\tilde\Lambda_t=0$)} \\
&\leq O_P(1)  \left\|   \frac{1}{pT}  \sum_{t=1}^T \tilde \Lambda_t' \bar U_{\cdot t}\right\|_2\max_{m,k\leq p}  \left\|\frac{1}{nT} \sum_{i=1}^n\sum_{s=1}^T U_{is,m} \tilde g_{sk}'  \right\|_2 \\
&= O_P\left(\frac{1}{\sqrt{npT}}\right)\sqrt{\frac{\log p}{nT}}.
\end{align*}
\begin{align}
\label{ee.9a5}
\begin{split}
C_5 &= \max_{m,k\leq p}\left\|\frac{1}{nT} \sum_{i=1}^n  \sum_{s=1}^T \frac{1}{pTn}\sum_{j=1}^n\widehat f_j\sum_{t=1}^T\tilde f_j'\tilde \Lambda_t'  (U_{it}  U_{is,m}-EU_{it}  U_{is,m}) \tilde g_{sk}'  \right\|_F \\
&\leq O_P(1)\max_{m,k\leq p}\left\|\frac{1}{nT} \sum_{i=1}^n  \sum_{s=1}^T \frac{1}{pT}  \sum_{t=1}^T \tilde \Lambda_t'  (U_{it}  U_{is,m}-EU_{it}  U_{is,m}) \tilde g_{sk}'  \right\|_F \\
&= O_P\left(\sqrt{\frac{\log p}{npT^2}}\right) 
\end{split}
\end{align}
using the same proof as that of Lemma  \ref{lc.5} (ii).
\begin{align*}
C_6 &= \max_{m,k\leq p}\left\|\frac{1}{nT} \sum_{i=1}^n  \sum_{s=1}^T \frac{1}{pTn}\sum_{j=1}^n\widehat f_j\sum_{t=1}^T\tilde f_j'\tilde \Lambda_t'  (EU_{it}  U_{is,m}) \tilde g_{sk}'  \right\|_F \\
&\leq O_P(1)
\max_{m,k\leq p}\left\|\frac{1}{nT} \sum_{i=1}^n  \sum_{s=1}^T \frac{1}{pT} \sum_{t=1}^T\tilde \Lambda_t'  (EU_{it}  U_{is,m}) \tilde g_{sk}'  \right\|_F \\
&=O_P\left(\frac{1}{pT}\right).
\end{align*}
\begin{align*}
 C_7 &= \max_{m,k\leq p}\left\|\frac{1}{nT} \sum_{i=1}^n  \sum_{s=1}^T \frac{1}{pTn}\sum_{j=1}^n\widehat f_j\sum_{t=1}^T\tilde f_i'\tilde \Lambda_t'\tilde U_{jt}  U_{is,m} \tilde g_{sk}'  \right\|_F \\
 &\leq \left\| \frac{1}{pTn}\sum_{j=1}^n\sum_{t=1}^T  \widehat f_j\tilde U_{jt}    '    \tilde \Lambda_t\right\|_2 \max_{m,k\leq p}\left\|\frac{1}{nT} \sum_{i=1}^n  \sum_{s=1}^T \tilde f_i      U_{is,m} \tilde g_{sk}'  \right\|_F \\
 &\leq O_P\left(\sqrt{\frac{\log p}{nT}}\right)\left( \| \frac{1}{pTn}\sum_{j=1}^n\sum_{t=1}^T \tilde  f_j\tilde U_{jt}    '    \tilde \Lambda_t\|_2+ \| \frac{1}{n}\sum_{j=1}^n   ( \widehat f_j-H'\tilde f_j)\tilde U_{jt,m} \|_2  \right) \\
 &= O_P\left(\sqrt{\frac{\log p}{nT}}\right)\left( \frac{1}{n}+   \frac{1}{T\sqrt{n}}+\frac{1}{T\sqrt{p}} +\sqrt{\frac{\log (pT)}{npT}} \right).
\end{align*} 
 Combining the above, we reach
 \begin{align*}
 \max_{m,k\leq p}&\left\|\frac{1}{nT} \sum_{i=1}^n  \sum_{t=1}^T(\widehat  f_i -H'\tilde f_i) \tilde U_{it,m} \tilde g_{tk}'  \right\|_F \\
&= O_P\left(   \frac{1}{\sqrt{n}} \sqrt{\frac{\log(pT)}{nT}}  
 +\sqrt{\frac{\log p}{nT}}(\frac{1}{\sqrt{pT}}+   \frac{1}{T\sqrt{n}} +\Delta_F)+\frac{1}{pT}\right) \\
&= O_P\left(   \frac{1}{\sqrt{n}} \sqrt{\frac{\log(pT)}{nT}}  
 +\sqrt{\frac{\log p}{nT}}(\frac{1}{\sqrt{pT}}+   \frac{1}{T\sqrt{n}} )+\frac{1}{pT}\right).
 \end{align*} 
 $\blacksquare$

\subsection{Proof of Proposition \ref{p4.1} (for \texorpdfstring{$\widehat F^*$}{F-hat} using the bootstrap data)} 

 (i) Similar to (\ref{ec.1}), it  can be proven that there is $\|H^*\|=O_{P^*}(1)$ and $\|V^{*-1}\|=O_{P^*}(1)$such that 
 $$
 \widehat f_i^*-H^{*'}\widehat f_i=V^{*-1} \sum_{l=1}^4 A^*_{il},
 $$
 where 
\begin{equation}\label{ee.9}
\begin{aligned}
A^*_{i1} &= \frac{1}{pTn}\sum_{j=1}^n\widehat f_j^*\sum_{t=1}^T (\tilde U_{it}^{*'}\tilde U_{jt}^*- E^*\tilde U_{it}^{*'}\tilde U_{jt}^*),  
& A^*_{i2} &= \frac{1}{pTn}\sum_{j=1}^n\widehat f_j^*\sum_{t=1}^T E^*\tilde U_{it}^{*'}\tilde U_{jt}^*, \\
A^*_{i3} &= \frac{1}{pTn}\sum_{j=1}^n\widehat f_j^*\sum_{t=1}^T\widehat f_j'\widehat \Lambda_t'\tilde U_{it}^*,
& A^*_{i4} &= \frac{1}{pTn}\sum_{j=1}^n\widehat f_j^*\sum_{t=1}^T\widehat f_i'\widehat\Lambda_t'\tilde U_{jt}^*.
\end{aligned}
\end{equation}
% A^*_{i2}=\frac{1}{pTn}\sum_{j=1}^n\widehat f_j^*\sum_{t=1}^T (U_{it}' U_{jt}-EU_{it}' U_{jt}),

We first treat $A^*_{i2} - A_{i4}^*$. 
Because $\tilde U_{it}^*$ and $\tilde U_{jt}^*$ are independent if $i\neq j$, we have
\begin{align*}
\frac{1}{n}\sum_i\|A_{i2}^*\|_2^2 &=  \frac{1}{n}\sum_i  \|\frac{1}{pTn}\sum_{j=1}^n\widehat f_j^*\sum_{t=1}^T E^*\tilde U_{it}^{*'}\tilde U_{jt}^*\|_2^2 \\
&=  \frac{1}{n}\sum_i  \|\frac{1}{pTn} \widehat f_i^*\sum_{t=1}^T E^*\tilde U_{it}^{*'}\tilde U_{it}^*\|_2^2 \\
&= O_{P^*}\left(\frac{1}{n^2}\right).
\end{align*}
  By Lemma \ref{le.10}, $  \frac{1}{n}\sum_i\|A_{i3}^*\|_2^2+  \frac{1}{n}\sum_i\|A_{i4}^*\|_2^2=O_{P^*}(\Delta_F^2)$.  Hence 
  \begin{equation}\label{ee.10a}
  \sum_{l=2}^4 \frac{1}{n}\sum_i\|A_{il}^*\|_2^2=O_{P^*}(\Delta_F^2).
  \end{equation}

Now we bound $\frac{1}{n}\sum_{i=1}^n\|A^*_{i1}\|_2^2$. A preliminary rate is provided by Lemma \ref{le.9} where we have that 
$$
\frac{1}{n}\sum_{i=1}^n\|A^*_{i1}\|_2^2=O_{P^*}\left(\Delta_F^2+\frac{1}{n}\right).
$$
However, this rate is not sharp due to the $O_{P^*}(n^{-1})$ term and can be improved.  Specifically, the proof of Lemma \ref{le.9} (iii) uses a Cauchy-Schwarz inequality and is not sharp for  terms involving $E\left[U_{it}'U_{jt}\right]$. To see this intuitively, consider a simple example where we bound 
$$
\frac{1}{n}\sum_{i=1}^n\|\frac{1}{nTp}\sum_{j=1}^n\sum_{t=1}^Tf_jEU_{it}'U_{jt}\|^2_2.
$$
Since $U_{it}$ and $U_{jt}$ are independent when $i\neq j$, this term may be simplified to 
$$
\frac{1}{n}\sum_{i=1}^n\|\frac{1}{nTp}\sum_{t=1}^Tf_iEU_{it}'U_{it}\|^2_2 = O_P\left(\frac{1}{n^2}\right).
$$ 
In contrast, using the Cauchy-Schwarz inequality gives
\begin{align*}
\frac{1}{n}\sum_{i=1}^n\|\frac{1}{nTp}\sum_{j=1}^n\sum_{t=1}^Tf_jEU_{it}'U_{jt}\|^2_2 &\leq \frac{1}{n}\sum_j \|\widehat f_j\|_2^2 \frac{1}{n^2}\sum_j\sum_{i=1}^n\left(\frac{1}{Tp}\sum_{t=1}^TEU_{it}'U_{jt}\right)^2 \\
&=O_P\left(\frac{1}{n}\right).
\end{align*}

Lemma \ref{le.9} (iii) does provide a useful preliminary rate to build upon.  Applying Lemma \ref{le.9} (iii) and (\ref{ee.10a}), we obtain a preliminary rate 
$$
\frac{1}{n}\sum_{i=1}^n\|\widehat f_i^*-H^{*'}\widehat f_i\|_2^2=O_{P^*}\left(\Delta_F^2+\frac{1}{n}\right).
$$
Our goal is to remove the term $\frac{1}{n}$ through improving the bound for $\frac{1}{n}\sum_{i=1}^n\|A^*_{i1}\|_2^2$. By the triangle inequality,
\begin{align*}
\frac{1}{n}\sum_{i=1}^n\|A^*_{i1}\|_2^2 &\leq \frac{2}{n}\sum_{i=1}^n\left\|\frac{1}{pTn}\sum_{j=1}^nH^{*'}\widehat f_j\sum_{t=1}^T (\tilde U_{it}^{*'}\tilde U_{jt}^*- E^*\tilde U_{it}^{*'}\tilde U_{jt}^*)\right\|_2^2 \\
& \qquad + \frac{2}{n}\sum_{i=1}^n\left\|\frac{1}{n}\sum_{j=1}^n(\widehat f_j^*-H^{*'}\widehat f_j)\frac{1}{pT}\sum_{t=1}^T (\tilde U_{it}^{*'}\tilde U_{jt}^*- E^*\tilde U_{it}^{*'}\tilde U_{jt}^*)\right\|_2^2 \\
&\leq^{(a)} O_{P^*}(\Delta_F^2) + O_{P^*}(\Delta_F^2)\frac{1}{n^2}\sum_{ij}\left(\frac{1}{pT}\sum_{t=1}^T (\tilde U_{it}^{*'}\tilde U_{jt}^*- E^*\tilde U_{it}^{*'}\tilde U_{jt}^*)\right)^2 \\
&=O_{P^*}(\Delta_F^2) 
\end{align*}
 where in (a) we used Lemma \ref{le.11} and the last equality follows from (\ref{ee.13}). Hence combining with (\ref{ee.10a}), we have $ \frac{1}{n}\sum_{i=1}^n\|\widehat f_i^*-H^{*'}\widehat f_i\|_2^2=O_{P^*}(\Delta_F^2).$
Thus, we have $\Delta_F^*=\Delta_F.$

(ii)  We now verify the conditions in Assumption \ref{a6.2}:
 $\sqrt{nT}|J|^2_0\Delta_F^{*2}=o(1)$,     $\Delta_{eg}^*=o(\frac{1}{\sqrt{nT}})$,  $ \Delta_{ud}^*=o(\sqrt{\frac{\log p}{nT}}) $,   $|J|_0^2 \sqrt{\log p }      \Delta_{ud}^*=o(1)$,    $  \Delta_{F}^{*2}=o(\frac{\log p}{T \log (pT)}), $     and     $\Delta_{\max}^2|J|_0^2T\Delta_{F}^{*2}=o(1)$.
 
$\Delta_F^{*}$:  We have previously proven that $\Delta_F^2=\frac{1}{n^2}+ 
 \frac{1}{nT^2}+\frac{1}{pT}.$  In addition, Lemma \ref{lc.8} gives $ \Delta_{\max}=\frac{1}{\sqrt{n}}  +\sqrt{\frac{\log n}{Tp}}$. Hence it is straightforward to verify the required conditions involving $\Delta_F^*$, given the assumption that $|J|_0^2\log n=o(p)$.

$\Delta^*_{ud}$:  $\Delta_{ud}^*$ is defined in Lemma \ref{le.13} which gives $\Delta_{ud}^*=  b_n+\Delta_{ud}$ for  
$$  
b_n=  \frac{1}{T}\sqrt{\frac{\log p\log (np)}{np}} + \sqrt{\frac{\log n}{npT}}  +\frac{\log p}{n\sqrt{T}} +\frac{1}{pT}+\sqrt{\frac{\log p}{npT}}+\frac{\sqrt{\log p}}{n}+\frac{\sqrt{|J|_0}}{nT}+\sqrt{\frac{|J|_0}{npT}}.
$$   
In the proof of Proposition \ref{p4.1}, we verified  $ \Delta_{ud}=o\left(\sqrt{\frac{\log p}{nT}}\right)$ and $|J|_0^2 \sqrt{\log p }\Delta_{ud}=o(1)$.
 It is also straightforward to verify that $b_n=o\left(\sqrt{\frac{\log p}{nT}}\right) $ and that $|J|_0^2 \sqrt{\log p } b_n=o(1)$ given that $\log n=o(p)$, $|J|^4_0n=o(p^2T)$, $|J|_0^2\log^3p=o(n)$, and $|J|_0^4=o(nT^3)$. In particular, we need to  verify $|J|_0^5\log p\left(\frac{1}{n^2T^2}+\frac{1}{npT}\right)=o(1)$. To verify this condition, we use $|J|^4_0n=o(p^2T)$ and $|J|_0^4=o(nT^3)$ to show
\begin{align*}
|J|_0^5\log p\left(\frac{1}{n^2T^2}+\frac{1}{npT}\right) &= |J|_0^3\log p\left(\frac{|J|_0^2}{n^2T^2}+\frac{|J|_0^2n^{1/2}}{n^{3/2}pT}\right) \\
&=o(1)|J|_0^3\log p\left(\frac{n^{1/2}T^{3/2}}{n^2T^2}+\frac{ pT^{1/2}}{n^{3/2}pT}\right) \\
&=o(1) \frac{|J|_0^3\log p}{n^{3/2}T^{1/2}} \\
&=o(1)(\frac{|J|_0\log p}{n^{1/2}})^3=o(1).
\end{align*}

$\Delta_{eg}^*$:  Note that for $\widehat g_{tm} \in \{\widehat\Lambda_t'\widehat\gamma_d, \widehat\Lambda_t'\widehat\gamma_y, \widehat\delta_{dt},\widehat\delta_{yt} , \widehat\lambda_{tm}\}$, we have 
$$
\omega_n^*=\max_{m\leq p}\frac{1}{T}\sum_{t=1}^T \|\widehat g_{tm}\|_2^2=O_P(|J|_0^2).
$$ 
Hence by Lemma \ref{le.13a}, $\Delta_{eg}^{*2}=\left(\frac{1}{n^2}+\frac{\log n}{npT}+\frac{\log n}{n^2T^2}+\frac{\log^{1/2}n}{n^2T^{1/2}}\right)|J|_0^2.$  Given $|J|_0^2\log n=o(p)$ and $|J|_0^2T=o(n)$, it is then straightforward to verify $ \Delta_{eg}^{*2}=o\left(\frac{1}{nT}\right)$ which follows by verifying $\frac{|J|_0^2\log n}{nT}=o(1)$. To see $\frac{|J|_0^2\log n}{nT}=o(1)$, note that we have, by $|J|_0^{4/3}=o(n^{1/3}T)$,
$$
\frac{|J|_0^2\log n}{nT}=\frac{|J|_0^{2/3}|J|_0^{4/3}\log n}{nT}=o(1)\frac{|J|_0^{2/3}n^{1/3}T\log n}{nT}=o(1)\frac{|J|_0^{2/3}\log n}{n^{2/3}}=o(1).
$$
$\blacksquare$

\begin{lemma}\label{le.13a}
In the bootstrap sampling space, let  $\tilde z_{it}^*=\widehat z_{it}w_i^Z$ where $\{w_i^Z\}_{i=1}^n$ are  i.i.d. with mean zero and bounded variance and independent of $\{w_i^U\}$ and $\widehat z_{it}=\widehat\eta_{it}$ or $\widehat z_{it}=\widehat\epsilon_{it}$.  In addition, let $\{\widehat g_{tm}\}$ be a deterministic sequence (in the bootstrap sampling space) of vectors with a  fixed dimension, $m\leq p$.  Then for  $\omega_n^*=\max_{m\leq p}\frac{1}{T}\sum_{t=1}^T \|\widehat g_{tm}\|_2^2,$
$$
\max_{m\leq p}\left\|  \frac{1}{nT}\sum_{i=1}^n\sum_{t=1}^T\widehat g_{tm}\tilde z_{it}^*(\widehat f^*_i-H^{*'}\widehat f_i)'\right\|_F^2=
O_{P^*}\left(\frac{1}{n^2}+\frac{\log n}{npT}+\frac{\log n}{n^2T^2}+\frac{\log^{1/2}n}{n^2T^{1/2}}\right)\omega_n^{*}
$$
where the term $O_{P^*}\left(\frac{1}{n^2}+\frac{\log n}{npT}+\frac{\log n}{n^2T^2}+\frac{\log^{1/2}n}{n^2T^{1/2}}\right)\omega_n^{*}$ defines $\Delta_{eg}^{*2}$.
\end{lemma}

\proof It follows from (\ref{ee.9}) that 
$$
\max_{m\leq p}\|  \frac{1}{nT}\sum_{i=1}^n\sum_{s=1}^T\tilde z_{is}^*(\widehat f^*_i-H^{*'}\widehat f_i)\widehat g_{sm}'\|_F\leq \sum_{l=1}^3\bar C_l O_P(1),
$$
where each term $\bar C_l$ is defined and bounded below.

First, we have 
\begin{eqnarray*}
\bar C_1 &=& \max_{m\leq p} \left\|  \frac{1}{nT}\sum_{i=1}^n\sum_{s=1}^T\tilde z_{is}^*\frac{1}{pTn}\sum_{j=1}^n\widehat f_j^*\sum_{t=1}^T \tilde U_{it}^{*'}\tilde U_{jt}^*\widehat g_{sm}' \right\|_F\cr
         &\leq&   \left(\frac{1}{pT}\sum_t\|\frac{1}{n}\sum_{j=1}^n\widehat f_j^* \tilde U_{jt}^{*'}\|_F^2\right)^{1/2}\left(\frac{1}{T^2p} \sum_{s, t}\|\frac{1}{n}\sum_{i=1}^n\tilde z_{is}^*\tilde U_{it}^*\|_2^2 \right)^{1/2}\omega_n^{*1/2}\cr
         &=&O_{P^*}\left(\Delta_F+\frac{1}{\sqrt{n}}\right)\frac{1}{\sqrt{n}}\omega_n^{*1/2}\cr
				&=&\omega_n^{1/2} O_{P^*}\left(\frac{1}{\sqrt{npT}}+\frac{1}{n}\right),
              \end{eqnarray*}
where we used  
\begin{align}\label{ee.15}
\begin{split}
\frac{1}{pT}\sum_t\|\frac{1}{n}\sum_{j=1}^n\widehat f_j^* \tilde U_{jt}^{*'}\|_F^2 &\leq O_{P^*}(\Delta_F^2) + O_{P^*}(1)\frac{1}{pT}\sum_t\|\frac{1}{n}\sum_{j=1}^n\widehat f_j \tilde U_{jt}^{*'}\|_F^2 \\
&= O_{P^*}\left(\Delta_F^2+\frac{1}{n}\right), \\
\frac{1}{T^2p} \sum_{s, t}\|\frac{1}{n}\sum_{i=1}^n\tilde z_{is}^*\tilde U_{it}^*\|_2^2 &=
\frac{1}{T^2p} \sum_{s, t}\|\frac{1}{n}\sum_{i=1}^n\widehat  z_{is}w_i^Zw_i^U\widehat  U_{it}\|_2^2 \\
&=O_{P^*}\left(\frac{1}{n}\right),
\end{split}
\end{align}
and $\Delta_F=\Delta_F^*$.

For the second term, we have by Lemma \ref{le.11} that
\begin{align*}
\bar C_2 &= \max_{m\leq p}\left\|  \frac{1}{nT}\sum_{i=1}^n\sum_{s=1}^T\tilde z_{is}^*
\frac{1}{pTn}\sum_{j=1}^n\widehat f_j^*\sum_{t=1}^T\widehat f_j'\widehat \Lambda_t'\tilde U_{it}^*  \widehat g_{sm}'\right\|_F \\
&\leq O_{P^*}(\omega_n^{*1/2})  \left( \frac{1}{T}\sum_{s=1}^T \| \frac{1}{pTn}    \sum_{i=1}^n\sum_{t=1}^T \widehat \Lambda_t'\tilde U_{it}^* \tilde z^*_{is}\|_2^2\right)^{1/2} \\
&= O_{P^*}\left(\sqrt{\frac{\log n}{npT}}+\frac{\sqrt{\log n}}{nT}+ \frac{1}{n}+\frac{1}{n}(\frac{\log n}{T})^{1/4}\right)\omega_n^{*1/2}.
\end{align*}
Finally, 
\begin{align*}
\bar C_3 &= \max_{m\leq p}\left\|  \frac{1}{nT}\sum_{i=1}^n\sum_{s=1}^T\tilde z_{is}^*\frac{1}{pTn}\sum_{j=1}^n\widehat f_j^*\sum_{t=1}^T\widehat f_i'\widehat \Lambda_t'\tilde U_{jt}^*\widehat g_{sm}'\right\|_F \\
&\leq \left(\frac{1}{T}\sum_{s=1}^T( \frac{1}{n}\sum_{i=1}^n\tilde z_{is}^* \widehat f_i'  )^2\right)^{1/2}\omega_n^{*1/2} \left\|\frac{1}{nTp}\sum_{j=1}^n\sum_{t=1}^T\widehat f_j^*\widehat \Lambda_t'\tilde U_{jt}^*\right\|_2 \\
&= O_{P^*}\left(\omega_n^{*1/2}\frac{1}{\sqrt{n}}\right)\left(\Delta_F^2+\sqrt{\frac{\Delta_F^2\log n}{n}}\right)
\end{align*}
where we used Lemma \ref{ld.1} to obtain
$\frac{1}{T}\sum_t\|\frac{1}{n}\sum_{i=1}^n \tilde z_{it}^* \widehat f_i\|_2^2=O_{P^*}\left(\frac{1}{n}\right)$.
 
Combining the above, we have
$$
\max_{m\leq p}\|  \frac{1}{nT}\sum_{i=1}^n\sum_{t=1}^T\widehat g_{tm}\tilde z_{it}^*(\widehat f^*_i-H^{*'}\widehat f_i)'\|_F^2=
O_{P^*}\left(\frac{1}{n^2}+\frac{\log n}{npT}+\frac{\log n}{n^2T^2}+\frac{\log^{1/2}n}{n^2T^{1/2}}\right)\omega_n^{*}.
$$
 $\blacksquare$

\begin{lemma}\label{le.13}
For $\widehat h_{tk} \in \{\widehat \delta_{yt},\widehat \delta_{dt}, \widehat\lambda_{tk}\}$, we have 
\begin{align*}
\max_{m, k\leq p}&\left\|\frac{1}{nT} \sum_{i=1}^n  \sum_{t=1}^T(\widehat f_i^*-H^{*'}\widehat f_i) \tilde U_{it,m}^*  \widehat h_{tk}' \right\|_F \\
&\leq O_{P^*}\left(   \frac{1}{T}\sqrt{\frac{\log p\log (np)}{np}} + \sqrt{\frac{\log n}{npT}}  +\frac{\log p}{n\sqrt{T}} +\frac{1}{pT}+\sqrt{\frac{\log p}{npT}}+\frac{\sqrt{\log p}}{n}+\frac{\sqrt{|J|_0}}{nT}+\sqrt{\frac{|J|_0}{npT}}\right) \\
&\qquad + O_{P^*}(\Delta_{ud})
\end{align*}
for $\Delta_{ud}$ defined as in Lemma \ref{lc.7}.  The term on the right-hand-side of the inequality defines $\Delta_{ud}^{*}$.
%$ \max_{m, k\leq p}\|\frac{1}{nT} \sum_{i=1}^n  \sum_{t=1}^T(\widehat f_i^*-H^{*'}\widehat f_i) \tilde U_{it,m}^*  \widehat h_{tk}'  \|_F$ is  upper bounded by ($\Delta_{ud}$ is defined as in Lemma \ref{lc.7}): $$  O_{P^*}(   \frac{1}{T}\sqrt{\frac{\log p\log (np)}{np}} + \sqrt{\frac{\log n}{npT}}  +\frac{\log p}{n\sqrt{T}} +\frac{1}{pT}+\sqrt{\frac{\log p}{npT}}+\frac{\sqrt{\log p}}{n}+\frac{\sqrt{|J|_0}}{nT}+\sqrt{\frac{|J|_0}{npT}})+O_{P^*}(\Delta_{ud}).$$   This gives $\Delta_{ud}^*$.
\end{lemma}

\proof We have $ \max_{m, k\leq p}\|\frac{1}{nT} \sum_{i=1}^n  \sum_{s=1}^T(\widehat f_i^*-H^{*'}\widehat f_i) \tilde U_{is,m}^*  \widehat h_{sk}'  \|_F\leq \sum_{l=1}^5D_l$, where each $D_l$ is defined and bounded below. 
\begin{align*}
D_1 &= \max_{m, k\leq p} \left\| \frac{1}{T}\sum_{t=1}^T\frac{1}{n}\sum_{j=1}^n\widehat f_j^* \tilde U_{jt}^{*'} \frac{1}{nTp} \sum_{i=1}^n\sum_{s=1}^T(\tilde U_{it}^* \tilde U_{is,m}^*-E^*\tilde U_{it}^* \tilde U_{is,m}^*)  \widehat h_{sk}'  \right\|_F \\
&\leq \left(\frac{1}{T}\sum_{t=1}^T\|\frac{1}{n}\sum_{j=1}^n\widehat f_j^* \tilde U_{jt}^{*'}\|_2^2\right)^{1/2}\\
&\qquad \times \max_{m, k\leq p} \left(\frac{1}{T}\sum_{t=1}^T\|\frac{1}{nTp} \sum_{i=1}^n\sum_{s=1}^T(\tilde U_{it}^* \tilde U_{is,m}^*-E^*\tilde U_{it}^* \tilde U_{is,m}^*)  \widehat h_{sk}'\|_F^2\right)^{1/2} \\
&= O_{P^*}\left(\sqrt{\frac{\log (pT)}{n}(\frac{\log n\log (pT)}{n}+\frac{\log(np)}{T})}\right)\left(\Delta_F+\frac{1}{\sqrt{n}} \right)
\end{align*}
where we used Lemma \ref{le.11} and equation (\ref{ee.15}) from the proof of Lemma \ref{le.13a} which gives
$$
\frac{1}{pT}\sum_t\|\frac{1}{n}\sum_{j=1}^n\widehat f_j^* \tilde U_{jt}^{*'}\|_F^2\leq  O_{P^*}\left(\Delta_F^2+\frac{1}{n}\right).$$
Next, 
\begin{eqnarray*}    
D_2 &=&    \max_{m, k\leq p}\left\|\sum_{l=1}^p\frac{1}{T}\sum_{t=1}^T\frac{1}{pn}\sum_{j=1}^n\widehat f_j^* \tilde U_{jt,l}^{*'}\frac{1}{nT} \sum_{i=1}^n  \sum_{s=1}^TE^*(\tilde U_{it,l}^* \tilde U_{is,m}^*)  \widehat h_{sk}'  \right\|_F\cr
&\leq & \max_{tl}\left\|\frac{1}{n}\sum_{j=1}^n\widehat f_j^* \tilde U_{jt,l}^{*'}\right\|_2
       \max_{m, k\leq p}\frac{1}{pT}\sum_{l=1}^p\sum_{t=1}^T \left\|\frac{1}{nT} \sum_{i=1}^n  \sum_{s=1}^TE^*(\tilde U_{it,l}^* \tilde U_{is,m}^*)  \widehat h_{sk}'  \right\|_F \cr
&=& O_{P^*}\left(\sqrt{ \frac{\log (pT)}{n}}+\Delta_F\right)\left(\Delta_F+\frac{\sqrt{\log(pT)\log p}}{n}+ \sqrt{\frac{\log (pT)}{nT}}\right)
\end{eqnarray*}
where we used Lemma \ref{le.11} and $\max_{mt} \| \frac{1}{n}\widehat F^{*'}\tilde U_{t, m}^*\|_2 = O_{P^*}\left(\sqrt{ \frac{\log (pT)}{n}}+\Delta_F\right)$ due to Lemma \ref{ld.1}.    
We then have
\begin{eqnarray*}    
D_3&=&    \max_{m, k\leq p}\|\frac{1}{nT} \sum_{i=1}^n  \sum_{s=1}^T \frac{1}{pTn}\sum_{j=1}^n\widehat f_j^*\sum_{t=1}^T\widehat f_j'\widehat \Lambda_t'(\tilde U_{it}^* \tilde U_{is,m}^* -E^*\tilde U_{it}^* \tilde U_{is,m}^*) \widehat h_{sk}'  \|_F\cr
&\leq& O_{P^*} (1) \max_{m, k\leq p}\|\frac{1}{nT} \sum_{i=1}^n  \sum_{s=1}^T \frac{1}{pT}    \sum_{t=1}^T  \widehat \Lambda_t'(\tilde U_{it}^* \tilde U_{is,m}^* -E^*\tilde U_{it}^* \tilde U_{is,m}^*) \widehat h_{sk}'  \|_F\cr
&=& O_{P^*}\left(\sqrt{\frac{\log p}{n}(\frac{\log n\log (pT)}{n}+\frac{\log(np)}{T})\Delta_F^2}\right)
\end{eqnarray*}
by Lemma \ref{le.11}.  We also have
\begin{eqnarray*}
D_4 &=& \max_{m, k\leq p}\|\frac{1}{nT} \sum_{i=1}^n  \sum_{s=1}^T \frac{1}{pTn}\sum_{j=1}^n\widehat f_j^*\sum_{t=1}^T\widehat f_j'\widehat \Lambda_t'E^*\tilde U_{it}^* \tilde U_{is,m}^*  \widehat h_{sk}'  \|_F\cr
&\leq& O_{P^*} (1) \max_{m, k\leq p}\|\frac{1}{nT} \sum_{i=1}^n  \sum_{s=1}^T \frac{1}{pT}    \sum_{t=1}^T  \widehat \Lambda_t'E^*\tilde U_{it}^* \tilde U_{is,m}^*  \widehat h_{sk}'  \|_F\cr
&=& O_{P^*} \left(\frac{1}{pT}+\sqrt{\frac{\log p}{npT}}+\frac{\sqrt{\log p}}{n}+\frac{\sqrt{|J|_0}}{nT}+\sqrt{\frac{|J|_0}{npT}}\right)+ O_{P^*}(\Delta_{ud})
\end{eqnarray*}
where the inequality follows from Lemma \ref{le.12} (iii). 
Finally,
\begin{align*}    
D_5 &= \max_{m, k\leq p}\left\|  \frac{1}{pTn}\sum_{j=1}^n\widehat f_j^*\sum_{t=1}^T\widehat\Lambda_t'\tilde U_{jt}^* \frac{1}{nT}\sum_{s=1}^T \sum_{i=1}^n\widehat f_i\tilde U_{is,m}^*  \widehat h_{sk}'  \right\|_F \\
&\leq \left\|  \frac{1}{pTn}\sum_{j=1}^n\widehat f_j^*\sum_{t=1}^T\widehat\Lambda_t'\tilde U_{jt}^*\right\|_2\max_{m, k\leq p}\left\| \frac{1}{nT}\sum_{s=1}^T \sum_{i=1}^n\widehat f_i\tilde U_{is,m}^*  \widehat h_{sk}'  \right\|_F \\
&=^{(a)} O_{P^*}\left(\Delta_F^2+\sqrt{\frac{\Delta_F^2\log n}{n}}\right)\max_{m, k\leq p} \left\| \frac{1}{nT}\sum_{s=1}^T \sum_{i=1}^n\widehat f_i\tilde U_{is,m}^*  \widehat h_{sk}'  \right\|_F \\
&= O_{P^*}\left(\Delta_F^2+\sqrt{\frac{\Delta_F^2\log n}{n}}\right)
\end{align*}
where equality (a) results by applying Lemma \ref{le.11} (iii).  Note that the upper bound achieved in the last equality is not sharp but is sufficient to verify Assumptions about $\Delta_{ud}^*$.
 
Combining the above terms, we reach 
\begin{align*}   
\max_{m, k\leq p}&\left\|\frac{1}{nT} \sum_{i=1}^n  \sum_{s=1}^T(\widehat f_i^*-H^{*'}\widehat f_i) \tilde U_{is,m}^*  \widehat h_{sk}' \right\|_F \\
&= O_{P^*}\left(\sqrt{\frac{\log (pT)}{n}(\frac{\log n\log (pT)}{n}+\frac{\log(np)}{T})}\right)\left(\Delta_F+\frac{1}{\sqrt{n}} \right) \\
& \qquad + O_{P^*}\left(\sqrt{ \frac{\log (pT)}{n}}+\Delta_F\right)\left(\Delta_F+\frac{\sqrt{\log(pT)\log p}}{n}+ \sqrt{\frac{\log (pT)}{nT}}\right) \\
& \qquad + O_{P^*}(\Delta_{ud}+\Delta_F\sqrt{\frac{\log n}{n}}) +
  O_{P^*}\left(\Delta_F\sqrt{\frac{\log p}{n}(\frac{\log n\log (pT)}{n}+\frac{\log(np)}{T})}\right) \\
& \qquad +O_{P^*}\left(\frac{1}{pT}+\sqrt{\frac{\log p}{npT}}+\frac{\sqrt{\log p}}{n}+\frac{\sqrt{|J|_0}}{nT}+\sqrt{\frac{|J|_0}{npT}}\right) \\
&= O_{P^*}\left(   \frac{1}{T}\sqrt{\frac{\log p\log (np)}{np}} + \sqrt{\frac{\log n}{npT}}  +\frac{\log p}{n\sqrt{T}} +\frac{1}{pT}+\sqrt{\frac{\log p}{npT}}+\frac{\sqrt{\log p}}{n}+\frac{\sqrt{|J|_0}}{nT}+\sqrt{\frac{|J|_0}{npT}}\right) \\
& \qquad +O_{P^*}(\Delta_{ud})
\end{align*}
where $\Delta_{ud}$ is defined as in Lemma \ref{lc.7}. $\blacksquare$

\footnotesize\linespread{.8}
\bibliographystyle{ims}
\bibliography{liaoBib_newest}

\clearpage

\begin{figure}
	\centering
		\includegraphics[width = \textwidth]{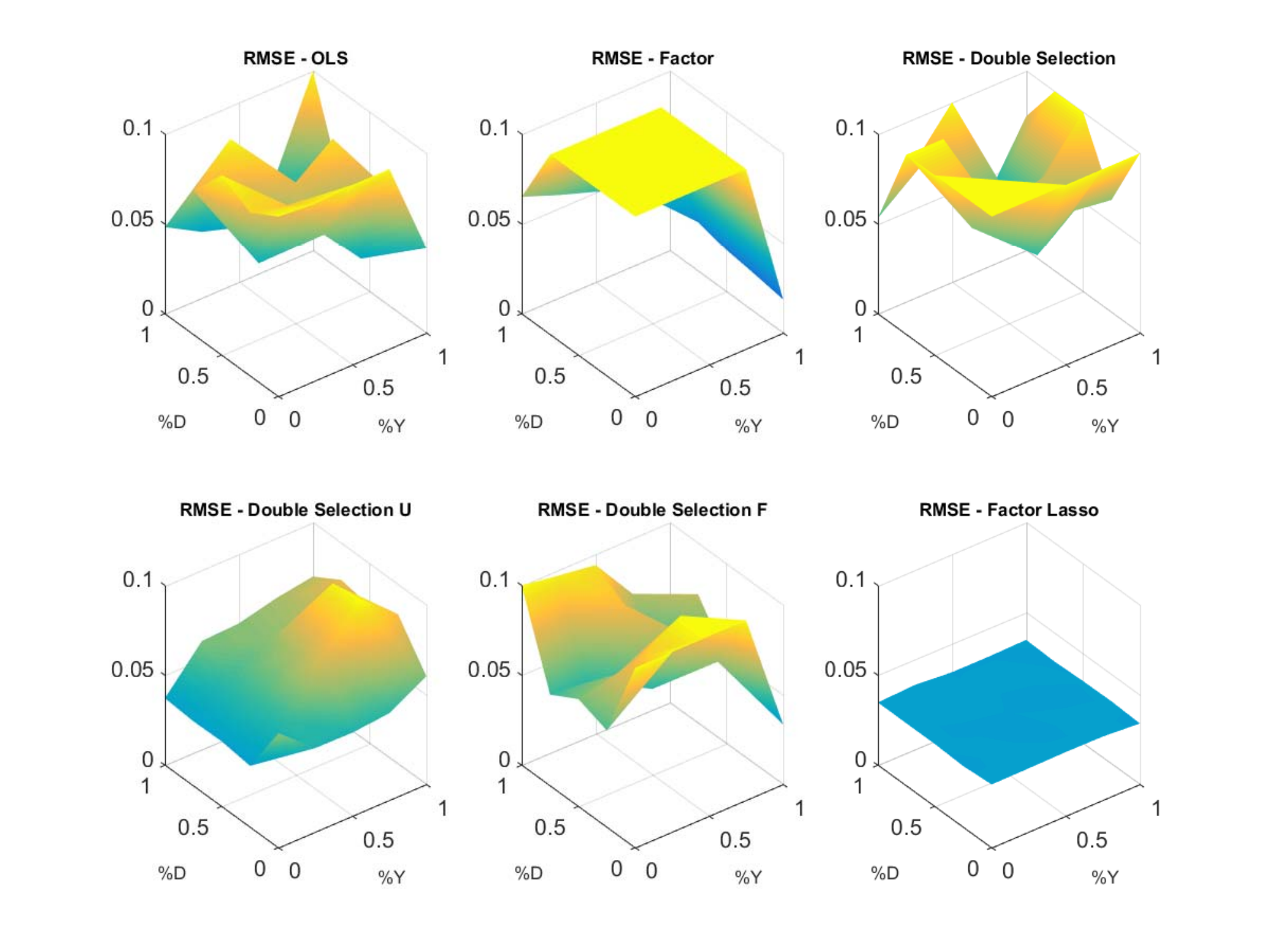}
	\caption{\footnotesize This figure shows the simulation RMSE of each of the estimators described in the text for estimating the coefficient of interest in a panel partial factor model. RMSE (truncated at 0.1) is shown in the vertical axis.  The horizontal axes give the fraction of the explanatory power in an infeasible regression of $Y$ on factors and factor residuals, ``\%Y,'' and the fraction of the explanatory power in an infeasible regression of $D$ on factors and factor residuals, ``\%D,'' where the infeasible regressions are described in the text.}
	\label{Fig: PPFM RMSE}
\end{figure}

\clearpage
     
\begin{figure}
	\centering
		\includegraphics[width = \textwidth]{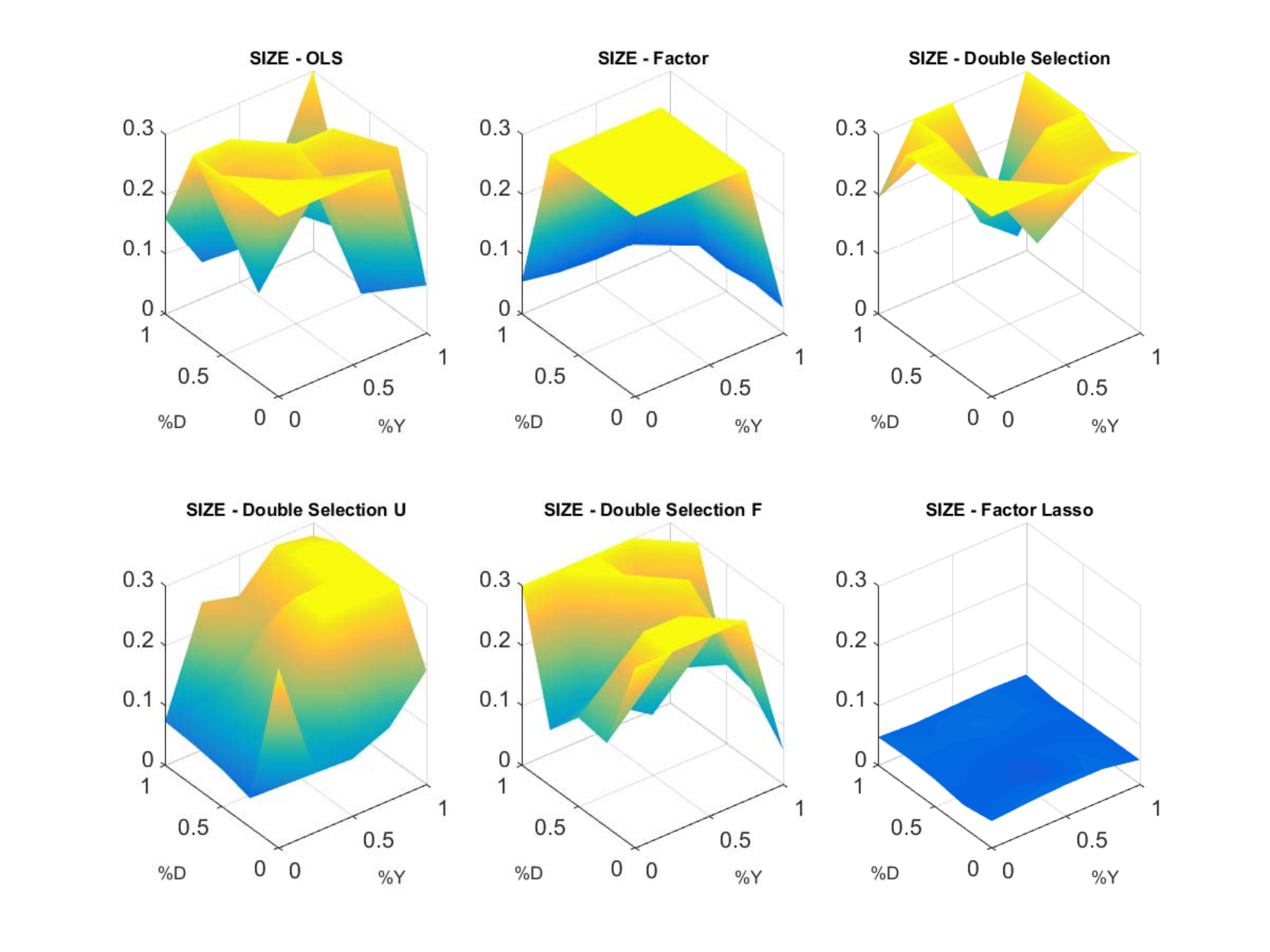}
	\caption{\footnotesize This figure shows the simulation size of 5\% level tests based on each of the estimators described in the text for the PPFM. Size (truncated at 0.3) is shown in the vertical axis.  The horizontal axes give the fraction of the explanatory power in an infeasible regression of $Y$ on factors and factor residuals, ``\%Y,'' and the fraction of the explanatory power in an infeasible regression of $D$ on factors and factor residuals, ``\%D,'' where the infeasible regressions are described in the text.}
	\label{Fig: PPFM SIZE}
\end{figure}

\clearpage 

\begin{figure}
	\centering
		\includegraphics[width = \textwidth]{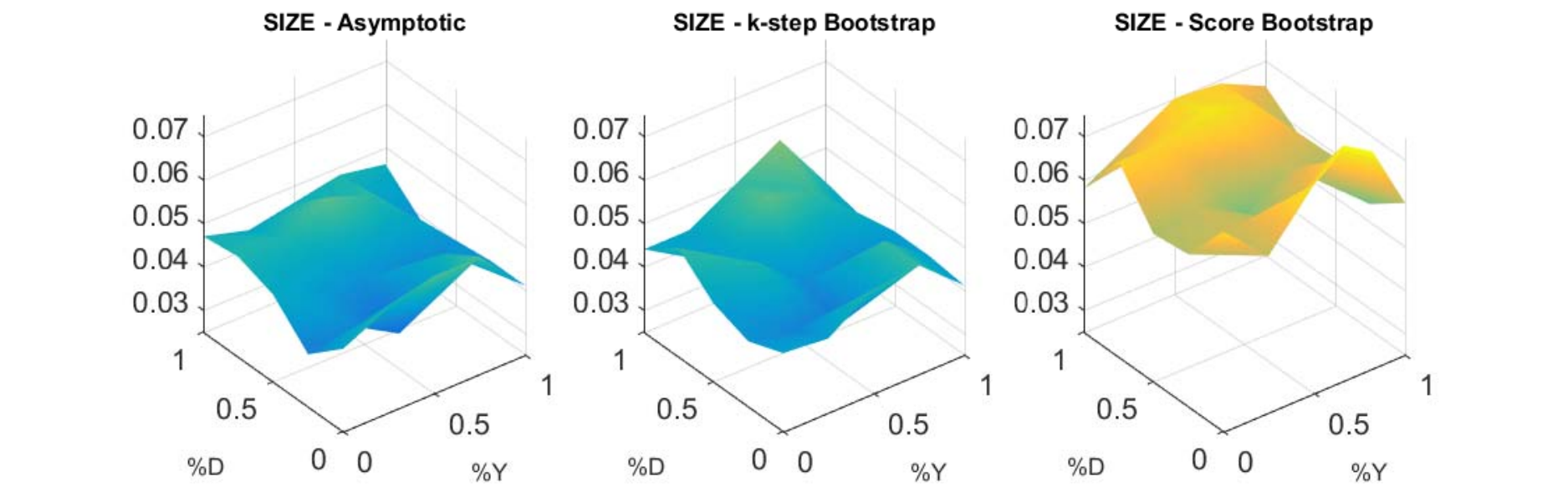}
	\caption{\footnotesize This figure shows the simulation size of 5\% level tests based on the factor-lasso estimator in the PPFM and the asymptotic Gaussian approximation, the k-step bootstrap, and a score based bootstrap.  Size is shown in the vertical axis.  The horizontal axes give the fraction of the explanatory power in an infeasible regression of $Y$ on factors and factor residuals, ``\%Y,'' and the fraction of the explanatory power in an infeasible regression of $D$ on factors and factor residuals, ``\%D,'' where the infeasible regressions are described in the text.}
	\label{Fig: PPFM BOOT}
\end{figure}

\clearpage

\begin{figure}
	\centering
		\includegraphics[width = \textwidth]{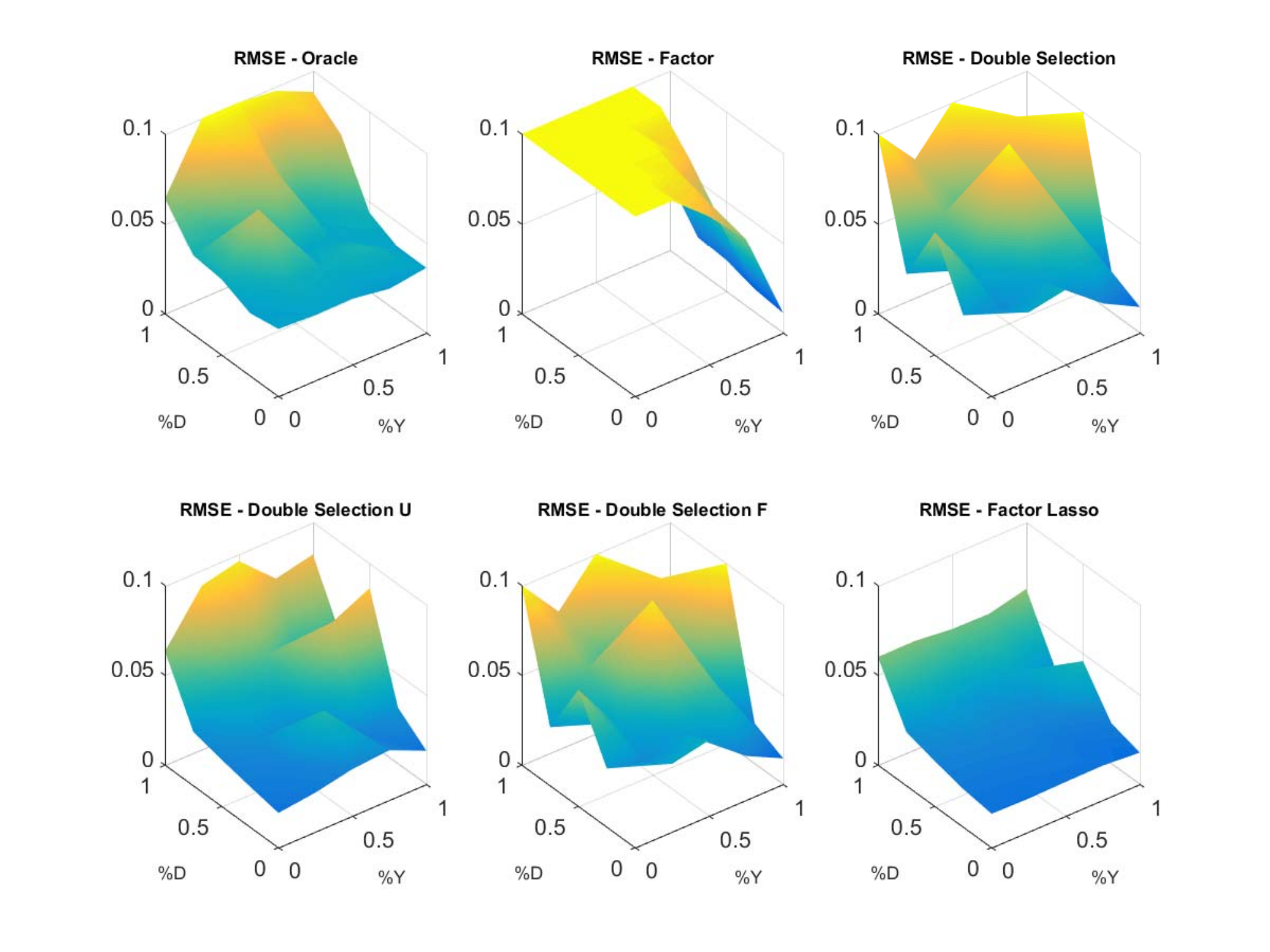}
	\caption{\footnotesize This figure shows the simulation RMSE of each of the estimators described in the text for estimating the coefficient of interest in an IV partial factor model. RMSE (truncated at 0.1) is shown in the vertical axis.  The horizontal axes give the fraction of the explanatory power in an infeasible regression of $Y$ on factors and factor residuals, ``\%Y,'' and the fraction of the explanatory power in an infeasible regression of $D$ on factors and factor residuals, ``\%D,'' where the infeasible regressions are described in the text.}
	\label{Fig: IV RMSE}
\end{figure}

\clearpage

\begin{figure}
	\centering
		\includegraphics[width = \textwidth]{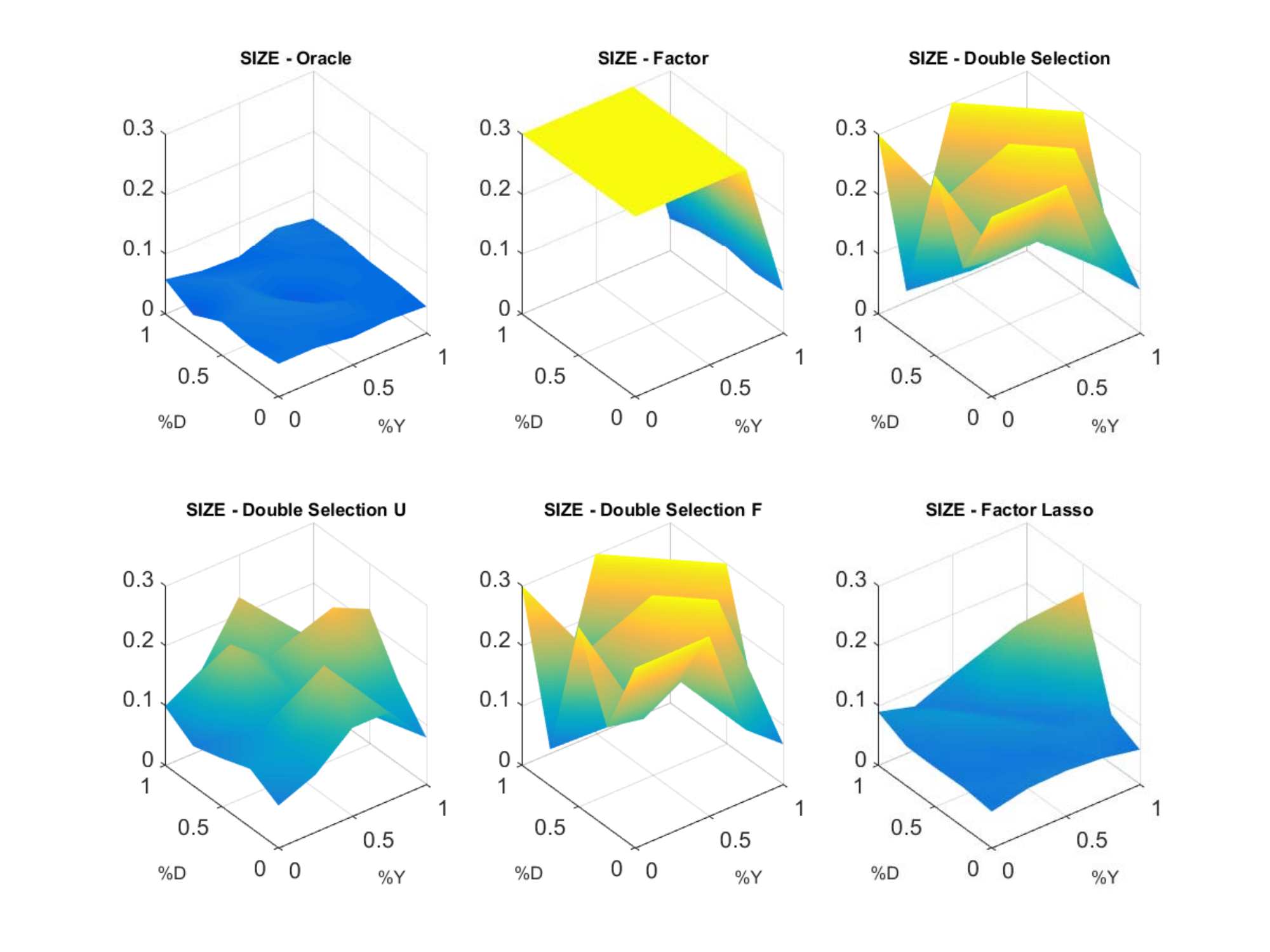}
	\caption{\footnotesize This figure shows the simulation size of 5\% level tests based on each of the estimators described in the text for the IV partial factor model. Size (truncated at 0.3) is shown in the vertical axis.  The horizontal axes give the fraction of the explanatory power in an infeasible regression of $Y$ on factors and factor residuals, ``\%Y,'' and the fraction of the explanatory power in an infeasible regression of $D$ on factors and factor residuals, ``\%D,'' where the infeasible regressions are described in the text.}
	\label{Fig: IV SIZE}
\end{figure}

\clearpage

\begin{figure}
	\centering
		\includegraphics[width = \textwidth]{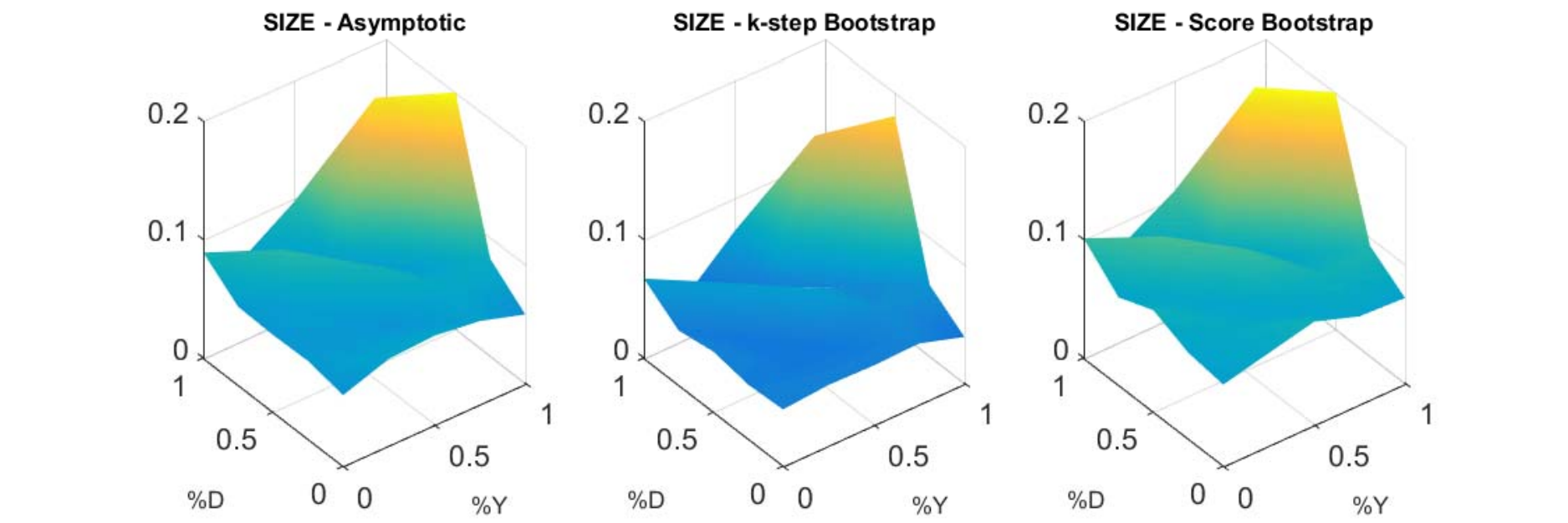}
	\caption{\footnotesize This figure shows the simulation size of 5\% level tests based on the factor-lasso estimator in the IV partial factor model and the asymptotic Gaussian approximation, the k-step bootstrap, and a score based bootstrap.  Size is shown in the vertical axis.  The horizontal axes give the fraction of the explanatory power in an infeasible regression of $Y$ on factors and factor residuals, ``\%Y,'' and the fraction of the explanatory power in an infeasible regression of $D$ on factors and factor residuals, ``\%D,'' where the infeasible regressions are described in the text.}
	\label{Fig: IV BOOT}
\end{figure}

\clearpage

\setlength{\tabcolsep}{10pt}
\begin{table}[h!]
\footnotesize{\caption{\footnotesize{Estimates of the Effect of Gun Prevalence on Homicide Rates}}\label{CLTableResults}
\begin{center}
\begin{tabular}{lccc}
\hline
  & Overall  & Gun & non-Gun  \\
\cline{2-4}
Cook and Ludwig (2006) Baseline & 0.086 (0.038)   & 0.173 (0.049)  & -0.033 (0.040) \\
Post Double Selection           & 0.062 (0.042)   & 0.138 (0.059)  & -0.055 (0.042) \\
                                & [-0.019,0.143]  & [0.036,0.240]  & [-0.139,0.029] \\
Factor                          & 0.104 (0.043)   & 0.210 (0.064)  & -0.022 (0.040) \\
                                & [0.019,0.189]   & [0.097,0.323]  & [-0.099,0.055] \\
Factor-Lasso                    & 0.069 (0.036)   & 0.167 (0.046)  & -0.048 (0.040) \\
                                & [0.000,0.138]   & [0.078,0.256]  & [-0.128,0.032] \\
\hline
\end{tabular}
\end{center}}
\begin{flushleft}
\footnotesize{This table presents estimates of the effect of gun ownership on homicide rates for a panel of 195 US Counties over the years 1980-1999.  The columns ``Overall'', ``Gun'', and ``non-Gun'' respectively report the estimated effect of gun prevalence on the log of the overall homicide rate, the log of the gun homicide rate, and the log of the non-gun homicide rate.  Each row corresponds to a different specification as described in the text.  In each specification, the outcome corresponding to the column label is regressed on lagged log(FSS) (a proxy for gun ownership) and additional covariates as described in the text.  Each specification includes a full set of year and county fixed effects.  Standard errors clustered by county are provided in parentheses.  k-step bootstrap 95\% confidence intervals are given in brackets.}
\end{flushleft}
\end{table}

\clearpage

\setlength{\tabcolsep}{10pt}
\begin{table}[h!]
\footnotesize{\caption{\footnotesize{Estimates of the First-Stage Relationship between Settler Mortality and Protection from Expropriation}}\label{AJRTableResults}
\begin{center}
\begin{tabular}{lccc}
\hline
  & $\widehat\pi$  & Estimated s.e. & Bootstrap C.I.  \\
\cline{2-4}
Latitude             & -0.549  &  (0.166)  & [-0.851,-0.246] \\
All Controls         & -0.218  &  (0.168)  & [-0.778,0.341]  \\
Double Selection     & -0.364  &  (0.178)  & [-0.885,0.158]  \\
Factor               & -0.475  &  (0.173)  & [-0.880,-0.070] \\
Factor-Lasso         & -0.353  &  (0.183)  & [-0.708.0.002]  \\
\hline
\end{tabular}
\end{center}}
\begin{flushleft}
\footnotesize{This table presents estimates of the coefficient on the instrument (Settler Mortality) in the first-stage regression of the endogenous variable from the \cite{acemoglu:colonial} example (Protection from Expropriation) on the instrument and geographic controls using different methods.  The row labeled ``Latitude'' uses the single variable distance from the equator to control for geography.  ``All Controls'' uses all 20 geographic controls without dimension reduction.  ``Double Selection'' uses the approach of \cite{belloni2014inference} to select important controls from among the 20 potential geography measures.  ``Factor'' reduces dimension through positing a conventional factor model.  ``Factor-Lasso'' makes use of the approach developed in this paper.  Point estimates from each method are provided in the column ``$\widehat\pi$'' and the associated estimated asymptotic standard errors are given in ``Estimated s.e.''.  The k-step bootstrap 95\% confidence interval is reported in ``Bootstrap C.I.''.}
\end{flushleft}
\end{table}

\end{document}